\documentclass[a4paper]{article}
\usepackage[]{graphicx}

\def\si {\mathop{\mathrm{si}}\nolimits}
\def\ci {\mathop{\mathrm{ci}}\nolimits}

\def\tr {\mathop{\rm tr}}

\title{Antiferromagnet-based nuclear spin model of scalable quantum register with inhomogeneous magnetic field}
\author{Alexander A.Kokin$^{1}$\footnote{E-mail: aakokin@mail.ru}~ and Vladimir A.Kokin$^{2}$}
\date{}
\begin{document}
\maketitle
\thanks{
$^{1}$Institute of Physics and Technology of RAS, 34, Nakhimovskii pr., 117218 Moscow, Russia;

$^{2}$Institute of Radioengineering and Electronics of RAS, 11, Mokhovaya str., 103907, Moscow, Russia
}
\begin{abstract} 
As a nuclear spin model of scalable quantum register, the one-dimensional chain of the magnetic atoms with nuclear spins 1/2 substituting the basic atoms in the plate of nuclear spin free easy-axis 3D antiferromagnet is considered. 
It is formulated the generalized antiferromagnet Hamiltonian in spin-wave approximation (low temperatures) considering the inhomogeneous external magnetic field, which is directed along the easy axis normally to plane of the plate and has a constant gradient along the nuclear spin chain. 
Assuming a weak gradient, the asymptotic expression for coefficients of unitary transformations to the diagonal form of antiferromagnet Hamiltonian is found. With this result the expression for indirect interspin coupling, which is due to hyperfine nuclear electron coupling in atoms and the virtual spin wave propagation in antiferromagnet ground state, was evaluated.

It is shown that the inhomogeneous magnetic field essentially modifies the characteristics of indirect interspin coupling. The indirect interaction essentially grows and even oscillates in relation to the interspin distance when the local field value in the middle point of two considered nuclear spin is close to the critical field for quantum phase transition of spin-flop type in bulk antiferromagnet or close to antiferromagnetic resonance.
Thus, the external magnetic field, its gradient, microwave frequency and power can play the role of control parameters for qubit states.

Finally, the one and two qubit states decoherence and longitudinal relaxation rate are caused by the interaction of nuclear spins with virtual spin waves in antiferromagnet ground state are calculated.

~



\end{abstract}

\section{Introduction}

In 1998 B.Kane proposed the scheme of large-scale (scalable) nuclear magnetic resonance (NMR) quantum computer with the register in the form of a regular chain of atoms $^{31}\mathrm{P}$ with nuclear spins 1/2 as qubits implanted into a near-surface layer of nuclear spin free substrate $^{28}$Si (Ref.\cite{1}). 
   \begin{figure}
   \begin{center}
   \begin{tabular}{c}
   \includegraphics{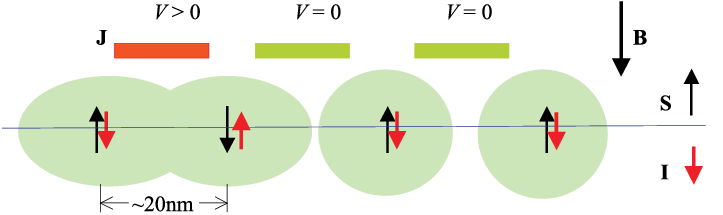}
   \end{tabular}
   \end{center}
   \caption[1]
   {\label{fig:1} 
The Kane's scheme of quantum register on nuclear spins of donor atoms $^{31}\mathrm{P}$.
\textbf{S} is electron spin,\textbf{ I} is nuclear spin (shot arrows),
\textbf{B} is external magnetic field, $V$ is electrical potential on gates \textbf{J}.
The gates \textbf{A} are not shown.
   }
   \end{figure}
It was assumed that the control of individual nuclear spin-qubits states is performed by electrical gates \textbf{A}, which change the electron-nucleus hyperfine interaction in atoms $^{31}\mathrm{P}$. The indirect nuclear spin coupling for neighboring donors is controlled by gates \textbf{J}, which change the overlap degree of donors wave functions (exchange integral). Two-qubit operations for far spatially separated qubits can be produced using SWAP quantum operations between neighboring qubits. The separation between neighboring donors in this scheme should be no more than 20 nm (Fig.~\ref{fig:1}).

The main difficulties encountered in realization of Kane's scheme were connected mainly with the necessity of using very low spin temperatures for nuclear spins,
using high precise technology for realization of quantum register structure and the complexity of quantum operation performing on states with random access to any qubits.
As an alternative to this semiconductor scheme,
a model for antiferromagnet-based NMR quantum cellular automata, which excludes the same difficulties of Kane's scheme, was proposed in our papers \cite{2,3}.

Authors of paper \cite{4} have proposed a variant of scalable quantum computer model, which does not use the gates of type \textbf{J}.
For quantum-information exchange between electron states of removed atoms $^{31}\mathrm{P}$ the implementation of spin wave propagation in the thin antiferromagnetic layer with easy axis of anisotropy grown on silicon layer with implanted donor atoms $^{31}\mathrm{P}$, was suggested.
For the local generation of spin wave at input of register and for signal detection in readout procedure the use of the scanning tunneling microscope was suggested.

\begin{figure}
   \begin{center}
   \begin{tabular}{c}
   \includegraphics{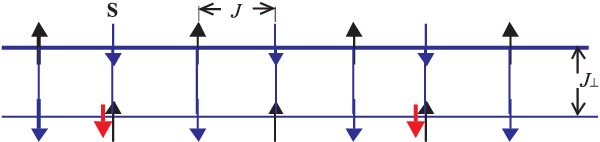}
   \end{tabular}
   \end{center}
   \caption[1]
   {\label{fig:2} 
The scheme of nuclear spin quantum register based on one dimension antiferromagnetic chain of two-lag ladder type [6]
   }
   \end{figure}

Another model of NMR quantum register based on two-leg ladder 1D antiferromagnet chain, where nuclear spins-qubits are placed in the inhomogeneous magnetic field and separated by a distance up to several tens of lattice constants, was proposed in papers \cite{5,6,7} (Fig.~\ref{fig:2}).
Relatively large inter-qubit distance in this case is effective both to diminish the nuclear dipole coupling between qubits and to give the considerable decoherence suppression.
It is assumed, that to achieve the required difference of resonance frequencies needed for individual access to individual qubits the gradient of external magnetic field along the spin chain can be used.
It is suggested also that two qubit operations on this register may be performed by switching the indirect interaction between longitudinal components of removed nuclear spins-qubits.
To do this, it is assumed to use of excitation of the spin wave packets by microwave pulses. The authors consider as the possible candidates for antiferromagnetic base a number of organic materials having energy gap for spin excitations.
They should consist of atoms of stable nuclear spin free isotopes like $^{12}\mathrm{C}$ (98.90\%), $^{16,18}$O (99.962\%),$^{14}$N (99.634\%) (in brackets their natural abundance is given). This also may be in particular same Haldane systems (Ref.\cite{6,7}).

At very low temperatures and in the absence of external magnon excitations the indirect interaction between longitudinal components of nuclear spins does not occur. However, the transverse part of interaction is different from zero and is determined by hyperfine nuclear spin and virtual magnon interaction. The theory of indirect nuclear interspin coupling in homogeneous isotropic antiferromagnet was the first given by T.Nakamura (Ref.\cite{8}) and than developed also in Refs.\cite{9,10,11}.

We have considered a model of NMR quantum register in the form of one-dimension chain of atoms with nuclear spins $I = 1/2$, placed on the thin plate of nuclear spin free 3D antiferromagnet with easy-axis anisotropy at low temperature. It is proposed that the external field is directed along the easy axis, which is normal to the plane of the antiferromagnet plate. In the case of homogeneous external magnetic field, the energy spectrum of spin waves (magnons) in easy axes antiferromagnet for field $B$ lower than the same critical value $B_{C}$ is characterized by the energy gap. We have shown (Ref.\cite{12}) that close to the critical point of spin-flop quantum phase transition the range of indirect nuclear spin coupling can run up to a great value. In the next papers \cite{13,14,15}, the previous model was extended into the case of external magnetic field with gradient along nuclear spin chain.

In present paper, we have presented the further investigations and development of this model, in which, as distinct from paper \cite{13}, the role of umklapp processes between distinct cells of reciprocal sublattice were accounted. Corrected and more refined analytical expressions were obtained and numerical calculations for the interspin indirect coupling were performed. It was shown that the character and magnitude of indirect coupling between nuclear spins essentially changes as the local external field in its middle point comes close to the critical field for homogeneous phase transition. The indirect interspin interaction at the antiferromagnetic resonance condition was also calculated. As a result, we have new fresh possibilities for control not only the individual nuclear spin resonance frequency, but also the interaction between separated spins without resort to many gate-controlling systems. The one-qubit and two-qubit decoherence and longitudinal relaxation rates are caused by the interaction of nuclear spins with virtual spin waves in antiferromagnet ground state were finally considered.

The paper is organized as follows. 
In Section 2 we discuss the structure of considered model with the Hamiltonian of easy axis antiferromagnet plate in inhomogeneous magnetic field.
In Section 3 we formulate the generalized antiferromagnet Hamiltonian in spin-wave approximation (low temperatures) with a constant gradient along the nuclear spin chain.
In Section 4 the diagonalization of antiferromagnet Hamiltonian is carried out.
In Section 5 asymptotic expressions for coefficients of unitary transformations to the diagonal form of antiferromagnet Hamiltonian are found for small gradient parameter.
In Section 6 we evaluate the expression for indirect coupling between transverse components of nuclear spins.
In Section 7 the indirect interaction, close to antiferromagnetic resonance condition, is evaluated. 
In Sections 8 and 9 the calculation of one and two qubit states decoherence and longitudinal relaxation rate are presented.

In Conclusion 10 we discuss the some prospects of considered quantum register model.
Some details of calculation are placed in the Appendixes.

\section{Spin Hamiltonian of easy-axis antiferromagnet plate in inhomogeneous magnetic field}

The antiferromagnet electron spin system of considered model plays here the role of an environment for nuclear spin quantum register,
whose interaction with spin wave leads on the one hand, to indirect coupling between nuclear spins and on the other hand to decoherence and relaxation processes of their states.

It may be proposed to use the natural antiferromagnetic crystals with easy-axis anisotropy as an antiferromagnet plate (or film, $d$ is its thickness).
As examples, they may be crystals $\mathrm{CeC}_{2}$ with tetragonal and $\mathrm{FeCO}_{3}$ (siderite) with trigonal symmetry.
The basic isotopes of these crystals $^{12}\mathrm{C}$, $^{56}\mathrm{Fe}$ (91.7\%), $^{16}$O, $^{140,142}\mathrm{Ce}$ (99.6\%) 
have no nuclear spins (in brackets percent isotopic abundance is given). 
To form the one-dimension nuclear spin chains the isotopic substitution atoms, such as $^{12}\mathrm{C}$ in corresponding crystal lattice sites, for isotopes $^{13}\mathrm{C}$ with nuclear spins 1/2, are proposed.
One would expect that period of such solid state NMR quantum registers may be much more than periods of crystal lattice.

The simple antiferromagnet model to be studied here consists of two incorporated to each other tetragonal magnetic sublattices \textbf{A} and \textbf{B} with $N = N_{\bot} N_{z}$ sites in each sublattice,
where $N_{\bot}= N_{x} N_{y} \gg 1$ are the sites numbers in plane of plate ($x$,$y$-axes) 
and $N_{z} > 1$ is the sites numbers in $z$ direction. 
The atom sites of sublattices are numbered respectively by numbers $j$ and $i$. 
Each sublattice constant in the plane of the plate is $a_{\bot}$ and along symmetry axis is $a_{z}$.

\begin{figure}
   \begin{center}
   \begin{tabular}{c}
   \includegraphics{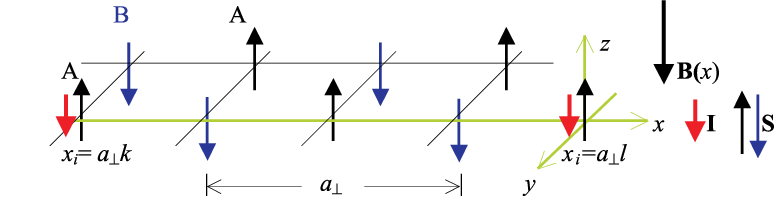}
   \end{tabular}
   \end{center}
   \caption[1]
   {\label{fig:3} 
The scheme of antiferromagnet based nuclear spin quantum register
 in external field lower than for spin-flop phase transition $B\left( {x} \right) \ll B_{C}$. 
The oriented (that is nonprecessing) arrows represent here the ground states of corresponding individual Bloch vectors.
The nuclear spins $\mathbf{I}_{\mathrm{A}}$ (shot red arrows) are contained here only in atoms of sublattice \textbf{A}
   }
   \end{figure}

The external magnetic field $B\left( {x} \right)$ in considered model (Fig.~\ref{fig:3}) is directed parallel to $z$-axis 
and to the easy axis. As calculations (Ref.\cite{16}) show, the field gradient $dB_{z} \left( {x} \right)/dx = G$ 
of the order of $1.4\,\mathrm{T}/\mu \mathrm{m}$, may be obtained 
by using dysprosium micro magnet with dimensions $10 \times 4 \times 400\,\mathrm{m}^{3}$ 
at a distance of $2.07\,\mathrm{m}$. We assume here that the field has somewhat lesser constant gradient 
$G\sim 0.1\,\mathrm{T}/\mu \mathrm{m}$ along nuclear spin chain ($x$-axis in plane of the plate). 
This value of the gradient corresponds to the difference of resonance frequencies of the order of $100\,\mathrm{kHz}$ 
for two nuclear spins, being separated by $100\,a_{\bot}$  $\left( a_{\bot}  \sim 1\,\mathrm{nm} \right)$.

The qubits number in quantum register will be limited by planar structure dimensions.
For example, the structure with linear dimension of the order of $10\,\,\mu \mathrm{m}$ 
will have 100 qubit register with period $L = 100 a_{\bot}$.

The starting electron spin Hamiltonian of 3D easy-axis antiferromagnet with interaction only between neighbouring atoms, which belong to the distinct sublattice, is represented for our model as 
\begin{equation}
\label{eq1}
H_{S} = \gamma_{S} \hbar \left( {\;\sum\limits_{i}^{N} {B\left( {x_{i}} \right)\;S_{\mathrm{A}z} \left( {\mathbf{r}_{i}} \right)} + \sum\limits_{j}^{N} {B\left( {x_{j}} \right)\;S_{\mathrm{B}z}} \left( {\mathbf{r}_{j}} \right)\;} \right) 
+ 
\end{equation}
$$ 
+ 2\gamma_{S} \hbar /Z\;\;\sum\limits_{i}^{N} {\;\sum\limits_{\delta}^{Z} {\{ \;B_{E} \;\mathbf{S}_{\mathrm{A}} \left( {\mathbf{r}_{i}} \right)\;\mathbf{S}_{\mathrm{B}} \left( {\mathbf{r}_{i} + \mathbf{r}_{\delta }} \right)\; + \;B_{\mathrm{A}} \;S_{\mathrm{A}z} \left( {\mathbf{r}_{i}} \right)\;S_{\mathrm{B}z} \left( {\mathbf{r}_{i} + \mathbf{r}_{\delta} } \right)\} }},  
$$
\noindent
where $\mathbf{S}_{\mathrm{A}} \left( {\mathbf{r}_{i}} \right)$ and $\mathbf{S}_{\mathrm{B}} \left( {\mathbf{r}_{i} + \mathbf{r}_{\delta} } \right)$ 
are electron spin operators ($S = 1/2$) for neighbouring sites of sublattices \textbf{A} and \textbf{B}, 
$B\left( {x_{i}} \right) = B + Gx_{i}$, $B$ is the field value at the origin of the coordinates 
$x_{i}$, $Z = 6$ is the number of neighbouring atoms for tetragonal sublattice, 
$\gamma_{S} = 175.88\,\mathrm{rad\,GHz/T}\,\,\,\,\left( {\gamma_{S} /2\pi = 28\,\mathrm{GHz/T}} \right)$ 
is electron spin gyromagnetic ratio, $\hbar = 1.054 \cdot 10^{-34}\,\mathrm{Js/rad}$.

Note, that the product of spin operators in Eq.(\ref{eq1}) 
written in matrix representation is common designated by symbol $\otimes$. 
In the following this symbol will be for brevity omitted.

In Eq.(\ref{eq1}) the parameters 
$B_{E} \sim 10 - 100\;\mathrm{T}$, 
$B_{\mathrm{A}} \sim 10^{-2} - 1\;\mathrm{T} > 0$, 
$B_{C} = \sqrt {2B_{E} B_{\mathrm{A}} + B_{\mathrm{A}}^{2}}$ are exchange field, anisotropy field and spin-flop field for easy-axis antiferromagnet (particularly, 
for $\mathrm{FeCO}_{3}$: $B_{E} = 35\;\mathrm{T}$, $B_{\mathrm{A}} = 3.3\;\mathrm{T}$, $B_{C} \approx 15.5\,\mathrm{T}$).

\section{Spin Hamiltonian of antiferromagnetic plate in spin wave approximation}

The radius vectors of positions for magnetic atoms (for example, for atoms \textbf{A})
will be presented as $\mathbf{r}_{i} \to \left( {a_{\bot} x_{i}, a_{\bot} y_{i,}, a_{z} z_{i}} \right)$,
where $\left( {x_{i} ,y_{i}} \right),z_{i} = \mathbf{r}_{\bot}$ , $z$ are now dimensionless radius-vector values. 
The dimensionless volume of elementary sublattice cell and of reciprocal sublattice cell in this case becomes equal to unity $\left( {v_{a} = 1} \right)$ and the plate dimensions in three perpendicular directions are coincident correspondingly with the site numbers $N_{x} ,N_{y} ,N_{z}$.

Because of large values $N_{x},N_{y}$, antiferromagnet spin states are essentially independent of boundary conditions at interface plate; it is convenient to use the Born-Karman periodic conditions for them.

The local spin operators of distinct sublattice atoms are expressed in terms of wave representation as follows:
$$ 
 S_{\mathrm{A}x} \left( {\mathbf{r}_{i}} \right) - iS_{\mathrm{A}y} \left( {\mathbf{r}_{i}} \right) = S_{\mathrm{A}}^{-} \left( {\mathbf{r}_{i}} \right) \equiv a_{i}^{+} = 1/\sqrt{N} \sum\limits_{\mathbf{q}} {a_{\mathbf{q}}^{+} \exp\left( {\;\,i\mathbf{q}\mathbf{r}_{i}} \right)} ,
$$
\begin{equation}
\label{eq2} 
 S_{\mathrm{A}x} \left( {\mathbf{r}_{i}} \right) + iS_{\mathrm{A}y} \left( {\mathbf{r}_{i}} \right) = S_{\mathrm{A}}^{+} \left( {\mathbf{r}_{i}} \right) \equiv a_{i}\;\, = 1/\sqrt{N} \sum\limits_{\mathbf{q}} {a_{\mathbf{q}}\;\, \exp\left( {  - i\mathbf{q}\mathbf{r}_{i}} \right)} ,
\end{equation}
$$ 
 S_{\mathrm{B}}^{-} \left( {\mathbf{r}_{j}} \right) \equiv b_{j}\;\, = 1/\sqrt{N} \sum\limits_{\mathbf{q}} {b_{\mathbf{q}}\;\, \exp\left( {  - i\mathbf{q}\mathbf{r}_{j}} \right)} , \,\,\,
 S_{\mathrm{B}}^{+} \left( {\mathbf{r}_{j}} \right) \equiv b_{j}^{+} = 1/\sqrt{N} \sum\limits_{\mathbf{q}} {b_{\mathbf{q}}^{+} \exp\left( {\;\,i\mathbf{q}\mathbf{r}_{j}} \right)} , 
$$
\noindent
with
\begin{equation}
\label{eq3}
S_{\mathrm{A}z} \left( {\mathbf{r}_{i}} \right) =\quad 1/2 - a_{i}^{+}  a_{i} =\quad 1/2 - \left( {1/N} \right)\sum\limits_{\mathbf{q},{\mathbf{q}}'} {a_{{\mathbf{q}}'}^{+}  a_{\mathbf{q}}}  \exp\left( {i\left( {{\mathbf{q}}' - \mathbf{q}} \right)\mathbf{r}_{i}} \right), 
\end{equation}
$$ 
S_{\mathrm{B}z} \left( {\mathbf{r}_{j}} \right) =    - 1/2 + b_{j}^{+}  b_{j} =    - 1/2 + \left( {1/N} \right)\sum\limits_{\mathbf{q},{\mathbf{q}}'} {b_{{\mathbf{q}}'}^{+}  b_{\mathbf{q}}}  \exp\left( {i\left( {{\mathbf{q}}' - \mathbf{q}} \right)\mathbf{r}_{j}} \right).
$$

The operators $a_{\mathbf{q}}^{+}$ and $a_{\mathbf{q}}$ act as operators of creation and annihilation of spin waves (magnons) 
with dimensionless wave vector $q$ (magnon quasi-moment) for sublattice \textbf{A}
and operators $b_{\mathbf{q}}^{+}$ and $b_{\mathbf{q}}$ accordingly for sublattice \textbf{B}.
These operators do not satisfy the Bose commutation relations.
However, it is known that by means of Holstein-Primakoff transformation they be converted to magnon operators with boselike commutation relations.

The operators in site representation $a_{i}^{+}a_{i}$, and $b_{j}^{+}b_{j}$ in the second terms in Eq.(\ref{eq3}) have the eigenvalues 0 and 1.
They describe the number of flopped spins for one site of sublattice. The mean electron spin state deviations
at sufficiently low temperature in vicinity of ground state $T \ll \gamma_{S} \hbar \left( {B_{C} - B} \right)/k_{\mathrm{B}}$ ($k_{\mathrm{B}}$ is Boltzmann constant) 
are small, that is $\langle 0|a_{i}^{+} a_{i} |0\rangle \ll 1$, $\langle 0|b_{i}^{+}  b_{i} |0\rangle \ll 1$.

In this case, one may belief that spin waves are created predominantly by the transverse components of electronic spins (therein lies the spin-wave approximation). Spin wave operators take approximately the boselike form without using the Holstein-Primakoff transformation. The commutation relations for local Pauli operators take form (Ref.\cite{17}):

\begin{equation}
\label{eq4}
 \left[ {a_{i} ,\;a_{{i}'}^{+}} \right] = \delta_{i,{i}'}, \quad\quad\quad\left[ {b_{i}, b_{{i}'}^{+} } \right] = \delta_{i,{i}'} , 
\end{equation}
$$ 
 \left[ {a_{i} ,\;a_{{i}'}} \right] = \left[ {a_{i} ,\;b_{{i}'}} \right] = \left[ {a_{i}, b_{{i}'}^{+} } \right] = 0,\quad\quad\left[ {b_{i} ,\;a_{{i}'}^{+} } \right] = 0\,.\
$$

Suggesting than that spin states modes have equal values at the interface of plate,
we will have for transverse component of wave vector for thin plate the discrete values
($d$ is the thickness of plate):
\begin{equation}
\label{eq5}
q_{z} \left( {n} \right) = 2\pi n_{z} /N_{z} ,\;\;n_{z} = 0,\;1,\;2,\,\dots ,\,N_{z} - 1,\,\,\,\,\,\,
N_{z} = d/a_{z} > 1,
\end{equation}
\noindent
that is these spin modes in $z$-direction are the standing wave.
Let us consider the plates with thickness $d$ whereby the nonzero transverse modes with $q_{z} \ne 0$ 
are not excited (conditions see Sec.~7) and the zero-order transverse spin mode with $q_{z} = 0$ 
is only populated. In this case, the spin waves are propagated only along the plate surface 
with two-dimensional wave vectors $\mathbf{q}_{\bot}  \equiv \left( {q_{x} ,q_{y}} \right)$, 
where $q_{x} ,\,\,q_{y} = 2\pi n_{x,y} /N_{x,y}$. 
For even $N_{x,y}$ we have values $n_{x,y} = 0,\; \pm 1,\; \pm 2,\dots ,\,\, \pm \left( {N_{x,y} /2 - 1} \right),\,\,N_{x,y} /2$,
that is components of wave vector sweep all states in one cell of reciprocal sublattice (the first Brillouin zone) with $0 \le \,\,|q_{x} |,\,\,\,|\,q_{y} |\,\, \le \pi$.
The sites of atoms in the sublattice \textbf{A} (and similarly in sublattice \textbf{B}) are determined now by two-dimensional vectors $\mathbf{r}_{\bot ,i} = \left( {x_{i} ,y_{i}} \right)$.

Let us go from three-dimensional magnon operators to two-dimensional operators 
$a_{\mathbf{q}}^{+}  = a_{\mathbf{q}_{\bot}}^{+} \delta_{q_{z} ,0}$,
$b_{\mathbf{q}}^{+}  = b_{\mathbf{q}_{\bot}}^{+} \delta_{q_{z} ,0}$. 
As a result, we come to a simplified quasi-two-dimensional Hamiltonian for the antiferromagnetic plate.

Then taking into account that the distribution of plane wave vectors $\mathbf{q}_{\bot}$ for $N_{x,y} \gg 1$ 
is practically continuous, we will go from the sums to integrals and from Kronecker's symbol 
to Dirac's $\delta-$function (Appendix 2 in Ref.\cite{17}):
\begin{equation}
\label{eq6}
 \sum\limits_{\mathbf{q}_{\bot},q_{z} = 0} {F_{\mathbf{q}_{\bot} } }  = 
 \sum\limits_{\mathbf{q}_{\bot} }  {F_{\mathbf{q}_{\bot} } }  \to N_{\bot}  /\left( {2\pi} \right)^{2}\int {F\left( {\mathbf{q}_{\bot} } \right)d\mathbf{q}_{\bot} }  ,\,\,\,\,\,\sum\limits_{\mathbf{q}_{\bot}}  {1} \to N_{\bot}  /\left( {2\pi} \right)^{2}\int {d\mathbf{q}_{\bot}  =}  N_{\bot}  
\end{equation}
$$ 
 \delta_{\mathbf{q}_{\bot},{\mathbf{q}}'_{\bot}}  \to \left( {2\pi} \right)^{2}/N_{\bot}  \;\delta \left( {\mathbf{q}_{\bot}  - {\mathbf{q}}'_{\bot} } \right). 
$$
In the general case, magnon quasi-moment $\mathbf{q}_{\bot}$ is defined modulo 
a vector of the reciprocal plane sublattice $\mathrm{Q}_{\bot}$. 
This essential property was ignored in our early works \cite{12,13}. 
Considering only $q_{z} = 0$ and by summing over $N$ sites of direct sublattice 
in terms of Hamiltonian (\ref{eq1}) after substitution of Eqs.(\ref{eq2}), (\ref{eq3}), we will have:
\begin{equation}
\label{eq7}
1/N\sum\limits_{i = 1}^{N} {\exp\left( {i\left( {{\mathbf{q}}' - \mathbf{q}} \right)\mathbf{r}_{i}}\right) }   =
 1/N_{\bot}  \sum\limits_{i = 1}^{N_{\bot} }  {\exp\left( {i\left( {{\mathbf{q}}'_{\bot}  - \mathbf{q}_{\bot} } \right)\mathbf{r}_{\bot ,i}} \right)}   = 
 \sum\limits_{\mathrm{Q}} {\delta_{\left( {{\mathbf{q}}'_{\bot}  - \mathbf{q}_{\bot} } \right),\,\mathrm{Q}_{\bot} } }  ,.
\end{equation}
\noindent
where $\mathrm{Q}_{\bot} = \left( {Q_{x} = 2\pi m_{x} ,\,\,Q_{y} = 2\pi m_{y}} \right)$, $m_{x,y} = 0,\, \pm \,1,\,\, \pm 2,\,\dots ,\,\, \pm \infty$.

Because the external magnetic field is inhomogeneous along $x$-axis, quasi-moment components $q_{x}$, unlike components $q_{y}$, $q_{z}$, are not integrals of motion and magnon states should be defined as superposition of states for $q_{x}$ and ${q}'_{x} = q_{x} + Q_{x}$. At the same time, displacement of components $q_{y}$ on $Q_{y}$ represents simply umklapp processes of states in other elementary cells of reciprocal sublattice, which represents the identical transformation and therefore then it will be assumed $Q_{y} = 0$. Thus in the following, only the values $\mathrm{Q}_{\bot}  = \left( {Q_{x} = 2\pi m_{x} ,\,\,Q_{y} = 0} \right)\,\,\,$will be considered. Taking in to account the above-mentioned remarks we will write Eq.(\ref{eq7}) in the form
\begin{equation}
\label{eq8}
\sum\limits_{\mathrm{Q}_{\bot} }  {\delta_{\left( {{\mathbf{q}}'_{\bot}  - \mathbf{q}_{\bot} } \right),\,\mathrm{Q}_{\bot} } }  = \sum\limits_{Q_{x}} {\,\,\,\delta_{\left( {{q}'_{x} - q_{x}} \right),\,Q_{x}}  \delta_{{q}'_{y} ,\,\,\,q_{y}} }  .
\end{equation}

Let us go also in the Eq.(\ref{eq8}) to continuous variables for $\mathbf{q}_{\bot}$ 
but retain discrete values for $Q_{x}$. We will obtain:
\begin{equation}
\label{eq9}
\sum\limits_{Q_{x}}  {\delta_{\left( {{q}'_{x} - q_{x}} \right),\,Q_{x}}  \delta_{{q}'_{y} ,\,\,\,q_{y}} }  \, \to \sum\limits_{\mathrm{Q}_{\bot} }  {\left( {2\pi} \right)^{2}/N_{\bot}  \delta \left( {{q}'_{x} - q_{x} - Q_{x}} \right)} \delta \left( {{q}'_{y} - q_{y}} \right).
\end{equation}
\noindent
and
\begin{equation}
\label{eq10}
1/N_{\bot}  \sum\limits_{i = 1}^{N_{\bot} }  {x_{i} \exp\left( {i\left( {{\mathbf{q}}'_{\bot}  - \mathbf{q}_{\bot} } \right)\mathbf{r}_{\bot i}} \right) } \to - i\left( {2\pi} \right)^{2}/N_{\bot}  \;\sum\limits_{\mathrm{Q}_{\bot} }  {\partial \delta \left( {{q}'_{x} - q_{x} - Q_{x}} \right)/\partial q_{x}}  \,\delta \left( {{q}'_{y} - q_{y}} \right).
\end{equation}

We will then go to magnon operators with continuous arguments $\mathbf{q}_{\bot}$.
For example, for creation operators of magnon state with wave vector $\mathbf{q}_{\bot}$ 
in the range from $\mathbf{q}_{\bot}$ to $\mathbf{q}_{\bot} + d\mathbf{q}_{\bot}$, we will write
\begin{equation}
\label{eq11}
a_{\mathbf{q}_{\bot} }^{+}  \to \,2\pi /\sqrt {N_{\bot} }  a^{+} \left( {\mathbf{q}_{\bot} } \right),\,\,\,\,b_{\mathbf{q}_{\bot} }^{+}  \to \,2\pi /\sqrt {N_{\bot}  } b^{+} \left( {\mathbf{q}_{\bot} } \right).
\end{equation}

By the use of Eqs.(\ref{eq7}), (\ref{eq9}) and (\ref{eq10}),
we will find the expressions of sums, in spin Hamiltonian (\ref{eq1}) 
involved (in a like manner for operators $b_{{\mathbf{q}}'_{\bot} }^{+}$, $b_{\mathbf{q}_{\bot}}$ and other operator pairs):
$$ 
\left( {1/N_{\bot} } \right)\sum\limits_{i = 1}^{N_{\bot}} 
  {\sum\limits_{\mathbf{q}_{\bot}  ,{\mathbf{q}}'_{\bot} }  {a_{{\mathbf{q}}'_{\bot} }^{+}  a_{\mathbf{q}_{\bot} } }  
  \exp\left( {i\left( {{\mathbf{q}}'_{\bot}  - \mathbf{q}_{\bot} } \right)\mathbf{r}_{\bot ,i}} \right)} = 
  \sum\limits_{\mathbf{q}_{\bot}  ,Q_{x}}  {a_{q_{x} + Q_{x} ,q_{y}}^{+}  a_{\mathbf{q}_{\bot} } } 
\to 
$$
$$ 
\to \int {\sum\limits_{Q_{x}}  {a^{+} \left( {q_{x} + Q_{x} ,q_{y}} \right)a\left( {\mathbf{q}_{\bot} } \right)d\mathbf{q}_{\bot} } }  ,   
$$
$$ 
\left( {1/N_{\bot} } \right)\sum\limits_{i = 1}^{N_{\bot} }  
  {\sum\limits_{\mathbf{q}_{\bot}  ,{\mathbf{q}}'_{\bot} }  {x_{i} a_{{\mathbf{q}}'_{\bot} }^{+}  a_{\mathbf{q}_{\bot} } }  
  \exp\left( {i\left( {{\mathbf{q}}'_{\bot} - \mathbf{q}_{\bot} } \right)\mathbf{r}_{\bot ,i}} \right)} 
\to
$$
\begin{equation}
\label{eq12}
\to  - i\int {\sum\limits_{Q_{x}}  {a^{+} \left( {q_{x} + Q_{x} ,q_{y}} \right)\partial a\left( {\mathbf{q}_{\bot} } \right)/\partial q_{x} d\mathbf{q}_{\bot} } }  ,  
\end{equation}
$$ 
\left( {1/N_{\bot} } \right)\sum\limits_{i = 1}^{N_{\bot} }  {\sum\limits_{\mathbf{q}_{\bot}  ,{\mathbf{q}}'_{\bot} }  {a_{{\mathbf{q}}'_{\bot} }^{+}  b_{- \mathbf{q}_{\bot} }^{+} }  \exp\left( {i\left( {{\mathbf{q}}'_{\bot}  + \mathbf{q}_{\bot} } \right) \mathbf{r}_{\bot ,i}} \right)
  1/6\left[ {\sum\limits_{\delta}^{4} {\exp\left( {i\left( {\mathbf{q}_{\bot}  \mathbf{r}_{\bot \delta} } \right)}\right)} + 2  }\right] } 
\to 
$$
$$ 
\to \int {\sum\limits_{Q_{x}}  {\gamma_{\mathbf{q}_{\bot} }  a^{+} \left( {q_{x} + Q_{x} ,q_{y}} \right)b^{+} \left( { - \mathbf{q}_{\bot} } \right)d\mathbf{q}_{\bot} } }  , \nonumber
$$
\noindent
where for simple tetragonal lattice ($Z = 6$, $\mathbf{r}_{\delta} = \left( {\pm 1/2,\,\,\, \pm 1/2,\,\,\, \pm 1/2} \right)$)
\begin{equation}
\label{eq13}
\gamma_{\mathbf{q}_{\bot} }  = 
\gamma_{\mathbf{q}_{\bot} } ^{*}  = 
\gamma_{\mathbf{q}_{\bot}  ,q_{z} = 0} = 
1/6 \left[ {\sum\limits_{\delta}^{4} {\exp\left( {i\left( {\mathbf{q}_{\bot}  \mathbf{r}_{\bot \delta} } \right)} \right)}+ 2 }\right] 
= \left( {\cos\left( {q_{x} /2} \right) + \cos\left( {q_{y} /2} \right) + 1} \right)/3.
\end{equation}

After passing in Eq.(\ref{eq1}) to dimensionless designations:

\begin{equation}
\label{eq14}
\gamma_{S} B_{E} = \omega_{E} ,\,\,\,\,0 < B_{\mathrm{A}} /B_{E} = b_{\mathrm{A}} < 1,\,\,\,\,\,\,0 < B/B_{E} = b,\,\,\,\,Ga_{\bot}  /B_{E} = g\sim \,10^{-5}\,,
\end{equation}
\noindent
we will obtain the expression for spin Hamiltonian of considered easy-axis antiferromagnet model
in the vicinity of ground state up to quadratic members
relative to operators $a\left( {\mathbf{q}_{\bot} } \right)$ and $b\left( {\mathbf{q}_{\bot} } \right)$
(spin-wave approximation) in the form:

$$ 
h_{S} = H_{S} /\left( {\hbar \omega_{E}} \right) = - N\left( {1 + b_{\mathrm{A}}} \right)/2 
+
$$
$$ 
 + \int {\sum\limits_{\mathrm{Q}_{\bot} }  {\{ \left( {1 + b_{\mathrm{A}} - b} \right)\;a^{+} \left( {\mathbf{q}_{\bot}  + \mathrm{Q}_{\bot} } \right)\,\,a\left( {\mathbf{q}_{\bot} } \right) +} } \left( {1 + b_{\mathrm{A}} + b} \right)\,\,b^{+} \left( { - \mathbf{q}_{\bot} } \right)\,b\left( { - \left( {\mathbf{q}_{\bot}  + \mathrm{Q}_{\bot} } \right)} \right) 
+ 
$$
\begin{equation}
\label{eq15}
+ ig\left[ {a^{+} \left( {\mathbf{q}_{\bot}  + \mathrm{Q}_{\bot} } \right)\,\partial a\left( {\mathbf{q}_{\bot} } \right)/\partial q_{x} - \left( {\partial b^{+} \left( { - \mathbf{q}_{\bot} } \right)/\partial q_{x}} \right)\,b\left( { - \left( {\mathbf{q}_{\bot}  + \mathrm{Q}_{\bot} } \right)} \right)} \right] 
+ 
\end{equation}
$$ 
+  \gamma_{\mathbf{q}_{\bot} }  \;\left[ {a^{+} \left( {\mathbf{q}_{\bot}  + \mathrm{Q}_{\bot} } \right)\,b^{+} \left( { - \mathbf{q}_{\bot} } \right) + a\left( {\mathbf{q}_{\bot} } \right)\,b\left( { - \left( {\mathbf{q}_{\bot}  + \mathrm{Q}_{\bot} } \right)} \right)} \right]\} \;d\mathbf{q}_{\bot}  .
$$

Hamiltonian (\ref{eq15}) takes into account here the correlations of magnon states for distinct cells 
of reciprocal sublattice, which are expressed as summation over all vector components $Q_{x}$.
In early works \cite{12,13} this circumstance was not considered.

With the help of Eq.(\ref{eq4}), (\ref{eq6}) and (\ref{eq7}) the commutation relations 
for operators $a_{{\mathbf{q}}'_{\bot} }^{+}$, $a_{\mathbf{q}_{\bot}}$ 
take the form (similar to operators $b_{-{\mathbf{q}}'_{\bot}}^{+}$, $b_{-\mathbf{q}_{\bot}}$):
\begin{equation}
\label{eq16}
\left[ {a_{\mathbf{q}_{\bot} }  ,a_{{\mathbf{q}}'_{\bot} }^{+} } \right] 
= \frac{{1}}{{N_{\bot} } }\sum\limits_{i,\,{i}'}^{N_{\bot} }  {\exp\left( {i\left( {{\mathbf{q}}'_{\bot}  \mathbf{r}_{\bot ,\,i} -  \mathbf{q}_{\bot}  \mathbf{r}_{\bot ,\,{i}'} } \right)} \right)\left[ {a_{i} ,a_{{i}'}^{+} } \right] }
= \sum\limits_{\,Q_{x}}  {\delta_{{q}'_{x} ,\,q_{x} + Q_{x} } \delta_{{q}'_{y} ,\,q_{y}} }  ,
\end{equation}
\noindent
or
\begin{equation}
\label{eq17}
\left[ {a_{\mathbf{q}_{\bot} }  ,a_{{q}'_{x} + {\mathrm{Q}}'_{x} ,{q}'_{y}}^{+} } \right]
= \sum\limits_{Q_{x}}  {\delta_{{q}'_{x} + {\mathrm{Q}}'_{x} , q_{x} + Q_{x}}  \delta_{{q}'_{y} ,\,q_{y}} }  .
\end{equation}

Going in Eq.(\ref{eq17}) to the continuous arguments $q_{x}$, $q_{y}$ 
for magnon operators we will obtain:
\begin{equation}
\label{eq18}
\left[ {a\left( {\mathbf{q}_{\bot} } \right),a^{+} \left( {{q}'_{x} + {\mathrm{Q}}'_{x} ,{q}'_{y}} \right)} \right] 
= \sum\limits_{Q_{x} - {\mathrm{Q}}'_{x}}  {\delta \left( {{q}'_{x} - q_{x} - Q_{x} + {\mathrm{Q}}'_{x}} \right)\delta \left( {{q}'_{y} - q_{y}} \right)} ,\; 
\end{equation}
$$ 
\left[ {b^{+} \left( { - \mathbf{q}_{\bot} } \right),\;b\left( { - \left( {{q}'_{x} + {\mathrm{Q}}'_{x}} \right), - {q}'_{y}} \right)} \right] 
= - \sum\limits_{Q_{x} - {\mathrm{Q}}'} {\delta \left( {{q}'_{x} - q_{x} - Q_{x} + {\mathrm{Q}}'_{x}} \right)\delta \left( {{q}'_{y} \, - q_{y}} \right)} . 
$$

Using the commutation relations (\ref{eq18}) and notation $Q_{x} - {\mathrm{Q}}'_{x} = {\mathrm{Q}}''_{x}$,
 we will obtain for the magnon operators in Heisenberg representation 
$a\left( {\tau ,q} \right)$ and $b^{+} \left( {\tau , - \mathbf{q}} \right)$ 
the following system of differential equations (here $\tau = \omega_{E} t$ is dimensionless time):
$$ 
i\partial a\,\,\left( {\tau ,\,\,\,\mathbf{q}_{\bot} } \right)/\partial \tau \;\, 
= \,\,\,\left[ {a\,\,\left( {\tau ,\,\,\,\mathbf{q}_{\bot} } \right),\,\,h_{S}} \right] 
= 
$$
\begin{equation}
\label{eq19}
= \,\sum\limits_{{\mathrm{Q}}''_{x}}  {\,\{ \left( {1 + b_{\mathrm{A}} - b + ig\;\partial /\partial q_{x}} \right)\;a\,\,\left( {\tau ,\,\,\,q_{x} + {\mathrm{Q}}''_{x} ,\,\,q_{y}} \right) + \gamma_{\mathbf{q}_{\bot} }  b^{+} \left( {\tau ,\,\, - \left( {q_{x} + {\mathrm{Q}}''_{x}} \right),\,\, - q_{y}} \right)} \} , 
\end{equation}
$$ 
i\partial b^{+} \left( {\tau , - \mathbf{q}_{\bot} } \right)/\partial \tau \; 
= \left[ {b^{+} \left( {\tau , - \mathbf{q}_{\bot} } \right),\,\,h_{S}} \right] 
=
$$
$$ 
= - \sum\limits_{{\mathrm{Q}}''_{x}}  {\{ \left( {1 + b_{\mathrm{A}} + b - ig\;\partial /\partial q_{x}} \right)\;b^{+} \left( {\tau , - \left( {q_{x} + {\mathrm{Q}}''_{x} } \right),\,\, - q_{y}} \right) + \gamma_{\mathbf{q}_{\bot} }  a\,\,\left( {\tau ,\,\,\,q_{x} + {\mathrm{Q}}''_{x} ,\,\,q_{y}} \right)\}}  .
$$

The inclusion of umklapp processes leads here to a system of engaging equation of motion for magnon operators
for different cells of reciprocal sublattice.

Note that the first equation passes here into the second equation with the changing
$a\left( {\tau ,\mathbf{q}_{\bot} } \right) \to \;b\left( {\tau , - \mathbf{q}_{\bot} } \right)$,
$b \to - b$ and with use of the Hermitian conjugation.
It is also not difficult to find the equations for the second pair of Hermit conjugated operators
$a^{+} \left( {\tau ,\mathbf{q}_{\bot} } \right)$, $b\left( {\tau , - \mathbf{q}_{\bot} } \right)$.

\section{Diagonalization of the antiferromagnetic spin Hamiltonian}

Let us perform the unitary transformation (a variant of Bogolubov-Tyablikov
transformation (Ref.\cite{17}, sec.~13) to new linearly independent creation and
annihilation operators $\xi^{+} \left( {\tau ,q_{y} ,E_{\pm} } \right)$ and $\xi \left( {\tau ,q_{y} ,E_{\pm} } \right)$ 
for two types of magnon states, which propagate along direction $x$-axis 
with dimensionless energy $E_{\pm} = E \pm b$, where $E$ - is a continuous energy parameter, 
and wave vector component $q_{y}$ in the range of $q_{y}$ to $q_{y} + dq_{y}$.

We will write out the following two of the four equations (two others are Hermitian conjugated):
\begin{equation}
\label{eq20}
 a\left( {\tau ,\,\,\mathbf{q}_{\bot} } \right)
 = \int {\left[ {u\left( {\mathbf{q}_{\bot}  ,E} \right)\xi \left( {\tau ,q_{y} ,E_{-} } \right) + v^{* }\left( {\mathbf{q}_{\bot}  ,E} \right)\xi^{+} \left( {\tau ,q_{y} ,E_{+} } \right)} \right]dE}, 
\end{equation}
$$ 
 b^{+} \left( {\tau , - \mathbf{q}_{\bot} } \right) 
 = \int {\left[ {v\left( {\mathbf{q}_{\bot}  ,E} \right)\xi \left( {\tau ,q_{y} ,E_{-} } \right) + u^{* }\left( {\mathbf{q}_{\bot}  ,E} \right)\xi^{+} \left( {\tau ,q_{y} ,E_{+} } \right)} \right]dE}.  
$$

The new operators must obey the boselike commutation relations:
$$ 
\left[ {\xi \left( {\tau ,q_{y} ,E_{\pm} } \right),\xi^{+} \left( {\tau ,{q}'_{y} ,{E}'_{\pm} } \right)} \right] = \delta \left( {q_{y} - {q}'_{y}} \right)\delta \left( {E - {E}'} \right), 
$$
\begin{equation}
\label{eq21}
\left[ {\xi \left( {\tau ,q_{y} ,E_{\pm} } \right),\xi \left( {\tau ,{q}'_{y} ,{E}'_{\pm} } \right)} \right] = \left[ {\xi^{+}\left( {\tau ,q_{y} ,E_{\pm} } \right),\xi^{+} \left( {\tau ,{q}'_{y} ,{E}'_{\pm} } \right)} \right] 
= 
\end{equation}
$$ 
= \left[ {\xi \left( {\tau ,q_{y} ,E_{\pm} } \right),\xi \left( {\tau ,{q}'_{y} ,{E}'_{\mp} } \right)} \right] = \left[ {\xi^{+} \left( {\tau ,q_{y} ,E_{\pm} } \right),\xi^{+} \left( {\tau ,{q}'_{y} ,{E}'_{\mp} } \right)} \right] = 0, 
$$
\noindent
and the following equations of motion
\begin{equation}
\label{eq22}
 i\partial /\partial \tau \;\xi \,\,\,\left( {\tau ,q_{y} ,E_{\pm} } \right) 
= \left[ {\,\xi \,\,\left( {\tau ,q_{y} ,E_{\pm} } \right),h_{S} \left( {\tau} \right)} \right] 
= E_{\pm}  \;\,\xi \,\,\left( {\tau ,q_{y} ,E_{\pm} } \right), 
\end{equation}
$$ 
 i\partial /\partial \tau \;\xi^{+} \left( {\tau ,q_{y} ,E_{\pm} } \right) 
= \left[ {\xi^{+} \left( {\tau ,q_{y} ,E_{\pm} } \right),h_{S} \left( {\tau} \right)} \right] 
= - E_{\pm}  \;\xi^{+} \left( {\tau ,q_{y} ,E_{\pm} } \right).
$$

Let us next retain in the right sides of Eqs.(\ref{eq18}) 
only members with $Q_{x} = {\mathrm{Q}}'_{x} = 0$, that is in the commutation relations 
the umklapp processes between distinct cells of reciprocal sublattice will be neglected (N-approximation). 
In this case in the right sides of Eqs.(\ref{eq19}) only terms with values ${\mathrm{Q}}''_{x} = 0$ are retained.

With Eqs.(\ref{eq21}) we will obtain in this approximation 
the first pair unitarity conditions for transformation coefficients:
\begin{equation}
\label{eq23}
\int {\left[ {u\left( {\mathbf{q}_{\bot}  ,E} \right)u^{*} \left( {{q}'_{x} ,q_{y} ,E} \right) - v^{*} \left( {\mathbf{q}_{\bot}  ,E} \right)v\left( {{q}'_{x} ,q_{y} ,E} \right)} \right]dE =}  \,\,\delta \left( {q_{x} - {q}'_{x}} \right), 
\end{equation}
$$ 
 \int {\left[ {v^{*} \left( {\mathbf{q}_{\bot}  ,E} \right)u\left( {{q}'_{x} ,q_{y} ,E} \right) - u\left( {\mathbf{q}_{\bot}  ,E} \right)v^{*} \left( {{q}'_{x} ,q_{y} ,E} \right)} \right]dE }=  \,0.
$$

Note that in this case due to terms, which contain the derivative $ig\;\partial /\partial q_{x}$, 
the equations of motion conserve the inhomogeneous external field dependence. 
The inclusion of umklapp processes would leads to corrections of higher-order for the parameter $g \ll 1$.

Substituting Eqs.(\ref{eq20}), (\ref{eq22}) to Eqs.(\ref{eq19}), 
and making under the integrals the coefficients ahead of operators 
$\xi_{-} \left( {\tau ,q_{y} ,E_{-} } \right)$ and 
$\xi_{+}^{+} \left( {\tau ,q_{y} ,E_{+} } \right)$ 
equal to zero, we obtain the system of four equations for determination 
of transformation coefficients 
$u\left( {\mathbf{q}_{\bot},E} \right)$, 
$v\left( {\mathbf{q}_{\bot},E} \right)$
and their Hermitian conjugated values:
$$ 
\left[ {ig\;\partial /\partial q_{x} - \left( {E - 1 - b_{\mathrm{A}}} \right)} \right]\;u\left( {\mathbf{q}_{\bot}  ,E} \right) + \gamma_{\mathbf{q}_{\bot} }  v\left( {\mathbf{q}_{\bot}  ,E} \right) = 0, 
$$
$$ 
\gamma_{\mathbf{q}_{\bot} }  u\left( {\mathbf{q}_{\bot}  ,E} \right) - \left[ {ig\;\partial /\partial q_{x} - \left( {E + 1 + b_{\mathrm{A}}} \right)} \right]\;v\left( {\mathbf{q}_{\bot}  ,E} \right) = 0, 
$$
\begin{equation}
\label{eq24}
\left[ {ig\;\partial /\partial q_{x} + \left( {E + 1 + b_{\mathrm{A}}} \right)} \right]\;v^{*} \left( {\mathbf{q}_{\bot}  ,E} \right) + \gamma_{\mathbf{q}_{\bot} }  u^{* }\left( {\mathbf{q}_{\bot}  ,E} \right) = 0, 
\end{equation}
$$ 
\gamma_{\mathbf{q}_{\bot} }  v^{*} \left( {\mathbf{q}_{\bot}  ,E} \right) - \left[ {ig\;\partial /\partial q_{x} + \left( {E - 1 - b_{\mathrm{A}}} \right)} \right]\;u^{*} \left( {\mathbf{q}_{\bot}  ,E} \right) = 0. 
$$

Note that the first pair of Eqs.(\ref{eq24}) transforms in the second pair by replacing $u\left( {\mathbf{q}_{\bot}  ,E} \right) \to v^{*} \left( {\mathbf{q}_{\bot}  , - E} \right)$, and $v\left( {\mathbf{q}_{\bot}  ,E} \right) \to u^{*} \left( {\mathbf{q}_{\bot}  , - E} \right)$. In this case, the condition (\ref{eq23}) is violated. Hence, it follows that solutions with $E < 0$ are not physical and should be eliminated. However, if the energy parameters $E$ have the common positive sign in both equation pairs, they change from one to the other by complex conjugation and by following replacing of transformation coefficients:
\begin{equation}
\label{eq25}
u\left( {\mathbf{q}_{\bot}  ,E} \right) \to u^{*} \left( {\mathbf{q}_{\bot}  ,E} \right),\,\,\,\,\,\nu \left( {\mathbf{q}_{\bot}  ,E} \right) \to v^{*} \left( {\mathbf{q}_{\bot}  ,E} \right).
\end{equation}

Therefore, we will deal next only with the-second pair of Eqs.(\ref{eq24}).

Remaining next in the context of the N-approximation; we will multiply the first equation from Eqs.(\ref{eq19}) to the left by $a^{+} \left( {\tau ,q_{x} + Q_{x} ,q_{y}} \right)$ and the second equation to the right by $b\left( {\tau , - \left( {q_{x} + Q_{x}} \right), - q_{y}} \right)$. Upon integration with respect $\mathbf{q}_{\bot}$ and summation over $Q_{x}$, the expression for antiferromagnetic spin Hamiltonian (\ref{eq15}) takes the form
\begin{equation}
\label{eq26}
h_{S} = h_{S} \left( {\tau} \right) = - N\left( {1 + b_{\mathrm{A}}} \right)/2 
+ 
\end{equation}
$$ 
+ \sum\limits_{Q_{x}} { \int { \{ ia^{+} \left( {\tau ,\mathbf{q}_{\bot}  + \mathrm{Q}_{\bot} } \right)\,\partial \,a\left( {\tau ,\mathbf{q}_{\bot} } \right)/\partial \tau - i\partial \,b^{+} \left( {\tau , - \mathbf{q}_{\bot} } \right)/\partial \tau \,b\left( {\tau , - \left( {\mathbf{q}_{\bot}  + \mathrm{Q}_{\bot} } \right)} \right)\} d\mathbf{q}_{\bot} } }  .
$$

Substituting Eqs.(\ref{eq20}), (\ref{eq22}) to Eq.(\ref{eq26}), we now obtain
$$ 
h_{S} \left( {\tau} \right) = 
- N\left( {1 + b_{\mathrm{A}}} \right)/2 
+ 
$$
$$ 
+ \sum\limits_{Q_{x}} { \int { \{ \left[ {u^{*} \left( {q_{x} + Q_{x} ,q_{y} ,E} \right)\xi^{+} \left( {\tau ,q_{y} ,E_{-} } \right)\, + v\left( {q_{x} + Q_{x} ,q_{y} ,E} \right)\xi \left( {\tau ,q_{y} ,E_{+} } \right)} \right]}}  
\cdot 
$$
\begin{equation}
\label{eq27}
\cdot \left[ {u\left( {\mathbf{q}_{\bot}  ,{E}'} \right){E}'_{-}  \xi \left( {\tau ,q_{y} ,{E}'_{-} } \right) - v^{* }\left( {\mathbf{q}_{\bot}  ,{E}'} \right){E}'_{+}  \xi^{+} \left( {\tau ,q_{y} ,{E}'_{+} } \right)} \right] 
- 
\end{equation}
$$ 
- \left[ {v\left( {\mathbf{q}_{\bot}  ,E} \right)E_{-}  \xi \left( {\tau ,q_{y} ,E_{-} } \right) - u^{*} \left( {\mathbf{q}_{\bot}  ,E} \right)E_{+}  \xi^{+} \left( {\tau ,q_{y} ,E_{+} } \right)} \right] 
\cdot 
$$
$$ 
\cdot \left[ {v^{*} \left( {q_{x} + Q_{x} ,q_{y} ,{E}'} \right)\xi^{+}\left( {\tau ,q_{y} ,{E}'_{-} } \right) + u\left( {q_{x} + Q_{x} ,q_{y} ,{E}'} \right)\xi \left( {\tau ,q_{y} ,{E}'_{+} } \right)} \right]\} d\mathbf{q}_{\bot}  dEd{E}',
$$
\noindent
where the transformation coefficients of type $u^{*} \left( {q_{x} + Q_{x} ,q_{y} ,E} \right)$ and $v^{*} \left( {q_{x} + Q_{x} ,q_{y} ,{E}'} \right)$ are derived by means of simple argument shift $\mathbf{q}_{\bot} \to q_{x} + Q_{x} ,q_{y}$ in coefficients $u^{*} \left( {\mathbf{q}_{\bot}  ,E} \right)$, $v^{*} \left( {\mathbf{q}_{\bot},E} \right)$. 
The spin Hamiltonian in the form (\ref{eq27}) accounts, like Eq.(\ref{eq15}), 
the magnon states correlations for different cells reciprocal sublattice.

We will transform next the part of expression (\ref{eq27}) containing operator product $\xi \left( {\tau ,q_{y} ,E_{+} } \right)\xi \left( {\tau ,q_{y} ,{E}'_{-} } \right)$ through variables ${E}'\, \leftrightarrow E$ permutation to the form:
$$ 
\sum\limits_{Q_{x}}  {\int {\left[ {{E}'_{-}  v\left( {q_{x} + Q_{x} ,q_{y} ,E} \right)u\left( {\mathbf{q}_{\bot}  ,{E}'} \right)\xi \left( {\tau ,q_{y} ,E_{+} } \right)\xi \left( {\tau ,q_{y} ,{E}'_{-} } \right)  - }\right.
}} 
$$
\begin{equation}
\label{eq28}
{{ \left.{
- E_{-}  v\left( {\mathbf{q}_{\bot}  ,E} \right)\,u\left( {q_{x} + Q_{x} ,q_{y} ,{E}'} \right)\xi \left( {\tau ,q_{y} ,E_{-} } \right)\xi \left( {\tau ,q_{y} ,{E}'_{+} } \right)} \right]} d\mathbf{q}_{\bot}  dEd{E}'} 
= 
\end{equation}
$$ 
= \int {{E}'_{-}  dEd{E}'\sum\limits_{Q_{x}}  {\int {\left[ {v\left( {q_{x} + Q_{x} ,q_{y} ,E} \right)u\left( {\mathbf{q}_{\bot}  ,{E}'} \right)\xi \left( {\tau ,q_{y} ,E_{+} } \right)\xi \left( {\tau ,q_{y} ,{E}'_{-} } \right)  - }\right.
}}} 
$$
$$ 
{{{ \left.{
- v\left( {\mathbf{q}_{\bot} ,{E}'} \right)\,u\left( {q_{x} + Q_{x} ,q_{y} ,E} \right)\xi \left( {\tau ,q_{y} ,{E}'_{-} } \right)\xi \left( {\tau ,q_{y} ,E_{+} } \right)} \right]d\mathbf{q}_{\bot} }}} .   
$$

Let us require now the fulfillment of one more condition from the second pair of unitarity transformation conditions
\begin{equation}
\label{eq29}
\int {\left[ {v\left( {q_{x} + Q_{x} ,q_{y} ,E} \right)\,\,u\left( {\mathbf{q}_{\bot} ,{E}'} \right)\, - v\left( {\mathbf{q}_{\bot},{E}'} \right)\,\,\,u\left( {q_{x} + Q_{x} ,q_{y} ,E} \right)} \right]d\mathbf{q}_{\bot} }  = 0,
\end{equation}
\noindent
which reduces to zero Eq.(\ref{eq28}). Instead of Eq.(\ref{eq27}) we will then obtain
$$ 
h_{S} \left( {\tau} \right) = - N\left( {1 + b_{\mathrm{A}}} \right)/2 
+ 
$$
$$ 
+ \sum\limits_{Q_{x}} { \int { \left\{ {E_{-} \left[ {u^{*} \left( {q_{x} + Q_{x} ,q_{y} ,{E}'} \right)u\left( {\mathbf{q}_{\bot}  ,E} \right)\xi^{+} \left( {\tau ,q_{y} ,{E}'_{-} } \right)\xi \left( {\tau ,q_{y} ,E_{-} } \right)\, -
}\right.}\right.}}
$$
\begin{equation}
\label{eq30}
\left.{ 
- v^{*} \left( {q_{x} + Q_{x} ,q_{y} ,{E}'} \right)v\left( {\mathbf{q}_{\bot}  ,E} \right)\xi \left( {\tau ,q_{y} ,E_{-} } \right)\xi^{+} \left( {\tau ,q_{y} ,{E}'_{-} } \right)} \right] + 
\end{equation}
$$ 
+ E_{+} \left[ {u^{*} \left( {\mathbf{q}_{\bot}  ,E} \right)u\left( {q_{x} + Q_{x} ,q_{y} ,{E}'} \right)\xi^{+}\left( {\tau ,q_{y} ,E_{+} } \right)\xi \left( {\tau ,q_{y} ,{E}'_{+} } \right) - 
}\right.
$$
$$ 
\left.{\left.{
- v^{*} \left( {\mathbf{q}_{\bot}  ,E} \right)v\left( {q_{x} + Q_{x} ,q_{y} ,{E}'} \right)\xi \left( {\tau ,q_{y} ,{E}'_{+} } \right)\xi^{+} \left( {\tau ,q_{y} ,E_{+} } \right)} \right]}\right\} d\mathbf{q}_{\bot} dEd{E}' . 
$$

Let us perform the magnon operators permutation in the second and the fourth rows of Eq.(\ref{eq30}). Using the first commutation relation in Eq.(\ref{eq21}), taking into account that
\begin{equation}
\label{eq31}
\lim\limits_{q_{y} \to {\mathbf{q}}'} \delta \left( {q_{y} - {q}'_{y}} \right) \to N_{y} /2\pi \,\,\,\,\,\,\,\,\,\,\,\left[ {\xi \left( {\tau ,q_{y} ,E_{\pm} } \right),\xi^{+} \left( {\tau ,q_{y} ,{E}'_{\pm} } \right)} \right] = \,N_{y} /2\pi \,\,\delta \left( {E - {E}'} \right)
\end{equation}
\noindent
and also the second condition from the second pair of unitarity transformation conditions
\begin{equation}
\label{eq32}
\sum\limits_{Q_{x}}  {\,\int {\left[ {u^{*} \left( {q_{x} + \mathrm{Q},q_{y} ,{E}'} \right)u\left( {\mathbf{q}_{\bot}  ,E} \right) - v^{*} \left( {q_{x} + Q_{x} ,q_{y} ,{E}'} \right)v\left( {\mathbf{q}_{\bot}  ,E} \right)} \right]dq_{x} =} }  \,\delta \left( {E - {E}'} \right),
\end{equation}

\noindent
we obtain for spin Hamiltonian (\ref{eq27}) diagonalized form:
$$ 
h_{S} \left( {\tau} \right) = - N\left( {1 + b_{\mathrm{A}}} \right)/2 
- 
$$
\begin{equation}
\label{eq33}
- \,N_{y} /\left( {2\pi} \right)\,\, \sum\limits_{Q_{x}} { \int { \left[ {E_{+}  v^{*} \left( {\mathbf{q}_{\bot}  ,E} \right)v\left( {q_{x} + Q_{x} ,q_{y} ,E} \right) + E_{-}  v^{*} \left( {q_{x} + Q_{x} ,q_{y} ,E }\right)v\left( {\mathbf{q}_{\bot}  ,E} \right)} \right]d\mathbf{q}_{\bot}} }   dE 
+ 
\end{equation}
$$ 
+ \int {\left[ {E_{+}  \xi^{+} \left( {\tau ,q_{y} ,E_{+}  } \right)\xi \left( {\tau ,q_{y} ,E_{+} } \right) + E_{-}  \xi^{+}\left( {\tau ,q_{y} ,E_{-} } \right)\xi \left( {\tau ,q_{y} ,E_{-} }\right) }\right]dq_{y} dE}. 
$$

The following transformation of the second term in Eq.(\ref{eq33}) will be given at the end of Section 5. The last term here describes magnon excitation energy. It is consistent with the above-considered equations of motion for magnon operators (\ref{eq22}). Because the quadratic form (\ref{eq33}) describes a stable state, close to ground state, it should be positively definite. That is, the energies of both types magnons should be $E_{\pm}  = E \pm b > 0$.

\section{Asymptotic solution of equation for the unitary transformation coefficients}

Let us go now to calculations of the transformation coefficients. We will use the second pair of equations (\ref{eq24}), in wich the value $v^{*} \left( {\mathbf{q}_{\bot}  ,E} \right)$ will be eliminated. Because the variable $q_{y}$ in Eqs.(\ref{eq24}) is considered as a constant parameter it will be not explicitly indicated in arguments and the new variable will be used
\begin{equation}
\label{eq34}
\zeta = \int\limits_{0}^{q_{x}}  {\gamma_{\mathbf{q}_{\bot} }  dq_{x}}  
\end{equation}

Then we will switch from the notations $u^{*} \left( {\mathbf{q}_{\bot},E} \right)$ and $\gamma_{\mathbf{q}_{\bot} } $ to $u^{*} \left( {\zeta ,E} \right)$ and $\gamma_{\varsigma}$. The partial derivatives with respect to $\zeta$ will be replaced by the ordinary derivatives. As a result, we obtain the following differential equation for the transformation coefficient $u^{*} \left( {\zeta ,E} \right)$
\begin{equation}
\label{eq35}
g^{2}d^{2}u^{*} \left( {\zeta ,E} \right)/d\zeta^{2}\; + p\left( {\zeta ,g} \right)\;gdu^{*} \left( {\zeta ,E} \right)/d\varsigma + r\left( {\zeta ,g} \right)\;u^{*} \left( {\zeta ,E} \right) = 0,
\end{equation}
\noindent
where
\begin{equation}
\label{eq36}
p\left( {\zeta ,g} \right) = p_{0} \left( {\zeta} \right) = - 2iE/\gamma_{\zeta}  , 
\end{equation}
$$ 
r\left( {\zeta ,g} \right) = \mathbf{r}_{0} \left( {\zeta} \right) + g\;\mathbf{r}_{1} \left( {\zeta} \right) = \left[ {\left( {1 + b_{\mathrm{A}}} \right)^{2} - E^{2} - \gamma_{\zeta}^{2}} \right] + ig\;\left( {1 + b_{\mathrm{A}} - E} \right)\; \cdot d\left( {1/\gamma_{\varsigma} } \right)/d\zeta .\, 
$$

Solution of Eq.(\ref{eq35}) will be presented in the form of an asymptotic expansion in terms of small parameter $g \ll 1$ (Ref.\cite{18}, sec.~7.1.6):
\begin{equation}
\label{eq37}
u^{*} \left( {\zeta ,E} \right) = \sum\limits_{n = 0}^{\infty}  {g^{n}\left( {A_{n} \left( {\zeta} \right)\exp\theta_{1} + B_{n} \left( {\zeta} \right)\exp\theta_{2}}\right) } ,\,\,\,\,\,d\theta_{i} /d\zeta = - \lambda_{i} \left( {\zeta} \right)/g,
\end{equation}
\noindent
where the variables $\theta_{i} ,\;\;\zeta$ are assumed independent from one another and $\lambda_{i} \left( {\zeta} \right)$ will be defined below.

The derivatives in Eq.(\ref{eq35}) are transformed as follows$\left( {{\lambda }'_{i} \equiv d\lambda_{i} /d\zeta} \right)$:
$$ 
g\;d/d\zeta = - \lambda_{1} \partial /\partial \theta_{1} - \lambda_{2} \partial /\partial \theta_{2} + g\partial /\partial \zeta , 
$$
\begin{equation}
\label{eq38}
g^{2}\;d^{2}/d\zeta^{2} = \lambda_{1}^{2} \partial^{2}/\partial \theta_{1}^{2} + 2\lambda_{1} \lambda_{2} \partial^{2}/\partial \theta_{1} \partial \theta_{2} + \lambda_{2}^{2} \partial^{2}/\partial \theta_{2}^{2} 
-
\end{equation}
$$ 
- 2g\lambda_{1} \partial^{2}/\partial \theta_{1} \partial \zeta - 2g\lambda_{2} \partial^{2}/\partial \theta_{2} \partial \zeta - g{\lambda }'_{1} \partial /\partial \theta_{1} - g{\lambda} '_{2} \partial /\partial \theta_{2} + g^{2}\partial^{2}/\partial \zeta^{2}. 
$$

The substitution of expression (\ref{eq37}) in Eq.(\ref{eq35}) and making the expressions ahead $g^{n}\exp\theta_{1,2}$ equal to zero gives two equations
\begin{equation}
\label{eq39}
\left( {\lambda_{1}^{2} + \lambda_{1} p_{0} \left( {\zeta} \right) + r\left( {\zeta ,g} \right)} \right)A_{n} - g\left( {2\lambda_{1} + p\left( {\zeta ,g} \right)} \right){A}'_{n} - g{\lambda} '_{1} A_{n} + g^{2}{{A}'}'_{n} = 0, 
\end{equation}
$$ 
\left( {\lambda_{2}^{2} + \lambda_{2} p_{0} \left( {\zeta} \right) + r\left( {\zeta ,g} \right)} \right)B_{n} - g\left( {2\lambda_{2} + p\left( {\zeta ,g} \right)} \right){B}'_{n} - g{\lambda} '_{2} B_{n} + g^{2}{{B}'}'_{n} = 0.
$$

By assembling then in Eq.(\ref{eq39}) the coefficients near the same powers of $g$, we will derive the equations for determination of values $A_{n} \left( {\zeta} \right)$, $B_{n} \left( {\zeta} \right)$. Coefficients ahead $g^{0}A_{0} \left( {\zeta} \right)$, $g^{0}B_{0} \left( {\zeta} \right)$, ($A_{0} ,\;B_{0} \ne 0$) give the equations for determination of $\lambda_{i} \equiv \lambda_{1,2}$:
\begin{equation}
\label{eq40}
\lambda_{i}^{2} + p_{0} \left( {\zeta} \right)\lambda_{i} + \mathbf{r}_{0} \left( {\zeta} \right) = 0,\,\,\,\,\,\,\,\,i = 1,2,
\end{equation}
\noindent
from where it follows
\begin{equation}
\label{eq41}
\lambda_{1,2} = i/\gamma_{\zeta} \left( { - E \pm E\left( {\zeta} \right)} \right),\;\;E\left( {\;\zeta} \right) = \;\sqrt {\left( {1 + b_{\mathrm{A}} } \right)^{2} - \gamma_{\zeta}^{2}}  .
\end{equation}

The zero-order quantities $A_{0} \left( {\zeta} \right)$, $B_{0} \left( {\zeta} \right)$ are determined by differential equations obtained by making the coefficients ahead of the first power factor $g$ in Eq.(\ref{eq39}) equal to zero (we will restrict next only by the first order):
\begin{equation}
\label{eq42}
\left( {2\lambda_{1} + p_{0}} \right){A}'_{0} + \left( {{\lambda} '_{1} + \mathbf{r}_{1}} \right)A_{0} = 0, 
\end{equation}
$$ 
\left( {2\lambda_{2} + p_{0}} \right){B}'_{0} + \left( {{\lambda} '_{2} + \mathbf{r}_{1}} \right)B_{0} = 0. 
$$

Solutions of equations (\ref{eq42}) have the form

\begin{equation}
\label{eq43}
A_{0} \left( {\zeta} \right)\exp\theta_{1} = A\exp I_{1} \left( {\zeta} \right),\,\,\,\,B_{0} \left( {\zeta} \right)\exp\theta_{2} = B\exp I_{2} \left( {\zeta} \right),
\end{equation}
\noindent
where
\begin{equation}
\label{eq44}
I_{1,2} \left( {\zeta} \right) = - \int\limits_{0}^{\zeta}  {\frac{{{\lambda} '_{1,2} + \mathbf{r}_{1}} }{{2\lambda_{1,2} + p_{0}} }d\zeta}  - \frac{{1}}{{g}}\int\limits_{0}^{\zeta}  {\lambda_{1,2} d\zeta}  \,,
\end{equation}
\begin{equation}
\label{eq45}
2\lambda_{1,2} + p_{0} = \pm 2iE\left( {\zeta} \right)/\gamma_{\zeta}  ,
\end{equation}
\begin{equation}
\label{eq46}
 - \frac{{{\lambda} '_{1,2} + \mathbf{r}_{1}} }{{2\lambda_{1,2} + p_{0}} }d\zeta 
= - 1/2\,d\left[ {\ln\left( {E\left( {\zeta} \right)/\gamma_{\zeta} } \right) \mp E\left( {\;\zeta} \right)/\gamma_{\zeta} } \right]
\end{equation}
\noindent
and constants of integration $A = A_{0} \left( {0} \right) = A\left( {E,q_{y}} \right),\;\,\,B = B_{0} \left( {0} \right) = B\left( {E,q_{y}} \right)$ depend here on $E$ and $q_{y}$, as on parameters.

Due to $E > 0$ the quantity $\lambda_{1}$ (the upper sign in formulas (\ref{eq41}), (\ref{eq45}), (\ref{eq46})) may have any small values. In the limit of $g \to 0$ we have $\lambda_{1} \to 0$, that is in the case of homogeneous field we obtain the known result or two branches of magnon spectrum:
\begin{equation}
\label{eq47}
E_{\pm}  = E \pm b \to E_{\pm} \left( {\mathbf{q}_{\bot}  ,\;b} \right) = E\left( {\zeta} \right) \pm \;b = \sqrt {\left( {1 + b_{\mathrm{A}}} \right)^{2} - \gamma_{\zeta}^{2}}  \pm \;b > 0,
\end{equation}
\noindent
whereas the going to the homogeneous case in expression for $\lambda_{2}$ leads formally to nonphysical negative value of parameter $E < 0$ and conditions (\ref{eq23}), (\ref{eq32}) are violated. We will choose out of two approximate solutions the solution $u^{*} \left( {\zeta ,E} \right) \approx A_{0} \left( {\zeta} \right)\exp\theta_{1} = A\exp I_{1} \left( {\zeta} \right)$ that corresponds to low branch of magnon spectrum with gap $E_{-}  = E\left( {\zeta} \right) - b > 0$.

\begin{figure}
   \begin{center}
   \begin{tabular}{c}
   \includegraphics{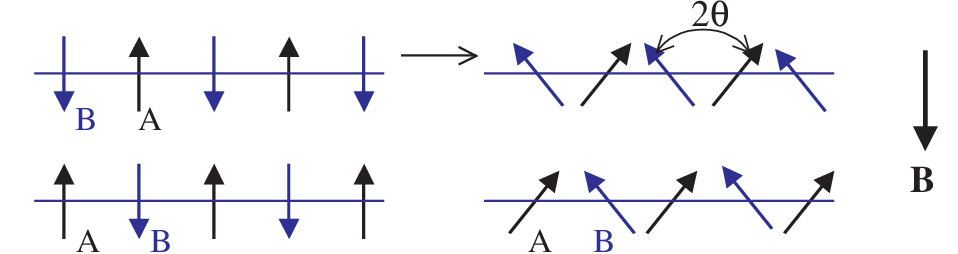}
   \end{tabular}
   \end{center}
   \caption[1]
   {\label{fig:4} 
The scheme of spin-flop phase transition in easy axis antiferromagnet at critical field $b = b_{C}$. For FeCO$_{3}$: $\cos\theta = b_{C} /2 \approx 0.22,\,\,\,2\theta \approx 155^{0}$.
   }
   \end{figure}

This gap disappears at the condition
$E\left( {\mathbf{q}_{\bot}  = 0} \right) = \sqrt {\left( {1 + b_{\mathrm{A}}} \right)^{2} - 1} = b_{C} \approx \sqrt { 2b_{\mathrm{A}} + b_{\mathrm{A}}^{2}}  = b$, 
which is consistent with known condition for orientational phase transition of spin-flop type in antiferromagnet at critical field $b = b_{C}$ (Ref.\cite{19}). The equilibrium state of two neighboring spins, belonging to different sublattice with zero total spin polarization passes into state with nonzero total spin polarization (Fig.~\ref{fig:4}).

Taking into account now that $E\left( {\zeta} \right) = E\left( {\mathbf{q}_{\bot}} \right) = \sqrt {1 + b_{C}^{2} - \gamma_{\mathbf{q}_{\bot} }^{2}}$, $E\left( {\;\zeta = 0} \right) = E\left( {q_{y}} \right)$, we will write the exponent $I_{1} \left( {\zeta} \right)$ from Eq.(\ref{eq43}) in the form

\begin{equation}
\label{eq48}
I_{1} \left( {\zeta} \right) = I_{1} \left( {\mathbf{q}_{\bot}  ,E} \right) 
=
\end{equation}
$$ 
= - 1/2 \ln\frac{{E\left( {\mathbf{q}_{\bot} } \right)\gamma_{q_{y}} } }{{E\left( {q_{y}} \right)\gamma_{\mathbf{q}_{\bot} } } } + 1/2\;\left[ {E\left( {\mathbf{q}_{\bot}  } \right)/\gamma_{\mathbf{q}_{\bot} }  - E\left( {q_{y}} \right)/\gamma_{q_{y}}  } \right] - \frac{{i}}{{g}}\left( {Eq_{x} - \int\limits_{0}^{q_{x}}  {E\left( {\mathbf{q}_{\bot} } \right)\;dq_{x}} } \right).
$$

To obtain the approximate expression for $I_{1} \left( {\mathbf{q}_{\bot},E} \right)$, we will perform for variables $q_{x}$ the extension transformation of type $q_{x} /g = \eta$ (Ref.\cite{18}, sec.~4.1). In the limit, $g \to 0$ the ``extended'' variable takes values $0 < \;|\eta |\; < \infty$ and asymptotic expression for Eq.(\ref{eq48}) takes the form
\begin{equation}
\label{eq49}
I_{1} \left( {\mathbf{q}_{\bot}  ,E} \right) = I_{1} \left( {g\eta ,q_{y} ,E} \right) = - i\,\left[ {E\eta - \int\limits_{0}^{\eta}  {E\left( {g\eta ,\;q_{y}} \right)\;d\eta} } \right].
\end{equation}

Turning next to the earlier ``no extended'' variables, we will obtain the following approximate expression
\begin{equation}
\label{eq50}
u^{*} \left( {\mathbf{q}_{\bot}  ,E} \right) = A\exp\left[ {\frac{{i}}{{g}}\left( {Eq_{x} - \int\limits_{0}^{q_{x}}  {E\left( {\mathbf{q}_{\bot} } \right)\;dq_{x}}  } \right)} \right] + O\left( {g} \right).
\end{equation}

Using next the recent equation in system (\ref{eq24}) for the second coefficient $v^{*} \left( {\eta ,q_{y} ,E} \right)$ we will find the following asymptotic expression
\begin{equation}
\label{eq51}
v^{*} \left( {\mathbf{q}_{\bot}  ,E} \right) = - \frac{{1}}{{\gamma_{\mathbf{q}_{\bot} } } }\;\left[ { - ig\;\partial /\partial q_{x} + \left( {\sqrt {1 + b_{C}^{2}}  - E} \right)} \right]\;u^{* }\left( {\mathbf{q}_{\bot}  ,E} \right) 
\approx 
\end{equation}
$$ 
\approx \frac{{1}}{{\gamma_{\mathbf{q}_{\bot} } } }\left[ { - \sqrt {1 + b_{C}^{2}}  + E} \right]\;A\exp\left[ {\frac{{i}}{{g}}\left( {Eq_{x} - \int\limits_{0}^{q_{x}}  {E\left( {\mathbf{q}_{\bot} } \right)\;dq_{x}} } \right)} \right] + O\left( {g} \right).
$$

The calculations of coefficient $A$ are placed in Appendix \textbf{A1}. We have
\begin{equation}
\label{eq52}
A^{2} = \frac{{1}}{{4\pi g}}\left( {\frac{{\sqrt {1 + b_{C}^{2}} } }{{E}} + 1} \right).
\end{equation}

Finally, for the unitarity transformation coefficients we will obtain the following expressions:
\begin{equation}
\label{eq53}
u^{*} \left( {\mathbf{q}_{\bot},E} \right) = \frac{{1}}{{\sqrt {4\pi g}} }\sqrt {\frac{{\sqrt {1 + b_{C}^{2}} } }{{E}} + 1\;} \cdot \exp\left[ {\frac{{i}}{{g}}\left( {Eq_{x} - \int\limits_{0}^{q_{x}}  {E\left( {\mathbf{q}_{\bot}} \right)\;dq_{x}} } \right)} \right] + O\left( {g} \right), 
\end{equation}
$$ 
v^{*} \left( {\mathbf{q}_{\bot},E} \right) = \frac{{1}}{{\sqrt {4\pi g}} }\sqrt {\frac{{\sqrt {1 + b_{C}^{2}} } }{{E}} - 1\;} \cdot \exp\left[ {\frac{{i}}{{g}}\left( {Eq_{x} - \int\limits_{0}^{q_{x}}  {E\left( {\mathbf{q}_{\bot}} \right)\;dq_{x}} } \right)} \right] + O\left( {g} \right). 
$$

The integration over variable $E$ in the second term of the right side of Eq.(\ref{eq33}) with the help of Eq.(A1.4) now gives
$$ 
\sum\limits_{Q_{x}}  {\int {\left[ {E_{+} \left( {v^{*} \left( {\mathbf{q}_{\bot} ,E} \right)v\left( {q_{x} + Q_{x} ,q_{y} ,E} \right)} \right) + E_{-}  \,\left( {v\left( {\mathbf{q}_{\bot}  ,E} \right)v^{*} \left( {q_{x} + Q_{x} ,q_{y} ,E} \right) } \right)} \right]dE}} 
\approx 
$$
\begin{equation}
\label{eq54}
\approx \frac{{1}}{{4\pi g}}\left( {\frac{{\sqrt {1 + b_{C}^{2}}  }}{{E}} - 1}            \right)\,\,\sum\limits_{Q_{x}}  {\int\limits_{- \infty }^{\infty}  {2E \exp\left[ {\frac{{i}}{{g}}\left( {E - E\left( {\mathbf{q}_{\bot} } \right)} \right) Q_{x}} \right] d\left( { {E}  - E\left( {\mathbf{q}_{\bot} } \right)} \right)}} + O\left( {g} \right) 
\approx 
\end{equation}
$$ 
\approx  \left( {\sqrt {1 + b_{C}^{2}}  - E\left( {\mathbf{q}_{\bot} } \right)} \right) \delta \left( {Q_{x}} \right)  + O\left( {g} \right) = \left( { \sqrt {1 + b_{C}^{2}} - E\left( {\mathbf{q}_{\bot} } \right) } \right) N_{x} /2\pi + O\left( {g} \right) ,
$$
\noindent
where we accounted that
\begin{equation}
\label{eq55}
\delta \left( {Q_{x}} \right) 
= \lim\limits_{{q}'_{x} - q_{x} \to 0} \delta \left( {{q}'_{x} - q_{x} - Q_{x}} \right) \to N_{x} /2\pi \,\delta_{Q_{x} ,0} = N_{x} /2\pi \,.
\end{equation}

As a result, the expression for the diagonalized Hamiltonian (\ref{eq33}) takes the following approximate asymptotic form:
\begin{equation}
\label{eq56}
h_{S} = - \sqrt {1 + b_{C}^{2}} \left( {N/2 + N_{\bot} } \right) + N_{\bot}  /\left( {2\pi} \right)^{2}\,\int {E\left( {\mathbf{q}_{\bot} } \right)d\mathbf{q}_{\bot}  }
+  
\end{equation}
$$ 
+ \int {\left[ {E_{+}  \xi^{+} \left( {\tau ,q_{y} ,E_{+} } \right) \xi \left( {\tau ,q_{y} ,E_{+} } \right) +  E_{-}  \xi^{+}\left( {\tau ,q_{y} ,E_{-} } \right) \xi \left( {\tau ,q_{y} ,E_{-} } \right)} \right]dEdq_{y}} + O\left( {g} \right).
$$

The first two terms in the right part of Eq.(\ref{eq56}) describe the ground state energy of antiferromagnet with regard to zero-point oscillations. The third term corresponds to the energy of zero oscillations inhomogeneous two-dimensional spin system.

\section{Indirect interspin interaction close to antiferromagnetic spin-flop phase transition}

Let us set off two substituted atoms with nuclear spins at sites $k$ and $l$ from others of substituted atoms in one-dimensional chain. They are arranged in the plane of plate $z_{k} = z_{l} = 0$ and separated along $x$-axis by $x_{lk} = x_{l} - x_{k} \equiv l - k$ distance. For simplicity, we assume that these atoms belong only to sublattice \textbf{A} (Fig.~\ref{fig:3}).

Due to the interaction with antiferromagnet spin system the nuclear spin of $k$-th atom excites a virtual spin wave, which is propagated along the plate and absorbed by nuclear spin of other atom $l$. As a result, the indirect interaction between them is produced. For the case of homogeneous external field, the indirect interaction was considered early in Ref.[8-10]. The contribution of magnetic dipole nuclear interaction with electron spins of neighboring atoms was considered in Ref.\cite{11}.

We will here account the inhomogeneity of external field with weak gradient along one-dimensional nuclear spin chain, which represents one-dimensional quantum register. Hamiltonian of interaction of $k$-th nuclear spin with external magnetic field and of hyperfine interaction with electron spin of own atom will be described by the following expression:
\begin{equation}
\label{eq57}
H_{IS} \left( {k} \right)/\left( {\hbar \omega_{E}} \right) 
= h_{IS} \left( {k} \right) 
= - \omega_{I} \left( {k} \right)I_{\mathrm{A}z} \left( {k} \right) + a\;\mathbf{I}_{\mathrm{A}} \left( {k} \right)\mathbf{S}_{\mathrm{A}} \left( {k} \right) 
= 
\end{equation}
$$ 
= - \left( {\omega_{I} \left( {k} \right) - aS_{\mathrm{A}z} \left( {k} \right)} \right)\;I_{\mathrm{A}z} \left( {k} \right) + a/2\;\left[ {I_{\mathrm{A}}^{+} \left( {k} \right)S_{\mathrm{A}}^{-} \left( {k} \right) + I_{\mathrm{A}}^{-} \left( {k} \right)S_{\mathrm{A}}^{+} \left( {k} \right)} \right],
$$
\noindent
where $I_{\mathrm{A}z}$, $I_{\mathrm{A}}^{\pm}  = I_{\mathrm{A}x} \pm iI_{\mathrm{A}y}$ are components of nuclear spin operators, $\omega_{I} \left( {k} \right) = \left( {\gamma_{I} /\gamma_{S}} \right)\left( {b + gk} \right)$ is resonance nuclear frequency in local field $b_{k} \equiv b + gk$, $\gamma_{I} /\gamma_{S} \sim 10^{-3}$, $a = A/\omega_{E} \sim 10^{-3}$ is isotropic dimensionless constant of hyperfine interaction, $A/2\pi \sim 100\,\mathrm{MHz}$.

Let us go now in Eq.(\ref{eq57}) to interaction representation. Keeping in mind that antiferromagnet is considered in the very low temperatures (close to ground sate) we will retain only linear dependence from electron spin operators $\left( {S_{\mathrm{A}z} \approx 1/2} \right)$:
\begin{equation}
\label{eq58}
h_{IS} \left( {\tau ,k} \right)
= - \left( {\omega_{I} \left( {k} \right) - a/2} \right)I_{\mathrm{A}z} \left( {k} \right) + \Delta h_{IS} \left( {\tau ,k} \right),
\end{equation}

\noindent
where Hamiltonian of nuclear and electron spins transverse hyperfine interaction of the $k$-th atom, considered as perturbation, has the form
\begin{equation}
\label{eq59}
\Delta h_{IS} \left( {\tau ,k} \right) 
= \exp\left( {ih_{S} \tau} \right)\Delta h_{IS} \left( {k} \right)\exp\left( {-ih_{S} \tau} \right) = a/2\;\left[ {I_{\mathrm{A}}^{+} \left( {k} \right)S_{\mathrm{A}}^{-} \left( {\tau ,k} \right) + I_{\mathrm{A}}^{-} \left( {k} \right)S_{\mathrm{A}}^{+} \left( {\tau ,k} \right)} \right].
\end{equation}

Taking into account Eqs.(\ref{eq8}),(\ref{eq11}),(\ref{eq22}), for the considered quasi-two-dimension antiferromagnetic structure we will have (in the following, index $\mathrm{A}$ will be omitted)
$$ 
S^{-} \left( {\tau ,k} \right) = \left( {S^{+} \left( {\tau ,k} \right)} \right)^{+} 
= 
$$
\begin{equation}
\label{eq60}
= \frac{{1}}{{\sqrt {N_{\bot} } } }\sum\limits_{\mathbf{q}_{\bot} }  {a_{\mathbf{q}_{\bot}  }^{+} \left( {\tau} \right)\exp\left( {iq_{x} k} \right)} \to \frac{{1}}{{\left( {2\pi} \right)}}\int {a^{+} \left( {\tau ,\mathbf{q}_{\bot} } \right)\exp\left( {iq_{x} k} \right)d\mathbf{q}_{\bot} } 
=  
\end{equation}
$$ 
= \frac{{1}}{{\left( {2\pi} \right)}}\int {\left[ {u^{*} \left( {\mathbf{q}_{\bot} ,E} \right)\,\exp\left( {iE_{-}  \tau} \right)\,\xi^{+} \left( {q_{y} ,E_{-} } \right) +  v\left( {\mathbf{q}_{\bot}  ,E} \right)\exp\left( { - iE_{+}  \tau \,} \right)\xi \left( {q_{y} ,E_{+} } \right)} \right]\exp\left( {iq_{x} k} \right)\,dEd\mathbf{q}_{\bot}}  . 
$$

Let us assume that hyperfine interactions of nuclear and electron 
spins of own atoms $k$ and $l$ are switched on adiabatically slowly 
at ${\tau}' = - \infty$, 
when the unperturbed density matrix for two nuclear spins in antiferromagnet 
is represented in the form of direct matrix product 
$\rho \left( { - \infty} \right) = \rho_{I} \left( { - \infty} \right)\rho_{S} \left( { - \infty} \right)$, 
and acts to moment ${\tau}'=\tau$.

We will write the equation for electron density matrix $\rho_{S} \left( {{\tau}'} \right)$ in interaction representation
\begin{equation}
\label{eq61}
i\partial \rho_{S} \left( {{\tau}'} \right)/\partial {\tau}' = \left[ {\left( {\Delta h_{IS} \left( {{\tau}',k} \right) + \Delta h_{IS} \left( {{\tau}',l} \right)} \right)\exp\left( { - s|\tau - {\tau}'|} \right),\,\,\,\,\rho_{S} \left( {{\tau}'} \right)} \right],
\end{equation}
\noindent
where $s$ is a small parameter, which characterizes here the rate of interaction switching $\left( {s \ll b_{C} - b} \right)$ in the adiabatic development processes from initial density matrix $\rho_{S} \left( { - \infty} \right)$ to matrix $\rho_{S} \left( {\tau} \right)$. It assumed the value of parameter $s$ is order of the magnon damping rate or width of antiferromagnetic resonance line, which for ideal crystal at low temperatures may be less than or of the order of several oersteds ($\sim 10^{-4}\,\mathrm{T}$) (Ref.\cite{19}).

It follows that in the first order of perturbation theory from Eq.(\ref{eq61}) we will have
\begin{equation}
\label{eq62}
\rho_{S} \left( {\tau} \right) \approx \rho_{S} \left( { - \infty} \right) - i\int\limits_{- \infty}^{\tau}  {\left[ {\left( {\Delta h_{IS} \left( {{\tau}',k} \right) + \Delta h_{IS} \left( {{\tau}',l} \right)} \right),\rho_{S} \left( { - \infty} \right)} \right]\exp\left( { - s\left( {\tau - {\tau}'} \right)} \right)d{\tau}'} ,
\end{equation}
\noindent
where in the limit of low temperatures only matrix element of non-perturbed density matrix $\rho_{S} \left( { - \infty} \right)$ for ground pure state $|0\rangle \langle 0|$ is nonzero, that corresponds to the absence of antiferromagnet magnon modes:
\begin{equation}
\label{eq63}
\langle 0|\xi^{+} \left( {q_{y} ,E_{\pm} } \right)\,\xi \left( {q_{y} ,E_{\pm} } \right)|0\rangle = 0.
\end{equation}

Using now Eq.(\ref{eq62}) and taking into account that partial trace over electron spin states

\noindent
$\tr\nolimits_{S} \rho_{S} \left( { - \infty} \right)\left( {\Delta h_{IS} \left( {\tau ,k} \right) + q\Delta h_{IS} \left( {\tau ,l} \right)} \right) = 0$,
we will obtain the second order correction to Hamiltonian of two nuclear spins as the mean value of their hyperfine interaction over electron ground state:
$$ 
h_{II} \left( {k,l} \right) = \tr\nolimits_{S} \rho_{S} \left( {\tau} \right)\left( {\Delta h_{IS} \left( {\tau,k} \right) + \Delta h_{IS} \left( {\tau ,l} \right)} \right) 
\approx 
$$
\begin{equation}
\label{eq64}
\approx - i\sum\limits_{j,{j}' = k,l} {\tr\nolimits_{S} \int\limits_{- \infty }^{\tau}  {\left[ {\Delta h_{IS} \left( {{\tau}',j} \right),\,\,\rho_{S} \left( { - \infty} \right)} \right]\;\Delta h_{IS} \left( {\tau ,{j}'} \right) \exp\left( {s\left( {{\tau}' - \tau} \right)} \right)d{\tau}'}}  
= 
\end{equation}
$$ 
= - i\sum\limits_{j,{j}' = k,l} {\int\limits_{- \infty}^{0} {\langle 0|\left[ {\Delta h_{IS} \left( {\tau,j} \right),\Delta h_{IS} \left( {\tau + {\tau}',{j}'} \right)} \right]|0\rangle  \exp\left( {s{\tau}'} \right)d{\tau}'}}  . 
$$

Taking into account that $\langle 0|S^{\pm} \left( {\tau ,j} \right),S^{\pm} \left( {\tau + {\tau}',{j}'} \right)|0\rangle = 0$, we will write
\begin{equation}
\label{eq65}
h_{II} \left( {k,l} \right) \approx - ia^{2}/4\sum\limits_{j,{j}' = k,l} {\int\limits_{- \infty}^{0} {\langle 0|\left[ {I^{+} \left( {j} \right)S^{-} \left( {\tau ,j} \right),\,\,I^{-} \left( {{j}'} \right)S^{+}\left( {\tau + {\tau}',{j}'} \right)} \right]|0\rangle \exp\left( {s{\tau }'} \right)d{\tau}' +} }  \quad \mathrm{H.c.}
\end{equation}

After cumbersome rearrangement of Eq.(\ref{eq65}) (see Appendix \textbf{A2}), we will obtain the expression for indirect interaction between two separated nuclear spins, belonging to common sublattice:
\begin{equation}
\label{eq66}
U\left( {k,l} \right) 
=
\end{equation}
$$
= \frac{{a^{2}}}{{2\left( {2\pi} \right)^{2}}}\mathrm{Re}\int {\left[ {\frac{{u^{*} \left( {q_{x} ,q_{y} ,E} \right)u\left( {{q}'_{x} ,q_{y} ,E} \right)}}{{E_{-}  + is}}} 
- \frac{{v\left( {q_{x} ,q_{y} ,E} \right)v^{*} \left( {{q}'_{x} ,q_{y} ,E} \right)}}{{E_{+}  - is}} \right]\exp\left[ {i\left( {q_{x} k - {q}'_{x} l} \right)} \right]\;dEd\mathbf{q}_{\bot}  d{q}'_{x} }, 
$$

Let us retain now in Eq.(\ref{eq66}) only major members, containing in dominator $E_{-}  = E - b$ (the energy of the low magnon mode) and change the variables of integration as follows:
$$ 
 q_{x} ,{q}'_{x} \to \left( {q_{x} - {q}'_{x}} \right) = q_{-}  ,\,\,\;\left( {q_{x} + {q}'_{x}} \right)/2 = q_{+}  ,\,\,\,\,\,\,\,\,dq_{x} d{q}'_{x} = dq_{-}  dq_{+}  , 
$$
\begin{equation}
\label{eq67}
  \,\,\,\,\,\,\,\,\,\,\,\,\,\,\,\,\,\,\,\,\,\,\,\,\,\,\,\,\,\,\,\left( {q_{x} k - {q}'_{x} l} \right) = q_{-} \left( {l + k} \right)/2 - \mathbf{q}_{+} \left( {l - k} \right) 
\end{equation}

Once again, we will use the transition in the exponent to ``extended'' variable $w = q_{-}  /g$ and go to limit $g \to 0$ in the preexponential factors that is we will take $q_{-}  \, = \,q_{x} - {q}'_{x} \; = \,gw$. This allows to perform the following changing $q_{+}  ,\,\,{q}'_{x} \to q_{x} ,$ and $dq_{-}  dq_{+ } \to gdwdq_{x}$. As a result, we will have in place of (\ref{eq66}) the expression
$$ 
 U\left( {k,l} \right) 
= 
$$
\begin{equation}
\label{eq68}
=
\frac{{a^{2}}}{{2\left( {2\pi} \right)^{2}}}g\mathrm{Re}\int {\frac{{|u\left( {\mathbf{q}_{\bot}  ,E} \right)|^{2}\exp\,\,\left[ {i\left( {E - E\left( {\mathbf{q}_{\bot} } \right)\; + g\left( {l + k} \right)/2} \right)\;w - iq_{x} \left( {l - k} \right)} \right]}}{{E_{-}  + is}}dE\,dwd\mathbf{q}_{\bot}  + O\left( {g} \right).} 
\end{equation}

Note here, that the value of denominator $E_{-}  = E - b$ in (\ref{eq68}) may be small but larger than some minimum value, required for implementation of permutation theory condition, which is determined by ignored weak nonlinear effects.

By extending limits of integrating over $w$ in Eq.(\ref{eq68}) to $\pm \infty$, we will obtain
\begin{equation}
\label{eq69}
\int\limits_{- \infty}^{\infty}  {\{ \exp\left[ {i\left( {E - E\left( {\mathbf{q}_{\bot} } \right)\; + g\left( {l + k} \right)/2} \right)\;w} \right]\} d} w \approx 2\pi \delta \left( {E - E\left( {\mathbf{q}_{\bot} } \right)\; + gk + g\left( {l - k} \right)/2} \right).
\end{equation}

By integrating then over $E$ and using Eq.(\ref{eq53}), as a first approximation of asymptotic expansion in terms of $g \ll 1$ we will have
\begin{equation}
\label{eq70}
U\left( {k,l} \right) = \frac{{a^{2}}}{{4\left( {2\pi} \right)^{2}}}\mathrm{Re}\int {\frac{{\sqrt {1 + b_{C}^{2}}  + E\left( {\mathbf{q}_{\bot} } \right)}}{{E\left( {\mathbf{q}_{\bot} } \right)}} \cdot \frac{{\cos\left( {q_{x} \left( {l - k} \right)} \right)}}{{E\left( {\mathbf{q}_{\bot} } \right) - b_{\left( {l + k} \right)/2} + is}}d\mathbf{q}_{\bot}  + O\left( {g} \right)} ,
\end{equation}
\noindent
where $b_{\left( {l + k} \right)/2} = b + g\left( {l + k} \right)/2 = b_{k} + g\left( {l - k} \right)/2$ is the local external field in the middle point $\left( {l + k} \right)/2$ of $l$-th and $k$-th atoms. It is further conveniently to choose the field value $b_{k} = b + gk$ at $k$-th site as the origin of inhomogeneous field reading.

In the following, we will simplify the expression for magnon spectrum 
\begin{equation}
\label{eq71}
E\left( {\mathbf{q}_{\bot} } \right) = \sqrt {1 + b_{C}^{2} - \gamma_{\mathbf{q}_{\bot} }^{2}} .
\end{equation}
\noindent
For this we will replace the initial $\mathbf{q}_{\bot}$ dependence of $\gamma_{\mathbf{q}_{\bot} }^{2}$ for all components of wave vector values $0 \le |q_{x} |,|q_{y} | < \pi$ by parabolic isotropic dependence. That is we perform in our model the following replacing
\begin{equation}
\label{eq72}
\gamma_{\mathbf{q}_{\bot} }^{2} = 
\left[ {1/3\,\left( {\cos\left( {q_{x} /2} \right) + \cos\left( {q_{y} /2} \right) + 1} \right)} \right]^{2} \to 
= 1 - \mathbf{q}_{\bot}^{2} /12 ,
\end{equation}
\noindent
wich coincide with the expression received by simple expansion for small values $\mathbf{q}_{\bot}^{2}$ of initial spectrum:
\begin{equation}
\label{eq73}
\gamma_{\mathbf{q}_{\bot} }^{2} \approx 1 - \mathbf{q}_{\bot}^{2} /12 + \dots .
\end{equation}

Because the small values of $\mathbf{q}_{\bot}^{2}$ are the most essential for the following, we will use the form (\ref{eq73}), extrapolating its in the expression under integral over all values $0 \le \mathbf{q}_{\bot}  < \pi$.

Using this simplification in the expression (\ref{eq71}), we will go to the new variable of integrating $\xi + b_{C} = E\left( {\mathbf{q}_{\bot} } \right) \approx \sqrt {b_{C}^{2} + \mathbf{q}_{\bot}^{2} /12}$ and $d\mathbf{q}_{\bot}  = \mathbf{q}_{\bot}  d\mathbf{q}_{\bot}  d\varphi = 12\left( {\xi + b_{C}} \right)d\xi d\varphi$. As a result, we will obtain (the symbol ${O}(g)$ will be then omitted) the expression, generalizing the known Nakamura's expression for antiferromagnet in homogeneous field \cite{7}:
\begin{equation}
\label{eq74}
\frac{{2\pi}}{{3a^{2}}}U\left( {k,l} \right) = V\left( {\Delta b_{k} ,l - k} \right) 
= 
\end{equation}
$$ 
= \int\limits_{0}^{\sqrt {b_{C}^{2} + \pi^{2}/12} - b_{C}}  {\frac{{\left( {\sqrt {1 + b_{C}^{2}}  + b_{C} + \xi} \right)\,\,\left( {\xi + \Delta b_{k} - g\left( {l - k} \right)/2} \right)J_{0} \left( {\sqrt {12\left( {\left( {\xi + b_{C}} \right)^{2} - b_{C}^{2}} \right)} \,\left( {l - k} \right)} \right)d\xi} }{{\left( {\xi + \Delta b_{k} - g\left( {l - k} \right)/2} \right)^{2} + s^{2}}}} , 
$$
\noindent
where the integral over $\varphi$ is expressed in terms of zero-order Bessel function of the first order (Ref.\cite{20}, Eq.~3.715-18.):
\begin{equation}
\label{eq75}
\int\limits_{0}^{2\pi}  {\cos\,\left[ {\mathbf{q}_{\bot} \left( {l - k} \right)\cos\varphi} \right]d\varphi = 2\pi \,J_{0} \left( {\mathbf{q}_{\bot} \left( {l - k} \right)} \right)} 
\end{equation}
\noindent
and it is denoted
\begin{equation}
\label{eq76}
24b_{C} \,\left( {b_{C} - b - g\left( {l + k} \right)/2} \right) = 24b_{C} \,\left( {\Delta b_{k} - g\left( {l - k} \right)/2} \right) = \mu_{\left( {l + k} \right)/2}^{2} \,,\,\,\,\,\,\,\,\,\,\Delta b_{k} = b_{C} - b - gk .
\end{equation}

For $\mu_{\left( {l + k} \right)/2}^{2} > 0$ the ranges of interqubit distance are
\begin{equation}
\label{eq77}
1 \le \,\,l - k\, \le 2\Delta b_{k} /g.
\end{equation}

For $\mu_{\left( {l + k} \right)/2}^{2} = - \nu_{\left( {l + k} \right)/2}^{2} < 0$ the ranges are
\begin{equation}
\label{eq78}
2\Delta b_{k} /g \le l - k\,
\end{equation}

The critical value of the middle point position $\left( {l + k} \right)/2 = b_{C}$ or $\Delta b_{k} = g\left( {l - k} \right)/2$, wherein the parameter $\mu_{\left( {l + k} \right)/2}^{2}$ changes sign and the dependence of indirect interaction of nuclear spins from interspin distance is qualitatively changed, we will denoted as ``turning point'', by analogy to problem on the barrier reflection of quantum mechanical particle in WKB-approximation (Ref.\cite{21}, sec.~46). Note, that further it will be conveniently to tune through parameter $\Delta b_{k} = b_{C} - b - gk$ the state of nuclear spin $k$ in quantum register and to use spins $l$ in the turning points, separated by distances $l - k\, = Ln_{k} = 2\Delta b_{k} /g$, where $L$ is period of chain, $n_{k} = 1,\,2,\,3,\dots $. To perform the two-qubit operations in quantum register it should switched on the interaction between considered nuclear spins. The way to do this is through the tuning of the spin-qubit state to the turning point state. This requires the fast as compared to operation and decoherence rate variation of external field.

Let us account now that major contribution to the integral in Eq.(\ref{eq74}) is given by values 
$\xi \ll 1$, 
$s \ll \Delta b_{k}$ and
$\pi^{2} \gg \mu_{\left( {l + k} \right)/2}^{2} > 0$,
whereby the denominator is small. Let us retain next under integrals only quadratic on $\mathbf{q}_{\bot}$ terms and assuming then the rapid convergence of the integral on upper limit, we will replace $\pi \to \infty$. As a result we will obtain for indirect interaction the approximate expression (Ref.\cite{20}, Eqs.~6.532-4, 9.561.5):
$$ 
\frac{{2\pi }}{{3a^{2}}}U\left( {k,l} \right) = V\left( {\Delta b_{k} ,l - k} \right) 
\approx 
$$
\begin{equation}
\label{eq79}
\approx 2\left( {\sqrt {1 + b_{C}^{2}}  + b_{C}} \right)\int\limits_{0}^{\infty}  {\frac{{J_{0} \left( {\mathbf{q}_{\bot} \left( {l - k} \right)} \right)\,\,\mathbf{q}_{\bot}  d\mathbf{q}_{\bot} } }{{\mathbf{q}_{\bot} ^{2} + \mu_{\left( {l + k} \right)/2}^{2}} }} + 2\int\limits_{0}^{\pi}  {\frac{{J_{0} \left( {\mathbf{q}_{\bot} \left( {l - k} \right)} \right)\,\,\mathbf{q}_{\bot}  d\mathbf{q}_{\bot}  }}{{24b_{C}} }} 
\approx 
\end{equation}
$$ 
\approx 2\left( {\sqrt {1 + b_{C}^{2}}  + b_{C}} \right)\,K_{0} \left( {\mu_{\left( {l + k} \right)/2} |l - k|} \right), 
$$
\noindent
where $K_{0} \left( {x} \right)$ is zero order Macdonald function.

Its asymptotic value for $\pi \left( {l - k} \right) > \mu_{\left( {l + k} \right)/2} |l - k|\, \gg 1$ is:
\begin{equation}
\label{eq80}
V\left( {\Delta b_{k} ,l - k} \right) \approx 2\left( {\sqrt {1 + b_{C}^{2} } + b_{C}} \right)\sqrt {\frac{{\pi} }{{2\mu_{\left( {l + k} \right)/2} |l - k|}}} \exp\left[ { - \left( {\mu_{\left( {l + k} \right)/2} |l - k|} \right)} \right]\;..
\end{equation}

As is seen from Eq.(\ref{eq80}), the parameter $\left( {\mu_{\left( {l + k} \right)/2}} \right)^{-1}$ presents the effective radius of indirect interaction between nuclear spins. In the case of homogeneous field it rapidly increases close to value for bulk spin-flop phase transition, when $b_{C} \to b$. In the case of inhomogeneous field effective radius increases at condition $\mu_{\left( {l + k} \right)/2}^{2} \to + 0$, that is when only local field in the middle point for the considered two spins is close to the value for bulk spin-flop phase transition $b + g\left( {l + k} \right)/2 \to b_{C}$ or $\Delta b_{k} + g\left( {l - k} \right)/2 \to 0$. At the same time, relation $b_{C} > b$ takes place and homogeneous phase transition does not occur. Note that if aforementioned condition is not fulfilled, the radius of indirect interaction remains of the order of lattice parameter.

To receive the real value $V\left( {\Delta b_{k} ,l - k} \right)$ for the case $\mu_{\left( {l + k} \right)/2}^{2} = - v_{\left( {l + k} \right)/2}^{2} < 0$, the lower limit of the first integral in Eq.(\ref{eq79}) should be replaced by $v_{\left( {l + k} \right)/2} > 0$. As a result, we will obtain (Ref.\cite{20}, Eqs.~3.753-4, 8.405-1, 8.407-2):
\begin{equation}
\label{eq81}
V\left( {\Delta b_{k} ,l - k} \right) \approx 2\left( {\sqrt {1 + b_{C}^{2}}  + b_{C}} \right)\,\mathrm{Re} K_{0} \left( { - i\nu_{\left( {l + k} \right)/2} \left( {l - k} \right)} \right) 
= 
\end{equation}
$$ 
= - \,\,2\left( {\sqrt {1 + b_{C}^{2}}  + b_{C}} \right)\,\,\frac{{\pi }}{{2}}N_{0} \left( {\nu_{\left( {l + k} \right)/2} \left( {l - k} \right)} \right), 
$$
\noindent
where $N_{0} \left( {x} \right)$ is zero order Neumann function (or zero order Weber function $Y_{0} \left( {x} \right)$). Using its asymptotic value for $\nu_{\left( {l + k} \right)/2} \left( {l - k} \right) = \sqrt {24b_{C} \left( {g\left( {l - k} \right)/2 - \Delta b_{k}} \right)} \left( {l - k} \right) \approx \sqrt {24b_{C} g\left( {l - k} \right)^{3}/2} \gg 1$ we will have
\begin{equation}
\label{eq82}
V\left( {\Delta b_{k} ,l - k} \right) \approx 2\left( {\sqrt {1 + b_{C}^{2} } + b_{C}} \right)\sqrt {\frac{{\pi} }{{2\nu_{\left( {l + k} \right)/2} |l - k|}}} \sin\left( {\nu_{\left( {l + k} \right)/2} |l - k| - \pi /4} \right).
\end{equation}

In this case, dependence of effective interaction from distance between spins takes oscillating character with the quasi-period, which is decreased with increasing of the distance $l - k$. As this take place, the interaction goes to zero at the sequence of distance $\sqrt {24b_{C} g\left( {l - k} \right)^{3}/2} \approx \left( {n + 1/4} \right)\pi$.

   \begin{figure}
   \begin{center}
   \begin{tabular}{c}
   ~~~~~~~~~~\includegraphics{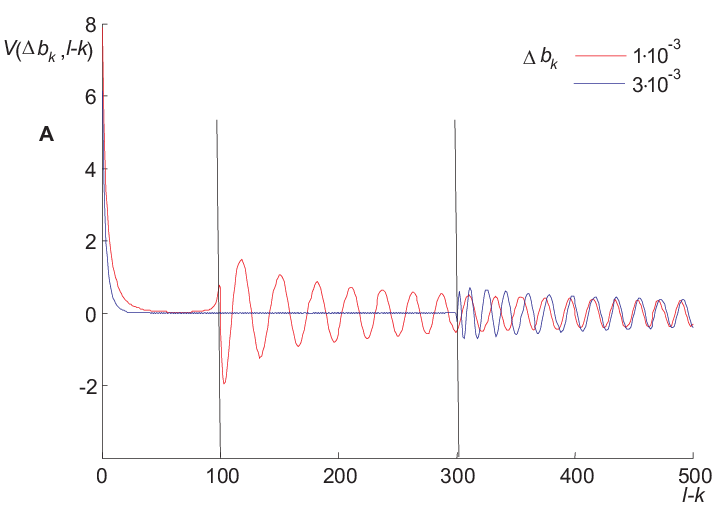}
\\
~~~~~~~~~~
   \includegraphics{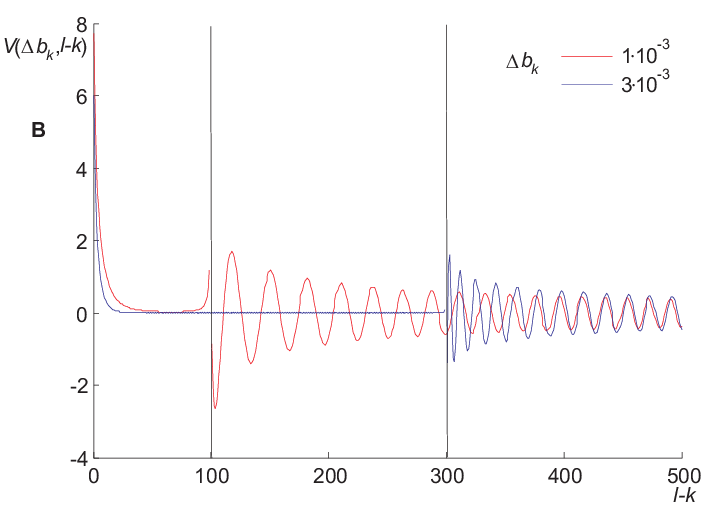}
   \end{tabular}
   \end{center}
   \caption[1]
   {\label{fig:5} 
Dimnsionless indirect interaction $V\left( {\Delta b_{k} ,l - k} \right)$ as a function of interqubit distance $l - k$ for ``exact'' expression (\ref{eq74}) (\textbf{A}) and for approximate analytical expressions (\ref{eq79}), (\ref{eq81}) (\textbf{B)}. The used numerical values of parameters are: $b_{C}^{2} = 1/4$, $g = 2 \cdot 10^{-5}$, $L = 100, \quad s = 10^{-5}$ and $\Delta b_{k} = b_{C} - b_{k} = 1 \cdot 10^{-3},\,\,\,3 \cdot 10^{-3}$. Note, that transition to value $g \ne 2 \cdot 10^{-5}$ corresponds here to the following modification of turning points positions (vertical lines): $Ln_{k} = 2\Delta b_{k} /g \to Ln_{k} \cdot 2 \cdot 10^{-5}/g.$
   }
   \end{figure}

As is seen in Fig.~\ref{fig:5}, in the case that value of damping $s = 10^{-5} \ll \Delta b_{k}$, the approximate distance dependence of indirect nuclear spin interaction (in the form of Eqs.(\ref{eq79}),(\ref{eq81})) outside of the turning points is in good agreement with exact relation (Eq.(\ref{eq74})).

In the limiting cases for $\mu_{\left( {l + k} \right)/2} \left( {l - k} \right) \ll 1$, $\nu_{\left( {l + k} \right)/2} \left( {l - k} \right) \ll 1$ we will have, respectively, the following logarithmic growing with decreasing $|l - k|$ asymptotic dependences of indirect interaction (Ref.\cite{20}, Eqs.~8.447-3, 8.444-1) (\textbf{C} = 0.577\dots  . is Euler constant):
\begin{equation}
\label{eq83}
V\left( {\Delta b_{k} ,l - k} \right) \approx 2\left( {\sqrt {1 + b_{C}^{2}}  + b_{C}} \right) \ln\left( {2/\left( {\mu_{\left( {l + k} \right)/2} \left( {l - k} \right)} \right)} \right),
\end{equation}
\begin{equation}
\label{eq84}
V\left( {\Delta b_{k} ,l - k} \right) \approx 2\left( {\sqrt {1 + b_{C}^{2} } + b_{C}} \right)\,\,\left[ {\ln\left( {2/\left( {\nu_{\left( {l + k} \right)/2} \left( {l - k} \right)} \right) }\right) - C} \right].
\end{equation}

Note however, that using the second order of perturbation theory for calculating the indirect interaction $V\left( {\Delta b_{k} ,l - k} \right)$ immediately close to the turning point is not sufficient and the account of magnon damping and relaxation effects are required. Therefore, minimal values $v_{\left( {l + k} \right)/2}$ and $\mu_{\left( {l + k} \right)/2}$ should be bonded by same small nonzero values. This values will be bound also the range of maximal distances wherein the expressions (\ref{eq79}) and (\ref{eq81}) are applicable and, respectively, the size of quantum register is estimated.

We will obtain the corresponding expression for $U\left( {k,k} \right)$ assuming $l = k$ and $s \ll \Delta b_{k}$ in Eq.(\ref{eq74}):
\begin{equation}
\label{eq85}
\frac{{2\pi} }{{3a^{2}}}U\left( {k,k} \right) = V\left( {\Delta b_{k} ,0} \right) = \int\limits_{0}^{\sqrt {b_{C}^{2} + \tau^{2}/12} - b_{C}}  {\frac{{\left( {\sqrt {1 + b_{C}^{2}}  + b_{C} + \xi} \right)\,d\xi} }{{\xi + \Delta b_{k}} }} 
= 
\end{equation}
$$ 
= \left( {\sqrt {1 + b_{C}^{2}}  + b_{C} - \Delta b_{k}} \right)\ln\frac{{\sqrt {b_{C}^{2} + \pi^{2}/12} - b_{C} + \Delta b_{k} }}{{\Delta b_{k}} } + \sqrt {b_{C}^{2} + \pi^{2}/12} - b_{C} . 
$$

It has relatively weak logarithmic dependence on the position of $k$-th nuclear spin.

Let us estimate, finally, the value of direct dipole interaction between nuclear spins-qubits in quantum register. Hamiltonian of transverse part of dipole interaction for nuclear spins $k$ and $l$ has the form
\begin{equation}
\label{eq86}
H_{dd} \left( {l - k} \right)/\hbar \omega_{E} = D\frac{{\left( {I_{y} \left( {l} \right)I_{y} \left( {k} \right) - 2I_{x} \left( {l} \right)I_{x} \left( {k} \right)} \right)}}{{|l - k|^{3}}},
\end{equation}
\noindent
where dimensionless ratio
\begin{equation}
\label{eq87}
D = \frac{{\gamma_{I} B_{I}} }{{\gamma_{S} B_{E}} },
\end{equation}
\noindent
in which
\begin{equation}
\label{eq88}
B_{I} = \frac{{\mu_{0}} }{{4\pi} }\frac{{\left( {\gamma_{I} \hbar} \right)}}{{a_{x}^{3}} }
\end{equation}
\noindent
is the dipole field produced by nucleus, which is located at a distance of the order of sublattice period $a_{x} \sim 1\,\mathrm{nm}$, $\mu_{0} = 0.4\pi \,\,\mathrm{T}^{2}\mathrm{cm}^{3}/\mathrm{J}$ is vacuum magnetic permeability. This field is of the order of $10^{-6}\,\mathrm{T}$, the ratio $D\sim 10^{-10} \ll \,\,a^{2}\sim 10^{-8}$.

The ratio of dipole interaction to the indirect interaction (in the form of Eq.(\ref{eq80}) is:
\begin{equation}
\label{eq89}
\frac{{D}}{{U\left( {k,l} \right)\,\,|k - l|^{3}}}\sim \frac{{\sqrt {2\pi}  D}}{{3a^{2}\left( {\sqrt {1 + b_{C}^{2}}  + b_{C}} \right)}}\frac{{\,\mu_{\left( {l + k} \right)/2}^{3}} }{{\,R^{5/2}\exp\left( { - R} \right)\,}},
\end{equation}
\noindent
where $R = \mu_{\left( {l + k} \right)/2} |l - k|\, > 0$.

\begin{figure}
   \begin{center}
   \begin{tabular}{c}
   \includegraphics{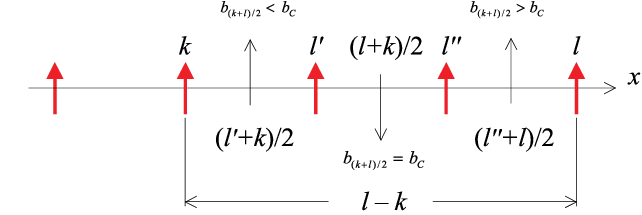}
   \end{tabular}
   \end{center}
   \caption[1]
   {\label{fig:6} 
The indirect interaction $V\left( {\Delta b_{k} ,l - k} \right)$ is large for qubit pairs $l,k$ marked by the local field close to middle point $b_{\left( {k + l} \right)/2} \sim b_{C}$ or for $1 \gg \mu_{\left( {l + k} \right)/2}^{2} > 0$ At the same time, this interaction is far less for other qubit pairs with $b_{\left( {k + l} \right)/2} < b_{C}$.
   }
   \end{figure}

Maximum of $R^{5/2}\exp\left( { - R} \right)$ obtained at $R = 2,5$ is 0,78. For the same value of $R$ and for distance $|k - l|\, > 100$, parameter $\mu_{\left( {l + k} \right)/2}$ corresponds to value $0 < \mu_{\left( {l + k} \right)/2} \ll 1$ or $0 < \,24b_{C} \left( {b_{C} - b - g\left( {l + k} \right)/2} \right) \ll 1\,.$ In the case, when the value of local field in middle point is close to the critical field of spin-flop phase transition, the indirect interaction becomes very large and dipole interaction for considered nuclear spins would be negligibly small in comparison with indirect interaction (Fig.~\ref{fig:6}). It follows that a possibility of considerably increasing of interaction arises for a definite pair of removed qubits as compared with interaction between all another pairs. Furthermore, if the local field in middle point is more than the critical field or $\,24b_{C} \left( {b_{C} - b - g\left( {l + k} \right)/2} \right) < 0,$ the indirect interaction takes oscillating character (see Fig.~\ref{fig:5}).

Let us assume for example, that register has 1000 nuclear spins-qubits that is the total distance $|l - k|$ in register with period 100 qubits is of the order of $l - k = 10^{5}$ and value of $\mu_{\left( {l + k} \right)/2}^{2}$, wherein the indirect interaction is essential, corresponds to turning point or $\Delta b_{k} = b_{C} - b_{k} \sim g\left( {l - k} \right)/2\sim 1$.

\section{Indirect interspin interaction close to antiferromagnetic resonance (AFR)}

The homogeneous antiferromagnetic resonance (AFR) phenomenon refers to the excitation in antiferromagmet by microwave field's homogeneous mode with $\mathrm{q} = 0$. Resonance in thin plate will be well resoluted when the energy difference of principal homogeneous transverse mode with $q_{z} = 0$ and of the next transverse modes with $q_{z} = 2\pi a_{z} /d$, will fit the condition
\begin{equation}
\label{eq90}
\sqrt {b_{C}^{2} + \left( {2\pi a_{z} /d} \right)^{2}/12} - b_{C} > \Delta \omega_{0} /\omega_{E} ,
\end{equation}
\noindent
where $\Delta \omega_{0}$ is AFR width. It follows that $b_{C} > \left( {2\pi a_{z} /d} \right)^{2}/24b_{C} > \Delta \omega_{0} /\omega_{E}$ that is thickness $d$ of large plate and correspondingly the site numbers of sublattice $N_{z}$, should be determined by inequalities
\begin{equation}
\label{eq91}
\frac{{\pi} }{{\sqrt {6b_{C}} } }\,\, < N_{z} = d/a_{z} < \pi \sqrt {\frac{{\omega_{E}} }{{6b_{C} \Delta \omega_{0}} }} ,
\end{equation}

Consequently, for example, if $\Delta \omega_{0} /\omega_{E} \sim 10^{-5}b_{C}$ the thickness should be in the range of micrometers. Note, that the condition permitting to restrict by only zero mode in the case of Sec.~2 differs from the condition (\ref{eq91}) by the change of the AFR width to the quantum states decoherence rate $1/T_{D}$.

Let us assume that the antiferromagnet plate interact with the homogeneous transverse microwave field left-hand circularly polarized relatively to external field (dimensionless amplitude $b_{\bot}  \ll 1$ and frequency $\omega$):
\begin{equation}
\label{eq92}
b_{x} \left( {\tau} \right) + ib_{y} \left( {\tau} \right) = b_{\bot}  \exp\left( {i\omega \tau} \right).
\end{equation}

The interaction Hamiltonian in interaction representation, accounting that directions of sublattice spin vectors have opposite sign (Ref.\cite{17}, sec.~4.2), has the form:
$$ 
 \Delta H_{S} \left( {\tau} \right)/\left( {\hbar \omega_{E}} \right) = 
 \Delta h_{S} \left( {\tau} \right) 
       = b_{\bot}  \exp\,\,\left( {\,i\omega \tau} \right)\,\,\,\sum\limits_{i = 1}^{N} {\{ S_{\mathrm{A}}^{+} \left( {\tau ,\mathbf{r}_{i}} \right) - 1/6\sum\limits_{\delta}^{6} {S_{\mathrm{B}}^{+} \left( {\tau ,\left( {\mathbf{r}_{i} + \mathbf{r}_{\delta}} \right) } \right)} \} }+ \mathrm{H.c.} 
\approx 
$$
\begin{equation}
\label{eq93}
\approx b_{\bot}  \exp\,\,\left( {\,i\omega \tau} \right)\,\,\,\sqrt {N_{\bot} }  \sum\limits_{q_{x} ,\,\,Q_{x} ,\,q_{y}}  {\delta_{q_{x} ,Q_{x}}  \delta_{q_{y} ,0} \left( {a_{\mathbf{q}_{\bot} } \left( {\tau} \right) - \gamma_{\mathbf{q}_{\bot} }  b_{- \mathbf{q}_{\bot} }^{+} \left( {\tau} \right)} \right)} + \mathrm{H.c.} 
\to 
\end{equation}
$$ 
\to \int {F\left( {E} \right)\exp\left( { - i\Delta \left( {E,\omega} \right)\tau} \right)\xi \left( {0,E_{-} } \right)\,dE\,\,} \, + \mathrm{H.c.}, 
$$
\noindent
where
$$ 
F\left( {E} \right) = 2\pi \rlap{--} {b}_{\bot}  \sum\limits_{Q_{x}}  {\left[ {u\left( {Q_{x} ,0,E} \right) - \gamma_{Q_{x}}  v\left( {Q_{x} ,0,E} \right)} \right] } 
=  
$$
\begin{equation}
\label{eq94}
= 2\pi \rlap{--} {b}_{\bot}  \sum\limits_{Q_{x}}  {\left[ {u\left( {0,E} \right) - 1/6\left( {\exp\left( {iQ_{x} /2} \right) + \exp\left( { - iQ_{x} /2} \right) + 4} \right)\,v\left( {0,E} \right)} \right]\exp\left( { - i\left( {E - b_{C}} \right)Q_{x} /g} \right)} 
\approx 
\end{equation}
$$ 
\approx \rlap{--} {b}_{\bot}  \sqrt {\frac{{g\pi b_{C}} }{{\sqrt {1 + b_{C}^{2}} } }} \delta \left( {E - b_{C}} \right) + O\left( {g} \right). 
$$

Here the relations (\ref{eq2}), (\ref{eq6}), (\ref{eq11}), (\ref{eq20}) are used and it is accounted only lower branch of magnon modes with energy $E_{-}$. To receive the final result in the Eq.(\ref{eq94}), we have used Eq.(\ref{eq13}) for $\gamma_{Q_{x}}$, Eq.(A1.4) for sum over $Q_{x} = 2\pi m$ and accounted also relation $g/2 \ll b_{C} < 1$. It follows from Eq.(\ref{eq93}) that magnon modes can be excite only with $q_{y} = 0$. The difference $\Delta \left( {E,\omega} \right) = E_{-}  - \omega = E - b - \omega$ plays here the role of energy gap for lower branch of magnon spectrum in the frame of rotating with microwave field reference. In the absence of gradient of external field, this gap is the frequency detuning of microwave field relatively to frequency of homogeneous AFR: $\omega_{0} = b_{C} - b$. It follows that AFR may be considered as a spin-flop phase transition in rotating frame of reference at critical field $b_{C} = b_{eff} \left( {\omega} \right) = \left( {b + \omega} \right)$. At the same time, the equilibrium phase transition at this field does not take place if $b_{C} > b$.

Let us use now the so called resonance approximation (Ref.\cite{21}, sec.~40), whereby the frequency detuning is assumed to be small $|\Delta \left( {E,\omega}  \right)|\, \ll \omega$ and accounted only two magnon states. One state corresponds to the homogeneous non-perturbed ground state $|0\rangle $ in the absence of magnons, when $\left\langle {0|\xi^{+}\left( {{q}'_{y} ,{E}'} \right)\xi \left( {q_{y} ,E} \right)|\left. {0} \right\rangle = 0} \right.$. The second state $\int\limits_{0}^{\delta E} {\int\limits_{0}^{\delta q_{y}}  {|q_{y} ,E\rangle dEdq_{y}} }$ corresponds to the excited magnon packet state in rotating frame of reference. It is characterized by the narrow energy interval $\delta E \ll b_{C} $ near $E = E\left( {q_{\bot}  = 0} \right) = b_{C}$ and by the minimal discrete interval $\delta q_{y} = 2\pi /N_{y} $ for wave vector component $q_{y} $ near $q_{y} = 0$. The ground and magnon pocket states will be normalized as follows:
\begin{equation}
\label{eq95}
\langle 0|0\rangle = 1  \textrm{ and } \int\limits_{0}^{\delta E} {\int\limits_{0}^{\delta q_{y}}  {\langle q_{y,} E|q_{y,} E\rangle dEdq_{y}} } = 1
\end{equation}

Let us represent by microwave field perturbed state vector $|\tau \rangle $ as superposition of this two steady states:
\begin{equation}
\label{eq96}
|\tau \rangle = |0\rangle \,\,c_{0} \left( {\tau}  \right) + \int\limits_{0}^{\delta E} {\int\limits_{0}^{\delta q_{y}}  {|q_{y} ,E\rangle \,\,c_{1} \left( {E,\,q_{y} ,\tau}  \right)\,dEdq_{y}} }  .
\end{equation}

We will next consider only two non-diagonal matrix elements for perturbation operator (\ref{eq93}), which have the forms:
\begin{equation}
\label{eq97}
\langle 0|\Delta h_{S} \left( {\tau}  \right)|q_{y} ,E\rangle = exp\left( { - i\left( {b_{C} - b - \omega}  \right)\tau}  \right)\int {F\left( {{E}'} \right)\,\langle 0|\xi \left( {q_{y} ,{E}'_{ -} }  \right)|q_{y} ,E\rangle d{E}'} , 
\end{equation}
$$ 
\langle q_{y} ,E|\Delta h_{S}^{ +}  \left( {\tau}  \right)|0\rangle = exp\left( {i\left( {b_{C} - b - \omega}  \right)\tau}  \right)\,\,\int {F^{ *} \left( {{E}'} \right)\,\langle q_{y} ,E|\xi ^{ +} \left( {q_{y} ,{E}'_{ - }}  \right)|0\rangle d{E}'} , 
$$
\noindent
where the frequency detuning value in this interval will be taken constant: $|\Delta \left( {E,\omega}  \right)| \approx b_{C} - b - \omega$.

By using the Schr\"odinger equation for interaction Hamiltonian (\ref{eq93}) (magnon relaxation processes for simplicity are here neglected)
\begin{equation}
\label{eq98}
i\partial |\tau \rangle /\partial \tau = \Delta h_{S} \left( {\tau}  \right)|\tau \rangle 
\end{equation}

\noindent
and accounting the conditions (\ref{eq95}), we will obtain the system of equations for determination of amplitudes $c_{0} \left( {\tau}  \right)\,$ and$\,\,c_{1} \left( {E,q_{y} ,\tau}  \right) \approx c_{1} \left( {b_{C} ,\tau}  \right)$:
\begin{equation}
\label{eq99}
 i\partial c_{0} \left( {\tau}  \right)/\partial \tau = \,b_{ \bot}  exp\left[ { - i\left( {b_{C} - b - \omega}  \right)\tau}  \right]\,\int {F\left( {{E}'} \right)d{E}'\int\limits_{0}^{\delta E} {\int\limits_{0}^{\delta q_{y}}  {\langle 0|\xi \left( {q_{y} ,{E}'_{ -} }  \right)|q_{y} ,E\rangle c_{1} \left( {E,q_{y} ,\tau}  \right)dEdq_{y}} } }  , 
\end{equation}
$$ 
i\partial c_{1} \left( {E,q_{y} ,\tau}  \right)/\partial \tau = b_{ \bot}  exp\left[ {i\left( {b_{C} - b - \omega}  \right)\tau}  \right]\,\,\int {F^{ *} \left( {{E}''} \right)\,d{E}''\langle q_{y} ,E|\xi ^{ +} \left( {q_{y} ,{E}''_{ -} }  \right)|0\rangle}  .c_{0} \left( {\tau}  \right),
$$
\noindent
which should be solved with the normalizing condition
\begin{equation}
\label{eq100}
|c_{0} \left( {\tau}  \right)|^{2} + \int\limits_{0}^{\delta E} {\int\limits_{0}^{\delta q_{y}}  {\langle q_{y} ,E|q_{y} ,E\rangle |c_{1} \left( {E,q_{y} ,\tau}  \right)|^{2}dEdq_{y}} }  \approx \,|c_{0} \left( {\tau}  \right)|^{2} + |c_{1} \left( {b_{C} ,\tau}  \right)|^{2} = 1
\end{equation}
\noindent
and initial condition $|c_{0} \left( {0} \right)|^{2} = 1$.

By eliminating next the amplitude $c_{1} \left( {E,\tau}  \right)$ from system (\ref{eq99}), we will obtain for determination of $c_{0} \left( {\tau}  \right)$ the following equation
\begin{equation}
\label{eq101}
d^{2}c_{0} \left( {\tau}  \right)/d\tau ^{2} + i\left( {b_{C} - b - \omega}  \right)dc_{0} \left( {\tau}  \right)/d\tau + |B|^{2}c_{0} \left( {\tau}  \right) = 0,
\end{equation}
\noindent
wherein
\begin{equation}
\label{eq102}
|B|^{2} = \int {F\left( {{E}'} \right)F^{ *} \left( {{E}''} \right)d{E}'d{E}''\,\,} \int\limits_{0}^{\delta E} {\int\limits_{0}^{\delta q_{y}}  {\langle 0|\xi \left( {q_{y} ,{E}'_{ -} }  \right)|q_{y} ,E\rangle \langle q_{y} ,E|\xi ^{ +} \left( {q_{y} ,{E}''_{ -} }  \right)|0\rangle \,\,dEdq_{y}} }  
= 
\end{equation}
$$ 
\,\,\,\,\,\,\,\,\,\,\,\, = \int {F\left( {{E}'} \right)F^{ *} \left( {{E}''} \right)} d{E}'d{E}''\,\,\delta q_{y} \langle 0|\xi \left( {0,{E}'_{ -} }  \right)\xi ^{ +} \left( {0,{E}''_{ -} }  \right)|0\rangle .  
$$

Let us use now expression (A2.4) 
$$
\left\langle {0|\left[ {\xi \left( {q_{y} ,{E}'_{ -} }  \right),\,\;\xi ^{ + }\left( {{q}'_{y} ,{E}''_{ -} }  \right)} \right]|\left. {0} \right\rangle}  \right. = \left\langle {0|\xi \left( {q_{y} ,{E}'_{ -} }  \right)\xi ^{ + }\left( {{q}'_{y} ,{E}''_{ -} }  \right)|\left. {0} \right\rangle}  \right. = \delta \left( {q_{y} - {q}'_{y}}  \right)\delta \left( {{E}' - {E}''} \right) 
$$
\noindent
and go according to Eqs.(6), (11) from continuous to discrete values $q_{y} $, namely, $\xi \left( {q_{y} ,E_{ -} }  \right) \to \sqrt {N_{y} /2\pi \,} \xi _{q_{y}}  \left( {E_{ -} }  \right)$ and $\delta \left( {q{}_{y} - {q}'_{y}}  \right) \to N_{y} /2\pi \,\delta _{q{}_{y},{q}'_{y}}  $ (compare with Eq.(\ref{eq31})). Taking then $q_{y} = 0$ and using Eq.(A1.6), we will obtain for the matrix element product in Eq.(\ref{eq102}) the expression:
\begin{equation}
\label{eq103}
 \lim\limits_{{E}' \to {E}'' = b_{C}}  \langle 0|\xi \left( {0,{E}'_{ -} }  \right)\xi ^{ +} \left( {0,{E}''_{ -} }  \right)|0\rangle \to \,\,N_{y} /2\pi \,\,\lim\limits_{{E}' \to {E}'' = b_{C}}  \langle 0|\xi _{0} \left( {{E}'_{ -} }  \right)\xi _{0}^{ +}  \left( {{E}''_{ -} }  \right)|0\rangle 
= 
\end{equation}
$$ 
= N_{y} /2\pi \,\,\lim\limits_{{E}' \to {E}'' = b_{C}}  \delta \left( {{E}' - {E}''} \right) = N_{y} /\left( {2\pi g} \right)\,.
$$

As a result, by using then Eqs.(\ref{eq94}), (\ref{eq102}), we obtain
\begin{equation}
\label{eq104}
|B|^{2} = \frac{{\pi b_{\bot}^{2} b_{C}} }{{\sqrt {1 + b_{C}^{2}} } } + O\left( {g} \right).
\end{equation}

Consequently, the weak inhomogeneity of external field $\left( {g \ne 0} \right)$ practically does not influence the value of parameter $B \ne 0$, which goes to zero in the absence of anisotropy $\left( {b_{C} = 0} \right)$.

By using Eqs.(\ref{eq100}), (\ref{eq101}), we will now obtain the following expressions for coefficients $c_{0} \left( {\tau} \right)$ and $c_{1} \left( {b_{C} ,\tau} \right)$:
$$ 
c_{0} \left( {\tau} \right) = \exp\left( { - i\left( {b_{C} - b - \omega} \right)\tau /2} \right)\,\,\left( {\cos\frac{{\Omega \left( {\omega} \right)\tau} }{{2}} - \frac{{i\left( {b_{C} - b - \omega} \right)}}{{\Omega }}\sin\frac{{\Omega \left( {\omega} \right)\tau} }{{2}}} \right)\,\,, 
$$
\begin{equation}
\label{eq105}
c_{1} \left( {b_{C} ,\tau} \right) = i\exp\left( {i\left( {b_{C} - b - \omega} \right)\tau /2} \right)\frac{{2B^{*} }}{{\Omega \left( {\omega} \right)}}\sin\frac{{\Omega \left( {\omega} \right)\tau} }{{2}}. 
\end{equation}

The probability that the antiferromagnet is in ground state at the time $\tau$ will be determined by expression
\begin{equation}
\label{eq106}
|c_{0} \left( {\tau} \right)|^{2} = 1 - \frac{{4|B|^{2}}}{{\left( {b_{C} - b - \omega} \right)^{2} + 4|B|^{2}}}\sin^{2}\frac{{\Omega \left( {\omega} \right)\tau} }{{2}},
\end{equation}
\noindent
that is antiferromagnet ground state at homogeneous AFR is modulated with Rabi frequency
\begin{equation}
\label{eq107}
\Omega \left( {\omega} \right) = \sqrt {\left( {b_{C} - b - \omega} \right)^{2} + 4|B|^{2}} .
\end{equation}

At exact AFR resonance $\left( {b_{C} - b = \omega} \right)$, the probability of being in ground state
\begin{equation}
\label{eq108}
|c_{0} \left( {\tau} \right)|^{2} = 1 - \sin^{2}\left( {|B|\tau} \right) = \cos^{2}\left( {|B|\tau} \right)
\end{equation}
\noindent
varies from unity to zero.
The elements of density matrix, corresponding to ground state in AFR condition, are transformed to the form
\begin{equation}
\label{eq109}
|0\rangle \langle 0| \to |c_{0} \left( {\tau} \right)|^{2}|0\rangle \langle 0| = \left( {1 - \frac{{2|B|^{2}}}{{\left( {b_{C} - b - \omega} \right)^{2} + 4|B|^{2}}}\left( {1 - \cos\Omega \left( {\omega} \right)\tau} \right)} \right)|0\rangle \langle 0|.
\end{equation}

To obtain the additional part of indirect interaction $\Delta V\left( {\Delta b_{k} ,l - k} \right)$ caused by this modulation (the second term in brackets of Eq.(\ref{eq109}) we will change in denominator of Eq.(\ref{eq68}) the energy of low branch $E - b$ into the more low-lying energy $E - b - \Omega \left( {\omega} \right)$ and multiply the matrix element of ground state by factor $\frac{{|B|^{2}}}{{\left( {b_{C} - b - \omega} \right)^{2} + 4|B|^{2}}}$. As a result, we will have
\begin{equation}
\label{eq110}
\Delta V\left( {\Delta b_{k} ,l - k} \right) 
\approx 
\end{equation}
$$ 
\approx - \frac{2|B|^{2}\left( {\sqrt {1 + b_{C}^{2}}  + b_{C}} \right)}{\left( {b_{C} - b - \omega} \right)^{2} + 4|B|^{2}}
 \cdot \left[ {V\left( {\Delta b_{k} ,l - k} \right) 
 - 1/2V\left( {\Delta b_{k} - \Omega \left( {\omega} \right),l - k} \right) 
 - 1/2V\left( {\Delta b_{k} + \Omega \left( {\omega} \right),l - k} \right)} \right]\,,
$$

The first factor in Eq.(\ref{eq110}) has sharp maximum at frequency of homogeneous AFR. The role of gradient of magnetic field appears mainly in dependence of indirect interaction on the distance between nuclear spins.

The positions of additional tuning points are defined here by following expression
\begin{equation}
\label{eq111}
\Delta b_{k} - g\left( {l - k} \right)/2 = \pm \sqrt {\left( {b_{C} - b - \omega} \right)^{2} + 4|B|^{2}} .
\end{equation}

As is seen from Eq.(\ref{eq110}), amplitude $b_{\bot}  \ll 1$ and frequency $\omega$ of microwave field are the additional parameters, controlling the indirect interqubit interaction. At constant external field and microwave power the new tuning points is achieved at microwave frequencies
\begin{equation}
\label{eq112}
\omega_{\pm} = b_{C} - b \pm \sqrt {\left( {\Delta b_{k} - g\left( {l - k} \right)/2} \right)^{2} - 4|B|^{2}} .
\end{equation}

Note, that qubit tuning to one of this turning points states may be performed, as distinct from relatively slow external field variation; by fast variation of microwave field frequency that has also of direct interest.

\section{Nonadiabatic decoherence and longitudinal relaxation of one qubit states in antiferromagnet-based NMR quantum register}

Let us consider the decoherence and relaxation processes of quantum state for single nuclear spin at position $k$ on axes $x$ of sublattice \textbf{A} are caused by its interaction with virtual magnon excitation in antiferromagnet. These processes will be described by transverse and longitudinal relative to external field (quantization axis) components of Bloch vector (Ref.\cite{11})
$$ 
P^{-} \left( {k,\tau} \right) = \left( {P^{+} \left( {k,\tau} \right)} \right)^{*}  
= P_{x} \left( {k,\tau} \right) - iP_{y} \left( {k,\tau} \right) 
= 2\,\tr \left( {I_{k}^{-}  \rho \left( {k,\tau} \right)} \right) 
= 2\,\tr\nolimits_{I} \left( {I_{k}^{-}  \rho_{I} \left( {k,\tau} \right)} \right) 
$$
\begin{equation}
\label{eq113}
P_{z} \left( {k,\tau} \right) = 2\,\tr\nolimits_{I} \left( {I_{z} \rho_{I} \left( {k,\tau } \right)} \right), 
\end{equation}
$$ 
\tr\nolimits_{I} I_{z}^{2} = 1/2,\,\,\,\,\,\,\,\,\,\tr\nolimits_{I} I^{+} I^{-}  = 1, 
$$
\noindent
where non-steady reduced nuclear spin density matrix in antiferromagnet is determined by
\begin{equation}
\label{eq114}
\rho_{I} \left( {k,\tau} \right) 
= \tr\nolimits_{S} \rho \left( {k,\tau} \right) 
= 1/2\left[ {1 + 2P_{z} \left( {k,\tau} \right)I_{kz} + P^{-} \left( {k,\tau } \right)I_{k}^{+}  + P^{+} \left( {k,\tau} \right)I_{k}^{-} } \right].
\end{equation}

The interaction of $k$-th nuclear spin with external magnetic field and magnon excitations will be described here by Hamiltonian with isotropic hyperfine interaction constant of the form

\begin{equation}
\label{eq115}
 h = h_{S} - \left( {\omega_{I} \left( {k} \right) - a\left( {1/2 - a_{k}^{+}  a_{k}} \right)} \right)I_{kz} + a/2\left( {I_{k}^{+}  S + I_{k}^{-}  S_{k}^{+} } \right) = 
\end{equation}
$$ 
 \,\, = h_{S} - \left( {\omega_{I} \left( {k} \right) - a\left( {1/2 - \langle 0|a_{k}^{+}  a_{k} |0\rangle} \right)} \right)I_{kz} \left( {k} \right) + \Delta h_{IS} \left( {k} \right)\,, 
$$

\noindent
where the average value of flopped electron spin number in ground state of antiferromagnet on $k$-th site known as ``spin contraction'' has the form
\begin{equation}
\label{eq116}
 \langle 0|a_{k}^{+}  a_{k} |0\rangle = \frac{{1}}{{\left( {2\pi}\right)^{2}}}\int {v\left( {{q}'_{x} ,q_{y} ,E} \right)v^{*} \left( {\mathbf{q}_{\bot}  ,E} \right)\exp\left( {i\left( {{q}'_{x} - q_{x}} \right)k} \right)d{q}'_{x} d\mathbf{q}_{\bot}  dE} 
\approx 
\end{equation}
$$ 
\,\,\,\,\,\,\,\,\,\,\,\,\,\,\,\,\,\,\,\,\,\,\,\,\,\,\,\,\,\,\,\,\, \approx \frac{{1}}{{\left( {2\pi} \right)^{2}}}\int {\frac{{|\sqrt {1 + b_{C}^{2}}  - E\left( {\mathbf{q}_{\bot} } \right)|}}{{2E\left( {\mathbf{q}_{\bot} } \right)}}} d\mathbf{q}_{\bot}  = \psi \ll 1.
$$

The perturbation Hamiltonian, corresponding to the relaxation and decoherence processes have form
\begin{equation}
\label{eq117}
\Delta h_{IS} \left( {k} \right) = \Delta h_{IS}^{\left( {1} \right)} \left( {k} \right) + \Delta h_{IS}^{\left( {2} \right)} \left( {k} \right),
\end{equation}
\noindent
where (see (\ref{eq59}))
\begin{equation}
\label{eq118}
\Delta h_{IS}^{\left( {1} \right)} \left( {k} \right) 
= a/2\left( {I_{k}^{+}  S_{k}^{-}  + I_{k}^{-}  S_{k}^{+} } \right) 
= 
\end{equation}
$$ 
= \frac{{aI_{k}^{+} } }{{2\left( {2\pi} \right)}}
  \int {\left( {u^{*} \left( {\mathbf{q}_{\bot}  ,E} \right)\,\xi^{+}\left( {q_{y} ,E_{-} } \right) + v\left( {\mathbf{q}_{\bot}  ,E} \right) \xi \left( {q_{y} ,E_{+} } \right)} \right)\exp\left( {iq_{x} k} \right)\,dEd\mathbf{q}_{\bot} } \, + \mathrm{H.c.}
$$

The second term in Eq.(\ref{eq117})
\begin{equation}
\label{eq119}
\Delta h_{IS}^{\left( {2} \right)} \left( {k} \right) = - aI_{kz} \,\left( {a_{k}^{+}  a_{k} - \langle 0|a_{k}^{+}  a_{k} |0\rangle} \right) \approx - aI_{kz} \left( {k} \right)\left( {\,S_{k}^{-}  S_{k}^{+}  - \psi} \right)
\end{equation}
\noindent
describes two-magnon interaction, which is similar to two-phonon interaction (Ref.\cite{11}, sec.~3.4). This mechanism causes the modulation of nuclear spin resonance frequency without changing its state (adiabatic decoherence). It leads to the temperature depending decoherence rate, which is negligible small at $\left( {b_{C} - b \gg \frac{{k_{\mathrm{B}} T}}{{\hbar \gamma_{S} B_{E} }}} \right)$, when the system is very close to ground state. The zero-point electron spin oscillations are not involved in this mechanism. Therefore, we will neglect next the contribution of terms $\Delta h_{IS}^{\left( {2} \right)} \left( {k} \right)$ and will use as the perturbation Hamiltonian the expression $\,\,\,\,\Delta h_{IS} \left( {k} \right) = \Delta h_{IS}^{\left( {1} \right)} \left( {k} \right)$. In this case, relaxation of transverse component of Bloch vector is accompanied by nuclear spin flopping (nonadiabatic decoherence). At the same time, the relaxation of longitudinal component of Bloch vector also occurs. Thus, these two processes may be considered here as one unified process of nuclear quantum state damping.

We will now assume that interaction of nuclear spin, which is initially at coherent state (with nonzero no diagonal elements of density matrix $\rho_{I} \left( {k,0} \right)$), with electron antiferromagnetic system in ground state, is switching on at the initial moment $\tau = 0$, when no perturbed density matrix is represented as direct product $\rho \left( {k,0} \right) = \rho_{I} \left( {k,0} \right)\rho_{S} \left( {0} \right) = \rho_{I} \left( {k,0} \right)|0\rangle \langle 0|$.

Let us go next to interaction representation for density matrix relatively to Hamiltonian $h_{0} \left( {k} \right) = h_{S} - \left( {\omega_{I} \left( {k} \right) - a/2} \right)\;I_{kz} $:
\begin{equation}
\label{eq120}
\rho_{\mathrm{in}} \left( {k,\tau} \right) = \exp\left( {ih_{0} \left( {k} \right)\tau} \right)\rho \left( {k,\tau} \right)\exp\left( { - ih_{0} \left( {k} \right)\tau} \right)
\end{equation}
\noindent
and to the equation for density matrix of nucleus-electron system
\begin{equation}
\label{eq121}
i\partial \rho_{\mathrm{in}} \left( {k,\tau} \right)/\partial \tau = \left[ {\Delta h_{IS} \left( {k,\tau} \right),\,\rho_{\mathrm{in}} \left( {k,\tau} \right)} \right],
\end{equation}
\noindent
where
$$ 
\Delta \,h_{IS} \left( {k,\tau} \right) 
= \exp\left( {i h_{0} \left( {k} \right)\tau} \right)\,\Delta h_{IS} \left( {k} \right)\exp\left( { - i h_{0} \left( {k} \right)\tau} \right) 
= 
$$
\begin{equation}
\label{eq122}
= a/2\,\,\exp\left( { - i\left( {\omega_{I} \left( {k} \right) - a/2} \right)\tau} \right)\,I_{k}^{+}  S_{k}^{-} \left( {\tau} \right) + \mathrm{H.c.} 
= 
\end{equation}
$$ 
= \frac{{a}}{{4\pi} }\exp\left( { - i\left( {\omega_{I} \left( {k} \right) - a/2} \right)\tau} \right)\,I_{k}^{+}  \int {\left[ {u^{*} \left( {\mathbf{q}_{\bot}  ,E} \right)\,\exp\left( { - iE_{-}  \tau} \right)\,\xi^{+} \left( {q_{y} ,E_{-} } \right)  + }\right. 
}
$$
$$ 
{ 
\left.{
+ v\left( {\mathbf{q}_{\bot}  ,E} \right)\exp\left( {iE_{+}  \tau \,} \right)\xi \left( {q_{y} ,E_{+} } \right)} \right]\exp\left( {iq_{x} k} \right)\,dEd\mathbf{q}_{\bot}}  + \mathrm{H.c.} 
$$

It is follows from Eq.(\ref{eq121}) in the second order of perturbation theory
\begin{equation}
\label{eq123}
i\partial \rho_{\mathrm{in}} \left( {k,\tau} \right)/\partial \tau 
\approx \left[ {\Delta h_{IS} \left( {k,\tau} \right),\rho \left( {k,0} \right)} \right] 
- 
\end{equation}
$$ 
- i\int\limits_{0}^{\tau}  {\left[ {\Delta h_{IS} \left( {\tau ,k} \right),\left[ { \Delta h_{IS} \left( {k,{\tau}'} \right),\rho \left( {k,0} \right)} \right]} \right]d{\tau} }'. 
$$

Let us write the derivative of expression (\ref{eq113}) with respect to time, by using Eq.(\ref{eq123}) and perform then the cyclic permutation under tracing. Finally, accounting the relation $\tr \left( {\left[ {I_{k}^{-}  ,\Delta h_{IS} \left( {k,\tau} \right)} \right]\rho_{S} \left( {0} \right)} \right) = 0$, we will find
$$ 
\partial \left[ {P^{-} \left( {k,\tau} \right)\exp\left( {i\left( {\omega_{I} \left( {k} \right) - a\left( {1/2} \right)} \right)\tau} \right)} \right]/\partial \tau = 2\tr \{ I^{-} \partial \rho_{\mathrm{in}} \left( {k,\tau} \right)/\partial \tau \} 
\approx 
$$
\begin{equation}
\label{eq124}
\approx - 2\tr \{ I_{k}^{-}  \int\limits_{0}^{\tau } {\left[ {\Delta h_{IS} \left( {k,\tau} \right),\left[ {\Delta h_{IS} \left( {k{\tau}'} \right),\rho \left( {k,0} \right)} \right]} \right]} \,\,\} d{\tau}' 
=
\end{equation}
$$ 
= - 2\tr \{ \int\limits_{0}^{\tau}  {\left[ {\left[ {I_{k}^{-}  ,\Delta h_{IS} \left( {k,\tau} \right)} \right],\,\,\Delta h_{IS} \left( {k,{\tau}'} \right)} \right]\rho \left( {k,0} \right)\}}  d{\tau}'. 
$$

By defining the Bloch vector transverse component in the form
\begin{equation}
\label{eq125}
P^{-} \left( {k,\tau} \right) = 
P^{-} \left( {k,0} \right)\exp\left[ { - i\left( {\omega_{I} \left( {k} \right) - a/2} \right)\tau - \Gamma_{\bot} \left( {k,\tau} \right)} \right],
\end{equation}
\noindent
we will obtain:
\begin{equation}
\label{eq126}
d\left[ {P^{-} \left( {k,\tau} \right)\exp\left( {i\left( {\omega_{I} \left( {k} \right) - a/2} \right)\tau} \right)} \right]/d\tau 
= 
\end{equation}
$$ 
= - d\Gamma_{\bot} \left( {k,\tau} \right)/d\tau \, \cdot P^{-} \left( {k,\tau} \right)\exp\left( {i\left( {\omega_{I} \left( {k} \right) - a/2} \right)\tau} \right), 
$$
\noindent
where $\mathrm{Re}\Gamma_{\bot} \left( {\tau} \right)$ is decoherence decrement and $\mathrm{Im}\Gamma_{\bot} \left( {\tau} \right)$ is phase shift.

We will represent the nuclear density matrix in the right part of Eq.(\ref{eq124}) by expression
\begin{equation}
\label{eq127}
\rho_{I} \left( {k,0} \right) \approx 1/2\{ 1 + 2P_{z} \left( {k,0} \right)I_{kz} + P^{-} \left( {k,0} \right)I_{k}^{+}  + P^{+} \left( {k,0} \right)I_{k}^{-}  \} |0\rangle \langle 0||.
\end{equation}

Using in Eq.(\ref{eq124}) the perturbation Hamiltonian (\ref{eq118}), ignoring the insignificant factors 

\noindent
$\exp\left( {\pm i\left( {\omega_{I} \left( {k} \right) - a/2} \right)\tau } \right)$ and taking in the context of second order of perturbation theory in the left-hand side of Eq.(\ref{eq126}) $P^{-} \left( {k,\tau} \right) \approx P^{-} \left( {k,0} \right)$, for the decoherence rate we will obtain
$$ 
d\mathrm{Re}\Gamma_{\bot} \left( {k,\tau} \right)/d\tau = 
\mathrm{Re} \tr\nolimits_{I} \int\limits_{0}^{\tau}  {\langle 0|\left[ {\left[ {I_{k}^{-}  ,\Delta h_{IS} \left( {k,\tau} \right)} \right],\,\,\Delta h_{IS} \left( {k,{\tau}'} \right)} \right]\,} I_{k}^{+}  |0\rangle d{\tau }' 
= 
$$
\begin{equation}
\label{eq128}
= \frac{{a^{2}}}{{4}}\mathrm{Re}\tr\nolimits_{I} \int\limits_{0}^{\tau}  {\langle 0|\left[ {\left[ {I_{k}^{-}  ,\left( {I_{k}^{+}  S_{k}^{-} \left( {\tau} \right) + I_{k}^{-}  S_{k}^{+} \left( {\tau} \right)} \right)} \right],\,\,\left( {I_{k}^{+}  S_{k}^{-} \left( {{\tau}'} \right) + I_{k}^{-}  S_{k}^{+} \left( {{\tau}'} \right)} \right)} \right]\,} I_{k}^{+}  |0\rangle d{\tau}' 
= 
\end{equation}
$$ 
= \frac{{a^{2}}}{{4}}2 \mathrm{Re}\int\limits_{0}^{\tau}  {\langle 0|S_{k}^{-} \left( {\tau} \right)S_{k}^{+} \left( {{\tau}'} \right) + S_{k}^{+} \left( {{\tau}'} \right)S_{k}^{-} \left( {\tau} \right)|0\rangle d{\tau}'} . 
$$
\noindent
where $\tr\nolimits_{I}$ is partial trace over nuclear spin states.

Let us restrict next again to low magnon excitation mode with energy $E_{-}$. 
Further, let us insert week magnon damping $\left( {E_{-} \to E_{-}  + is,\,\,\,\,\,E_{-}  \gg \,s > 0} \right)$ and make a set of rearrangements, which are similarly to that in Eqs.(A2.3),(A2.6). We will obtain
\begin{equation}
\label{eq129}
\mathrm{Re} d\Gamma_{\bot} \left( {k,\tau} \right)/d\tau 
\approx 
\end{equation}
$$ 
\approx \frac{{a^{2}}}{{\left( {4\pi} \right)^{2}}}\mathrm{Re} 2\int {\int\limits_{0}^{\tau}  {u^{*} \left( {\mathbf{q}_{\bot}  ,E} \right)u\left( {{q}'_{x} ,q_{y} ,E} \right)\exp\left( { - \left( {iE_{-}  - s} \right)\left( {{\tau}' - \tau} \right)} \right)\exp\left( {i\left( {q_{x} - {q}'_{x}} \right)k} \right)d{q}'_{x} d\mathbf{q}_{\bot}  dEd{\tau}'}}  .
$$

Upon integrating over ${\tau}'$, we will have
\begin{equation}
\label{eq130}
\mathrm{Re} d\Gamma_{\bot} \left( {k,\tau} \right)/d\tau 
\approx 
\end{equation}
$$ 
\approx \frac{{a^{2}}}{{\left( {4\pi} \right)^{2}}}\mathrm{Re} 2\int {u^{*} \left( {\mathbf{q}_{\bot}  ,E} \right)u\left( {{q}'_{x} ,q_{y} ,E} \right)\exp\left( {i\left( {q_{x} - {q}'_{x}} \right)k} \right)\frac{{1 - \exp\left( {\left( {iE_{-}  - s} \right)\tau} \right)\;}}{{ - iE_{-}  + s}}dEd\mathbf{q}_{\bot}  d{q}'_{x}}.
$$

By using the expression $E\left( {\mathbf{q}_{\bot} } \right) \approx \sqrt {b_{C}^{2} + \mathbf{q}_{\bot}^{2} /12}$, we write
\begin{equation}
\label{eq131}
d\mathbf{q}_{\bot}  = 2\pi \mathbf{q}_{\bot}  d\mathbf{q}_{\bot}  = 24\pi E\left( {\mathbf{q}_{\bot} } \right)dE\left( {\mathbf{q}_{\bot} } \right),\,\,\,\,\,\,0 < \mathbf{q}_{\bot}  \le \pi
\end{equation}
\noindent
and take notations $\Delta b_{k} = b_{C} - b - gk,\,\,\,\,\,\,\,\xi = E\left( {\mathbf{q}_{\bot} } \right) - b_{C}$. Upon integrating over $\,\,E$ and ${q}'_{x}$ (similarly to Eqs. (\ref{eq67})-(\ref{eq70})), we will transform the expression (\ref{eq130}) for decoherence rate to the following form
\begin{equation}
\label{eq132}
\mathrm{Re} d\Gamma_{\bot} \left( {k,\tau} \right)/d\tau = \frac{{3a^{2}}}{{2\pi} }R_{\bot} \left( {\Delta b_{k} ,\tau} \right) = \frac{{3a^{2}}}{{2\pi} }\,\,\int\limits_{0}^{\sqrt {b_{C}^{2} + \pi^{2}/12} - b_{C}}  {\left( {\sqrt {1 + b_{C}^{2}}  + b_{C} + \xi} \right)\,\,Y\left( {\xi , + \Delta b_{k} ,\tau} \right)\,d\xi}  ,
\end{equation}
\noindent
where
\begin{equation}
\label{eq133}
Y\left( {\xi + \Delta b_{k} ,\tau} \right) 
= \mathrm{Re}\frac{{1 - \exp\left[ {\left( {i\left( {\xi + \Delta b_{k}} \right) - s} \right)\tau} \right]}}{{ - \left( {i\left( {\xi + \Delta b_{k}} \right) - s} \right)}} 
= 
\end{equation}
$$ 
= \frac{{\left( {\xi + \Delta b_{k}} \right)\sin\left( {\left( {\xi + \Delta b_{k}} \right)\tau} \right)\exp\left( { - s\tau} \right)}}{{\left( {\xi + \Delta b_{k}} \right)^{2} + s^{2}}} + s\frac{{1 - \cos\left( {\left( {\xi + \Delta b_{k}} \right)\tau} \right)\exp\left( { - s\tau} \right)}}{{\left( {\xi + \Delta b_{k}} \right)^{2} + s^{2}}} 
$$
\noindent
and
\begin{equation}
\label{eq134}
R_{\bot} \left( {\Delta b_{k} ,\tau} \right) = \,\mathrm{Re} i\int\limits_{t_{1}}^{t_{2}}  {\left( {\sqrt {1 + b_{C}^{2}}  + b_{C} - \Delta b_{k} + i\left( {t/\tau - s} \right)} \right)} \frac{{1 - \exp\left( { - t} \right)}}{{t}}dt,
\end{equation}
\noindent
with
\begin{equation}
\label{eq135}
t_{1} = \left( { - i\Delta b_{k} + s} \right)\tau ,\,\,\,\,t_{2} = \left( { - i \left( {\sqrt {b_{C}^{2} + \pi^{2}/12} - b_{C} + \Delta b_{k}} \right) + s} \right)\tau .
\end{equation}

Upon integrating over $t$, we have
$$ 
 R_{\bot} \left( {\Delta b_{k} ,\tau} \right) 
= \mathrm{Re}\{ 
 i\left( {\sqrt {1 + b_{C}^{2}}  + b_{C} - \Delta b_{k} - is} \right) 
  \left[ {\ln\frac{ -i\left( {\sqrt {b_{C}^{2} + \pi^{2}/12} - b_{C} + \Delta b_{k}} \right) + s}{-i\Delta b_{k} + s}
}\right.
-
$$
\begin{equation}
\label{eq136}
\left.{
-
\left( {
  E_{1} \left( {\left( { - i\Delta b_{k}                                                          + s} \right)\tau} \right) -
  E_{1} \left( {\left( { - i\left({\sqrt {b_{C}^{2} + \pi^{2}/12} - b_{C} + \Delta b_{k}} \right) + s} \right)\tau} \right)
} \right)
} \right] 
+ 
\end{equation}
$$ 
+ 1/\tau \exp\left( { - s\tau + i\Delta b_{k} \tau} \right)\left( {\exp\left( {\left( {i\sqrt {b_{C}^{2} + \pi^{2}/12} - b_{C}} \right)\tau} \right) - 1} \right)\} , 
$$
\noindent
where
\begin{equation}
\label{eq136a}
E_{1} \left( {x} \right) = \int\limits_{x}^{\infty}  {\frac{{\exp\left( {-t} \right)}}{{t}}} dt,\,\,\,\,|\arg x| < \pi 
\end{equation}
\noindent
is exponential integral.

For the case of $\Delta b_{k} \gg s > 0$, omitting $s^{2}$ in dominator, we will have the explicit approximate expression
$$ 
R_{\bot} \left( {\Delta b_{k} ,\tau} \right) \approx \left\{ { \left( {\sqrt {1 + b_{C}^{2}}  + b_{C} - \Delta b_{k}} \right)\left[ {\si\left( {\Delta b_{k} \tau} \right) - \si\left( {\left( {\sqrt {b_{C}^{2} + \pi^{2}/12} - b_{C} + \Delta b_{k}} \right)\tau} \right)} \right] 
} \right.
+ 
$$
$$ 
\left. {
+ 1/\tau \,\,\left[ {\cos\left( {\Delta b_{k} \tau} \right) - \cos\left( {\left( {\sqrt {b_{C}^{2} + \pi^{2}/12} - b_{C} + \Delta b_{k}} \right)\tau} \right)} \right] } \right\} \,\,\exp\left( { - s\tau} \right) 
+ 
$$
\begin{equation}
\label{eq137}
+ s\,\,\left( {\sqrt {1 + b_{C}^{2}}  + b_{C} - \Delta b_{k}} \right)\left\{{ \frac{{\sqrt {b_{C}^{2} + \pi^{2}/12} - b_{C}} }{{\Delta b_{k} \left( {\sqrt {b_{C}^{2} + \pi^{2}/12} - b_{C} + \Delta b_{k}} \right)}} 
} \right. 
- 
\end{equation}
$$ 
\left. {
-
\left[ {\frac{{\cos\left( {\Delta b_{k} \tau} \right)}}{{\Delta b_{k}} } - \frac{{\cos\left( {\left( {\sqrt {b_{C}^{2} + \pi^{2}/12} - b_{C} + \Delta b_{k}} \right)\tau} \right)}}{{\sqrt {b_{C}^{2} + \pi^{2}/12} - b_{C} + \Delta b_{k}} } 
}\right.}\right.
+ 
$$
$$ 
\left.{\left.{
+ \tau \left( {\si\left( {\Delta b_{k} \tau} \right) - \si\left( {\left( {\sqrt {b_{C}^{2} + \pi^{2}/12} - b_{C} + \Delta b_{k}} \right)\tau} \right)} \right)} \right]\,\,\,\exp\left( { - s\tau} \right)}\right\} 
+ 
$$
$$ 
+ s\,\left\{{\ln\frac{{\left( {\sqrt {b_{C}^{2} + \pi^{2}/12} - b_{C} + \Delta b_{k}} \right)}}{{\Delta b_{k}} } + \,\left[ {\ci\left( {\Delta b_{k} \tau} \right) - \ci\left( {\left( {\sqrt {b_{C}^{2} + \pi^{2}/12} - b_{C} + \Delta b_{k}} \right)\tau} \right)} \right]\,\,\,\exp\left( { - s\tau} \right)}\right\} > 0, 
$$
\noindent
where 
$\si\left( {x} \right) = - \int\limits_{x}^{\infty}  {\frac{{\sin t}}{{t}}dt}$, 
$\ci\left( {x} \right) = - \int\limits_{x}^{\infty}  {\frac{{\cos t}}{{t}}dt}$ 
are sine-integral and cosine-integral, and it is acounted, that $\lim\limits_{_{\tau \to 0}} \left[ {\ci\left( {a\tau} \right) - \ci\left( {b\tau} \right)} \right] = \ln\frac{{a}}{{b}}$.

\begin{figure}
   \begin{center}
   \begin{tabular}{c}
   \includegraphics{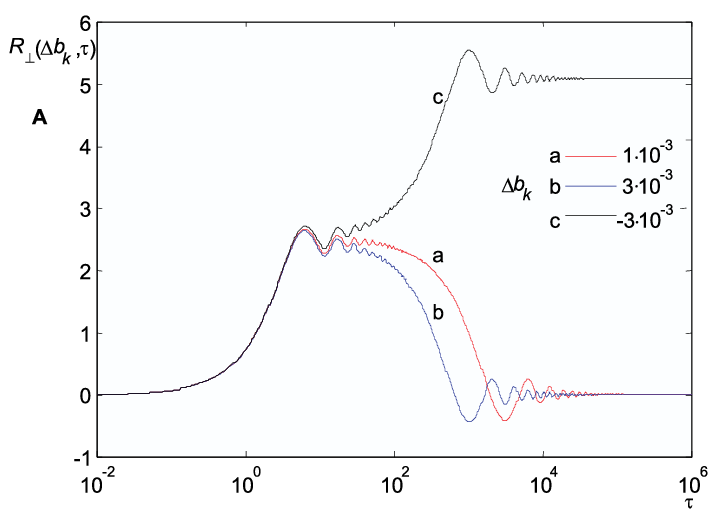}
\\
   \includegraphics{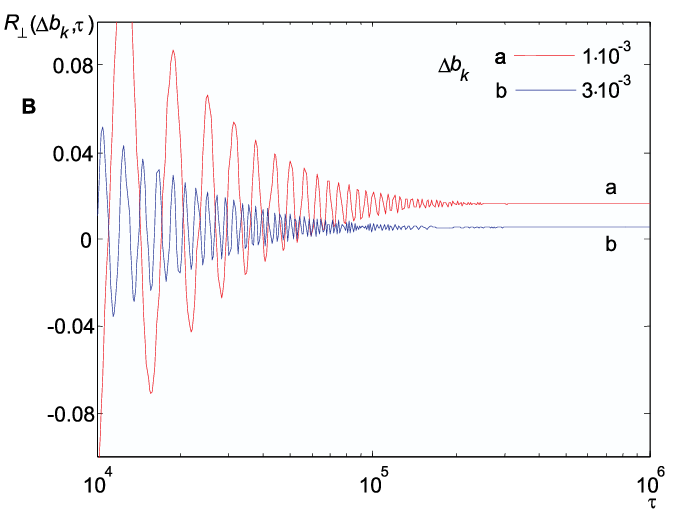}
   \end{tabular}
   \end{center}
   \caption[1]
   {\label{fig:7} 
\textbf{A} The $\tau-$dependence of $R_{\bot} \left( {\Delta b_{k} ,\tau} \right) \quad \left( {\tau = t\omega_{E}} \right)$, for, $b_{C}^{2} = 1/4$, $s = 10^{-5}$ and $\Delta b_{k} = b_{C} - b_{k} = - 3 \cdot 10^{-3},\,\,1 \cdot 10^{-3},\,\,3 \cdot 10^{-3}$. The case \textbf{B} corresponds to the increased values of $R_{\bot} \left( {\Delta b_{k} ,\tau} \right)$ only for $\Delta b_{k} = 1 \cdot 10^{-3},\,\,3 \cdot 10^{-3}$ in the range of $10^{-4} < \tau < 10^{-6}$.
   }
   \end{figure}

We notice (Fig.~\ref{fig:7}) that oscillating part of the rate of decoherence tends fast (microseconds) to constant.

Taking then $\tau \to \infty \,\,\,\left( {\tau \gg 1/s\sim 10^{5}} \right)$, we will obtain for inverse time of decoherence (it corresponds in NMR science to transverce relaxation rate $1/T_{2}$)
\begin{equation}
\label{eq138}
1/T_{D} = \omega_{E} \mathrm{Re} d\Gamma_{\bot} \left( {k,\infty} \right)/d\tau = \omega_{E} \frac{{3a^{2}}}{{2\pi} }R_{\bot} \left( {b_{k} ,\infty} \right) 
= 
\end{equation}
$$ 
= \omega_{E} s\frac{{3a^{2}}}{{2\pi} }\,\,\{ \frac{{\left( {\sqrt {1 + b_{C}^{2}}  + b_{C} - \Delta b_{k}} \right)\sqrt {b_{C}^{2} + \pi^{2}/12} - b_{C}} }{{\Delta b_{k} \sqrt {b_{C}^{2} + \pi^{2}/12} - b_{C} + \Delta b_{k}} } + \ln\frac{{\sqrt {b_{C}^{2} + \pi^{2}/12} - b_{C} + \Delta b_{k} }}{{\Delta b_{k}} }\} ,
$$

For values 
$a^{2}\sim 10^{-6}$, 
$s \ll \Delta b_{k}$, 
$\omega_{E} /2\pi \sim 10^{11}\,\mathrm{Hz}$, decoherence time in sedconds is 
$T_{D} \sim \Delta b_{k}/s$.

It follows that decoherence time caused by one-magnon nonadiabatic processes near turning points is fast decreasing (for $\Delta b_{k} \le 10^{-3}$, $s \sim 10^{-3}$ decoherence time may be less than milliseconds). Therefore, quantum operation, such as two-qubit operation, near these points should be carried out in the more short times.

Between other mechanisms of decoherence it should be pointed out the adiabatic mechanism that is determined by magnetic interaction of nuclear spins-qubits with electron and nuclear spins of impurity atoms (Ref.\cite{11}, sec.~5.4). The suppressing of this mechanism calls for appropriate cleaning of substrate from impurity atoms and using very low spin temperatures for nuclear spins.

Note that character of decoherence rate essentially depends on the anisotropy of antiferromagnet (through value $b_{C} $) and from inhomogeneity of external field (trough parameter $\Delta b_{k} $).

The expression for frequency shift agrees at with Eq.(\ref{eq85}) for $U\left( {k,k} \right) \quad \left( {s \ll \Delta b_{k}} \right)$:
$$ 
\mathrm{Im} d\Gamma_{\bot} \left( {k,\infty} \right)/d\tau 
= 
$$
\begin{equation}
\label{eq139}
= - \frac{{3a^{2}}}{{2\pi} }\{ \left( {\sqrt {1 + b_{C}^{2}}  + b_{C} - \Delta b_{k}} \right)\ln\frac{{\sqrt {b_{C}^{2} + \pi^{2}/12} - b_{C} + \Delta b_{k}} }{{\Delta b_{k}} } + \sqrt {b_{C}^{2} + \pi^{2}/12} - b_{C} \} 
\approx 
\end{equation}
$$ 
\approx - \frac{{3a^{2}}}{{2\pi }}\left( {\sqrt {1 + b_{C}^{2}}  + b_{C}} \right)\ln\frac{{\sqrt {b_{C}^{2} + \pi^{2}/12} - b_{C}} }{{\Delta b_{k}} } \approx - \,\,U\left( {k,k} \right). 
$$

Let us consider next the relaxation of Bloch vector's longitudinal component. We will write:
$$ 
\partial P_{z} \left( {k,\tau} \right)/\partial \tau = 2\tr I_{kz} \partial \rho_{\mathrm{in}} \left( {k,\tau} \right)/\partial \tau 
= 
$$
\begin{equation}
\label{eq140}
= - 2\tr \left( {I_{kz} \int\limits_{0}^{\tau}  {\left[ {\Delta h_{IS} \left( {k,\tau} \right),\left[ {\Delta h_{IS} \left( {k{\tau}'} \right),\rho_{\mathrm{in}} \left( {k,{\tau}'} \right)} \right]} \right]} d{\tau}'} \right) \approx 
\end{equation}
$$ 
\approx - 2 \mathrm{Re} \tr\nolimits_{I} \left( {\int\limits_{0}^{\tau}  {\left[ {\left[ {I_{kz} ,\Delta h_{IS} \left( {k,\tau} \right)} \right],\,\,\Delta h_{IS} \left( {k,{\tau}'} \right)} \right]\rho_{I} \left( {k,0} \right)} |0\rangle \langle 0|d{\tau}'} \right) . 
$$

Taking $P_{z} \left( {k,\tau} \right) = P_{z} \left( {k,0} \right)\exp\left( { - \Gamma_{||} \left( {k,\tau} \right)} \right)$, we will obtain in the second order of the perturbation theory:
\begin{equation}
\label{eq141}
dP_{z} \left( {k,\tau} \right)/d\tau \approx - d\Gamma_{||} \left( {k,\tau } \right)/d\tau \, \cdot P_{z} \left( {k,0} \right),
\end{equation}
\noindent
where the rate of the longitudinal relaxation is
$$ 
d\Gamma_{||} \left( {k,\tau} \right)/d\tau = \left( {d\Gamma_{||} \left( {k,\tau } \right)/d\tau} \right)^{*}  
= 
$$
\begin{equation}
\label{eq142}
= 2\mathrm{Re}\tr\nolimits_{I} \int\limits_{0}^{\tau}  {\langle 0|\left[ {\left[ {I_{z} ,\Delta h_{IS} \left( {k,\tau} \right)} \right],\,\,\Delta h_{IS} \left( {k,{\tau}'} \right)} \right]\,I_{z} |0\rangle}  d{\tau}' 
= 
\end{equation}
$$ 
= \frac{{a^{2}}}{{4}}2\mathrm{Re}\tr\nolimits_{I} \int\limits_{0}^{\tau}  {\langle 0|\left[ {\left[ {I_{z} ,\left( {I_{k}^{+}  S_{k}^{-} \left( {\tau} \right) + I_{k}^{-}  S_{k}^{+} \left( {\tau} \right)} \right)} \right],\,\,\left( {I_{k}^{+}  S_{k}^{-} \left( {{\tau}'} \right) + I_{k}^{-}  S_{k}^{+} \left( {{\tau}'} \right)} \right)} \right]\,I_{z} |0\rangle}  d{\tau}' 
$$

Similar to the foregoing calculation (Eq.(\ref{eq128})) we will have
\begin{equation}
\label{eq143}
d\Gamma_{||} \left( {k,\tau} \right)/d\tau 
= \frac{{a^{2}}}{{4}}2\mathrm{Re}\int\limits_{0}^{\tau}  {\langle 0|S_{k}^{+} \left( {\tau} \right)S_{k}^{-} \left( {{\tau}'} \right) + S_{k}^{-} \left( {\tau} \right)S_{k}^{+} \left( {{\tau}'} \right)|0\rangle}  d{\tau}' 
= 
\end{equation}
$$ 
= \mathrm{Re} d\Gamma_{\bot} \left( {k,\tau} \right)/d\tau , 
$$

\noindent
whence it follows that for the considered mechanisms the rate of relaxation of longitudinal component is equal practically to the rate of relaxation of transverse component (decoherence rate).

\section{Decoherence of pair entanglement of quantum states in NMR quantum register}

The arbitrary state of pair removed spins-qubits in quantum register with zero Bloch vector values 
is described by the following reduced density matrix of nuclear spin system 
($\alpha ,\beta = x,y,z$) (Ref.\cite{11}, Sec.~2.5-2.7):
\begin{equation}
\label{eq144}
\rho_{I} \left( {l,k,\tau} \right) = 
\tr\limits_{m \ne l,k} \rho_{I} \left( {1,\dots ,m,\dots N,\tau} \right) = 
1/4\{ 1 + \sum\limits_{\alpha ,\beta}  {4G_{\alpha ,\beta} \left( {l,k,\tau} \right)\left( {I_{l,\alpha}  \otimes I_{k\beta} } \right)\}} .
\end{equation}

Let us consider here the non-steady reduced density matrix of two nuclear state 
\begin{equation}
\label{eq145}
\rho_{I} \left( {l,k,\tau} \right) = 
1/4\,\{ 1 + 4G_{z,z} \left( {l,k,\tau} \right)\left( {I_{l,z} I_{k,z}} \right) + G^{+ , -} \left( {l,k,\tau} \right)\left( {I_{l}^{-}  I_{k}^{+} } \right) + G^{-, + }\left( {l,k,\tau} \right)\left( {I_{l}^{+}  I_{k}^{-} } \right)\} ,
\end{equation}
\noindent
where diagonal and non-diagonal elements of the matrix are determined by magnitudes
\begin{equation}
\label{eq146}
G^{+,-} \left( {l,k,\tau} \right) 
= \left( {G^{-,+} \left( {l,k,\tau } \right)} \right)^{*} 
= 4\tr\nolimits_{I}\left( {I_{l}^{+}  I_{k}^{-} } \right)\,\,\rho_{I} \left( {l,k,\tau} \right). 
\end{equation}
$$ 
G_{z,z} \left( {l,k,\tau} \right)
= 4\tr\nolimits_{I}\left( {I_{l,z} I_{k,z}} \right)\,\rho_{I} \left( {l,k,\tau} \right), 
$$

The Bloch vectors for this state Eq.(\ref{eq145}) are 
$\mathbf{P}\left( {k,\tau} \right) 
= 2\tr\nolimits_{I} {\mathbf{I}_{k} \rho_{I} \left( {l,k,\tau} \right) } 
= \mathbf{P}\left( {l,\tau} \right) = 0$.

In the interaction representation relative Hamiltonian
\begin{equation}
\label{eq147}
h_{0} \left( {l,k} \right) = h_{S} - \left( {\left( {\omega_{I} \left( {l} \right) - a/2} \right)\;I_{z} \left( {l} \right) + \left( {\omega_{I} \left( {k} \right) - a/2} \right)\;I_{z} \left( {k} \right)} \right)
\end{equation}
\noindent
the reduced density matrix has the form
\begin{equation}
\label{eq148}
\rho_{\mathrm{in}} \left( {l,k,} \right) = \exp\left( {ih_{0} \left( {l,k} \right)\tau} \right)\rho \left( {l,k,\tau} \right)\exp\left( { - ih_{0} \left( {l,k} \right)\tau} \right).
\end{equation}

Let there be the pure triplet entangled state of two removed spins $l$ and $k$ with zero total $z$-projection $I = 1$, $M = 0$, which belongs to the same sublattice, realized by the certain external action in the initial moment $\tau = 0$. It will be described by reducer state vector $|1,0\rangle = \sqrt {1/2} \left( {\,| \uparrow \downarrow \rangle + | \downarrow \uparrow \rangle} \right)$ and density matrix:
\begin{equation}
\label{eq149}
\rho_{I} \left( {l,k,0} \right) = |1,0\rangle \langle 1,0| = \frac{{1}}{{2}}\;\left| {\;{
\begin{array}{*{20}c}
 {0} \hfill & {0} \hfill & {0} \hfill & {0} \hfill \\
 {0} \hfill & {1} \hfill & {1} \hfill & {0} \hfill \\
 {0} \hfill & {1} \hfill & {1} \hfill & {0} \hfill \\
 {0} \hfill & {0} \hfill & {0} \hfill & {0} \hfill \\
\end{array}}
 \;} \right| = 1/4\{ 1 - 4\left( {I_{l,z} I_{k,z}} \right) + 2\left( {I_{l}^{-}  I_{k}^{+} } \right) + 2\left( {I_{l}^{+}  I_{k}^{-}} \right)\} ,
\end{equation}
\noindent
with the following tensor components $G_{z,z} \left( {l,k,0} \right) = - 1$, $G^{+,-} \left( {l,k,0} \right) = G^{-,+} \left( {l,k,0} \right) = 2$. 
The concurrence of this entangled state has maximum value $C= 1$.

Let us suppose now (as previously in Sec.~8), that the interaction of nuclear spins in ground coherent state (\ref{eq149}) with magnons is switching on at initial moment $\tau = 0$, when non-disturbed density matrix has the form of direct product: $\rho \left( {0} \right) = \rho_{I} \left( {l,k,0} \right)\rho_{S} \left( {0} \right) = \rho_{I} \left( {l,k,0} \right)|0\rangle \langle 0|$.

We will write the equation for density matrix in the interaction representation $\rho_{\mathrm{in}} \left( {l,k,\tau} \right)$:
\begin{equation}
\label{eq150}
i\partial \rho_{\mathrm{in}} \left( {l,k,\tau} \right)/\partial \tau = \left[ {\Delta h_{IS} \left( {l,k,\tau} \right),\,\,\,\,\rho_{\mathrm{in}} \left( {l,k,\tau} \right)} \right].
\end{equation}

The evolution of such two-qubit state in considered model is described by the following perturbation Hamiltonian in interaction representation
\begin{equation}
\label{eq151}
\Delta h_{IS} \left( {l,k,\tau} \right) 
= \left( {\Delta h_{IS} \left( {l,\tau} \right) + \Delta h_{IS} \left( {k,\tau} \right)} \right) 
= 
\end{equation}
$$ 
= \exp\left( {ih_{0} \left( {l,k} \right)\tau} \right)\left( {\Delta h_{IS} \left( {l} \right) + \Delta h_{IS} \left( {k} \right)} \right)\exp\left( { - ih_{0} \left( {l,k} \right)\tau } \right), 
$$
\noindent
where values $\Delta h_{IS} \left( {l,\tau} \right),\Delta h_{IS} \left( {k,\tau} \right)$ are determined by expression (\ref{eq122}).

From Eq.(\ref{eq150}) for density matrix in the second theory of perturbation theory it is follows the equation
$$ 
i\partial \rho_{\mathrm{in}} \left( {l,k,\tau} \right)/\partial \tau 
= \left[ {\Delta h_{IS} \left( {l,k,\tau} \right),\rho_{\mathrm{in}} \left( {l,k,0} \right)|0\rangle \langle 0|} \right] 
- 
$$
\begin{equation}
\label{eq152}
- i\int\limits_{0}^{\tau}  {\left[ {\Delta h_{IS} \left( {l,k,\tau} \right),\left[ {\Delta h_{IS} \left( {l,k,{\tau}'} \right),\rho_{I} \left( {l,k,0} \right)|0\rangle \langle 0|} \right]} \right]} d{\tau}'. 
\end{equation}

By using Eq.(\ref{eq152}) and making the cyclic permutation, we will obtain the equations for the elements of reduced density matrix (\ref{eq145}):
\begin{equation}
\label{eq153}
\partial G_{z,z} \left( {l,k,\tau} \right)/\partial \tau 
= 4\tr\nolimits_{I} \left( {I_{l,z} I_{k,z} } \right)\partial \rho_{\mathrm{in}} \left( {l,k,\tau} \right)/\partial \tau 
\approx 
\end{equation}
$$ 
\approx - 4\tr\nolimits_{I} \int\limits_{0}^{\tau}  {\langle 0|\left[ {\left[ {\left( {I_{l,z} I_{k,z}} \right),\Delta h_{IS} \left( {l,k,\tau} \right)} \right],\Delta h_{IS} \left( {l,k,{\tau}'} \right)} \right]\rho_{I} \left( {l,k,0} \right)\,|0\rangle}  d{\tau}', 
$$
\noindent
and
\begin{equation}
\label{eq154}
\partial G^{+,-} \left( {l,k,\tau} \right)/\partial \tau 
= 4\tr \,\left( {I_{l}^{+}  I_{k}^{-} } \right)\partial \,\rho_{\mathrm{in}} \left( {l,k,\tau} \right)/\partial \tau \approx 
\end{equation}
$$ 
\approx - 4\tr\nolimits_{I}\int\limits_{0}^{\tau}  {\langle 0|\left[ {\left[ {\left( {I_{l}^{+}  I_{k}^{-} } \right),\Delta h_{IS} \left( {l,k,\tau} \right)} \right],\Delta h_{IS} \left( {l,k,{\tau}'} \right)} \right]\rho_{I} \left( {l,k,0} \right)|0\rangle}  \,d{\tau}'. 
$$

Taking into account Eq.(\ref{eq149}), let us define the tensor longitudinal component in the form
\begin{equation}
\label{eq155}
G_{z,z} \left( {l,k,\tau} \right)= - \exp\left[ { - \Gamma_{||} \left( {l,k,\tau} \right)} \right]
\end{equation}
\noindent
and the tensor transverse component in the form
\begin{equation}
\label{eq156}
G^{+ , -} \left( {l,k,\tau} \right) = 2\exp\left[ {i\left( {\omega_{I} \left( {l} \right) - \omega_{I} \left( {k} \right)} \right)\tau - \Gamma_{\bot} \left( {l,k,\tau} \right)} \right].
\end{equation}

To obtain the expressions for the rates of pair relaxation of longitudinal and transverse components in the context of second order of perturbation theory, we will write:
\begin{equation}
\label{eq157}
\partial G_{z,z} \left( {l,k,\tau} \right)/\partial \tau = d\Gamma_{||} \left( {l,k,\tau} \right)/d\tau = d\Gamma_{||}^{*} \left( {l,k,\tau} \right)/d\tau 
= 
\end{equation}
$$ 
= - 4\tr\nolimits_{I} \int\limits_{0}^{\tau}  {\langle 0|\left[ {\left[ {\left( {I_{l,z} I_{k,z}} \right),\Delta h_{IS} \left( {l,k,\tau} \right)} \right],\Delta h_{IS} \left( {l,k,{\tau}'} \right)} \right]\left( { - \left( {I_{l,z} I_{k,z}} \right) + 1/2\left( {I_{l}^{+}  I_{k}^{-} } \right) + 1/2\left( {I_{l}^{-}  I_{k}^{+} } \right)} \right)\,|0\rangle d{\tau}'} 
= 
$$
$$ 
= d\Gamma_{||} \left( {l,\tau} \right)/d\tau + d\Gamma_{||} \left( {k,\tau} \right)/d\tau + d\tilde {\Gamma}_{||} \left( {k,l - k,\tau} \right)/d\tau
$$
\noindent
and
\begin{equation}
\label{eq158}
\mathrm{Re}\partial G^{+ , -} \left( {l,k,\tau} \right)/\partial \tau 
= - 2\mathrm{Re} d\Gamma_{\bot} \left( {l,k,\tau} \right)/d\tau 
= 
\end{equation}
$$ 
= - 4\mathrm{Re}\tr\nolimits_{I} \int\limits_{0}^{\tau}  {\langle 0|\left[ {\left[ {\left( {I_{l}^{+}  I_{k}^{-} } \right),\Delta h_{IS} \left( {l,k,\tau} \right)} \right],\Delta h_{IS} \left( {l,k,{\tau}'} \right)} \right]} \left( { - \left( {I_{l,z} I_{k,z}} \right) + 1/2\left( {I_{l}^{+}  I_{k}^{-} } \right) + 1/2\left( {I_{l}^{-}  I_{k}^{+} } \right)} \right)\,|0\rangle \,d{\tau}' 
= 
$$
$$ 
= - 2\mathrm{Re} d\Gamma_{\bot} \left( {l,\tau} \right)/d\tau 
  - 2\mathrm{Re} d\Gamma_{\bot} \left( {k,\tau} \right)/d\tau 
  - 2\mathrm{Re} d\tilde {\Gamma}_{\bot} \left( {k,l - k,\tau} \right)/d\tau ,
$$
\noindent
where the correlation part of longitudinal two spin relaxation rate is
$$ 
d\tilde {\Gamma}_{||} \left( {k,l - k,\tau} \right)/d\tau
=
$$
\begin{equation}
\label{eq159}
= - 4\tr\nolimits_{I} \int\limits_{0}^{\tau}  {\langle 0|\left[ {\left[ {I_{l,z} ,\Delta h_{IS} \left( {l,\tau} \right)} \right]I_{k,z} ,\Delta h_{IS} \left( {k,{\tau}'} \right)} \right]} \left( {\left( {I_{l}^{+}  I_{k}^{-}} \right) + \left( {I_{l}^{-}  I_{k}^{+} } \right)} \right)\,|0\rangle \,d{\tau}' 
= 
\end{equation}
$$ 
= - \frac{{a^{2}}}{{4}}4\mathrm{Re}\int\limits_{0}^{\tau}  {\langle 0|\left[ {S_{k}^{+} \left( {{\tau}'} \right),S_{l}^{-} \left( {\tau} \right)} \right]|0\rangle}  \,d{\tau}'
$$
\noindent
and the correlation part of transverse two spin decoherence rate is
$$
\mathrm{Re} d\tilde {\Gamma}_{\bot} \left( {k,l - k,\tau} \right)/d\tau 
= 
$$ 
\begin{equation}
\label{eq160}
= - 2\mathrm{Re}\tr\nolimits_{I} \int\limits_{0}^{\tau}  {\langle 0|\left[ {\left[ {I_{l}^{+} ,\Delta h_{IS} \left( {l,\tau} \right)} \right]I_{k}^{-} ,\Delta h_{IS} \left( {k,{\tau}'} \right)} \right]\left( {I_{l,z} I_{k,z}} \right)}  
+ 
\end{equation}
$$
+ \left[ {\left[ {I_{k}^{-}  ,\Delta h_{IS} \left( {k,\tau} \right)} \right]I_{l}^{+}  ,\Delta h_{IS} \left( {l,{\tau}'} \right)} \right]\left( {I_{l,z} I_{k,z}} \right)|0\rangle d{\tau}' 
= 
$$
$$ 
= - \frac{{a^{2}}}{{4}}2\mathrm{Re}\int\limits_{0}^{\tau}  {\langle 0|\left[ {S_{k}^{-} \left( {{\tau}'} \right),S_{l}^{+} \left( {\tau} \right)} \right] + \left[ {S_{k}^{+} \left( {{\tau}'} \right),S_{l}^{-} \left( {\tau} \right)} \right]|0\rangle}  \,d{\tau}' = - \frac{{a^{2}}}{{4}}4\mathrm{Re}\int\limits_{0}^{\tau}  {\langle 0|\left[ {S_{k}^{+} \left( {{\tau}'} \right),S_{l}^{-} \left( {\tau} \right)} \right]|0\rangle}  \,d{\tau}' .
$$
\noindent
So, the correlation part of decoherence rate is equal to the correlation part of longitudinal relaxation rate.

Let us rework next the expression (\ref{eq160}) similarly as it was made in Sec.~6, accounting only low magnon excitation mode with energy $E_{-}$:
$$ 
\mathrm{Re} d\tilde {\Gamma}_{\bot} \left( {k,l - k,\tau} \right)/d\tau
\approx
$$
$$ 
\approx \frac{{a^{2}}}{{\left( {4\pi} \right)^{2}}}2\mathrm{Re}\int {\int\limits_{o}^{\tau}  {u^{*} \left( {\mathbf{q}_{\bot}  ,E} \right)u\left( {{q}'_{x} ,q_{y} ,E} \right)\exp\left( { - \left( {iE_{-}  - s} \right)\left( {{\tau}' - \tau} \right)} \right)\exp\left[ {i\left( {q_{x} k - {q}'_{x} l} \right)} \right]dEd\mathbf{q}_{\bot}  d{q}'_{x} d{\tau}'}}  
= 
$$
\begin{equation}
\label{eq161}
= \frac{{a^{2}}}{{\left( {4\pi} \right)^{2}}}2\mathrm{Re}\int {u^{*} \left( {\mathbf{q}_{\bot}  ,E} \right)u\left( {{q}'_{x} ,q_{y} ,E} \right)\frac{{1 - \exp\left( {\left( {iE_{-}  - s} \right)\tau} \right)}}{{ - \left( {iE_{-}  - s} \right)}}\exp\left[ {i\left( {q_{x} k - {q}'_{x} l} \right)} \right]dEd\mathbf{q}_{\bot}  d{q}'_{x}}  
= 
\end{equation}
$$ 
= \frac{{a^{2}}}{{\left( {4\pi} \right)^{2}}}\int {\frac{{\sqrt {1 + b_{C}^{2}}  + E\left( {\mathbf{q}_{\bot} } \right)}}{{E\left( {\mathbf{q}_{\bot} } \right)}} }  \left\{ { \frac{{ - \left( {E\left( {\mathbf{q}_{\bot} } \right) - b_{\left( {l + k} \right)/2}} \right)\sin\left( {q_{x} \left( {l - k} \right)} \right) + s \cos\left( {q_{x} \left( {l - k} \right)} \right)}}{{\left( {E\left( {\mathbf{q}_{\bot} } \right) - b_{\left( {l + k} \right)/2}} \right)^{2} + s^{2}}}
}\right. 
+ 
$$
$$ 
+
\left[ {
\frac{{\left( {E\left( {\mathbf{q}_{\bot} } \right) - b_{\left( {l + k} \right)/2}} \right)\sin\left( {\left( {E\left( {\mathbf{q}_{\bot} } \right) - b_{\left( {l + k} \right)/2}} \right)\tau - q_{x} \left( {l - k} \right)} \right) }}
{{\left( {E\left( {\mathbf{q}_{\bot} } \right) - b_{\left( {l + k} \right)/2}} \right)^{2} + s^{2}}}
} \right. 
-
$$
$$ 
\left.{ \left.{
- \frac{{ s \cos\left( {\left( {E\left( {\mathbf{q}_{\bot} } \right) - b_{\left( {l + k} \right)/2}} \right)\tau - q_{x} \left( {l - k} \right)} \right) }}
{{\left( {E\left( {\mathbf{q}_{\bot} } \right) - b_{\left( {l + k} \right)/2}} \right)^{2} + s^{2}}}
} \right]
\exp\left( { - s\tau} \right)} \right\} d\mathbf{q}_{\bot}  . 
$$

\begin{figure}
   \begin{center}
   \begin{tabular}{c}
   \includegraphics{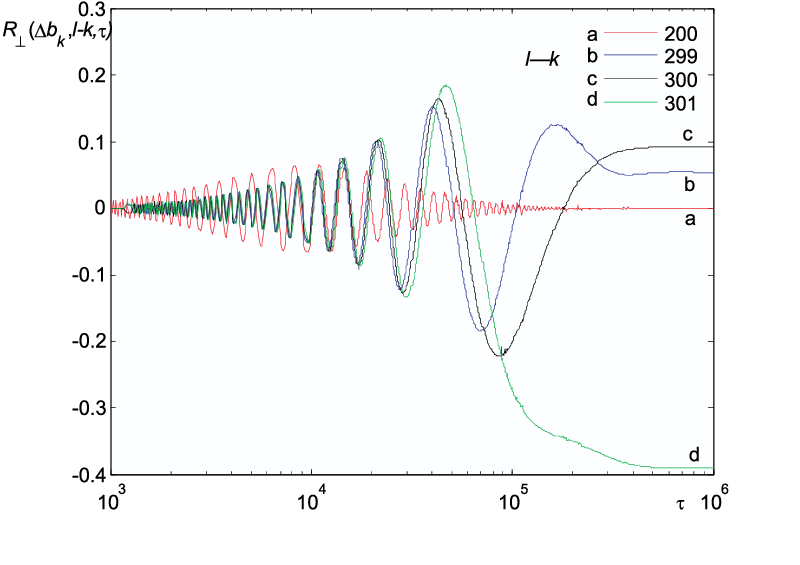}
   \end{tabular}
   \end{center}
   \caption[1]
   {\label{fig:8} 
The $\tau-$dependence for the correlation part of dimensionless decoherence rate $R_{\bot} \left( {\Delta b_{k} ,l - k,\tau} \right)$ for values $\Delta b_{k} = 3.10^{-3}$, $s = 10^{-5}$, $l - k = 200$ (away frow turning point), $l - k = 299, 300, 301$ (in the vicinity of the turning point).
   }
   \end{figure}

Taking next in to account that 
$\int\limits_{0}^{2\pi}  {\sin\left[ {\mathbf{q}_{\bot} \cos \varphi \left( {l - k} \right)} \right]d\varphi} = 0$ 
and introducing again the variable $\xi = E\left( {\mathbf{q}_{\bot} } \right) - b_{C}$, we will write $E\left( {\mathbf{q}_{\bot} } \right) - b_{\left( {l + k} \right)/2} = E\left( {\mathbf{q}_{\bot} } \right) - b_{k} - g\left( {l - k} \right)/2 = \xi + \Delta b_{k} - g\left( {l - k} \right)/2$ and for the correration part of decoherence rate we obtain (Fig.~\ref{fig:8})
\begin{equation}
\label{eq162}
\mathrm{Re} d\tilde {\Gamma}_{\bot} \left( {k,l - k,\tau} \right)/d\tau\, 
\approx \frac{{3a^{2}}}{{2\pi} }R_{\bot} \left( {\Delta b_{k} ,l - k,\tau} \right) 
=
\end{equation}
$$ 
= \int\limits_{0}^{\sqrt {b_{C}^{2} + \pi^{2}/12} - b_{C}}  {\left( {\sqrt {1 + b_{C}^{2}}  + b_{C} + \xi} \right)\,\,Y\left( {\xi + \Delta b_{k} - g\left( {l - k} \right)/2,\,\tau} \right)\,\,J_{0} \left( {\sqrt {12\left[ {\left( {b_{C} + \xi} \right)^{2} - b_{C}^{2}} \right]} \left( {l - k} \right)} \right)d\xi}  . 
$$

The decoherence rates of entangled qubit pair are due to decoherence of one spin states $l$, $k$ and also to the correlation between nuclear spins $l$, $k$. The diagonal and non-diagonal elements of density matrix $1 + G_{z,z} \left( {l,k,\tau} \right)$ and $G^{+ , -} \left( {l,k,\tau} \right)$ 
are decreased with full rates 
$\mathrm{Re} d\Gamma_{||} \left( {l,k,\tau} \right)/d\tau$ and 
$\mathrm{Re} d\Gamma_{\bot} \left( {l,k,\tau} \right)/d\tau$. 
Note, that the asymptotic value for correlation part of decoherence rate (Fig.~\ref{fig:8}) may change sign, if $\Delta b_{k} - g\left( {l - k} \right)/2 < 0$, that is for states after ``turning point'' (Eq.(\ref{eq78})).

In the case being considered the concurrence for entangled two-qubit state can be obtained by using the Wootters formula (Ref.\cite{22})
\begin{equation}
\label{eq163}
C\left( {l,k,\tau} \right) 
= 1/2 \max \{ |G^{+ , -} \left( {l,k,\tau} \right)| - \left( {1 + G_{z,z} \left( {l,k,\tau} \right)} \right);\,\,0\} .
\end{equation}

Taking into account that 
$\mathrm{Re}\Gamma_{\bot} \left( {k,\tau} \right) = \Gamma_{||} \left( {k,\tau} \right)$,  
$|G^{+ , -} \left( {l,k,\tau} \right)| \approx 2\exp\left( { - \mathrm{Re}\Gamma_{\bot} \left( {l,k,\tau} \right)} \right)$,
$G_{z,z} \left( {l,k,\tau} \right) \approx - \exp\left( { - \Gamma_{||} \left( {l,k,\tau} \right)} \right)$ and also Eqs (\ref{eq155}), (\ref{eq156}), (\ref{eq158}), we will obtain
\begin{equation}
\label{eq164}
C\left( {l,k,\tau} \right) = 1/2\{ 3\exp\left( { - \mathrm{Re}\Gamma_{\bot} \left( {l,\tau} \right) - \mathrm{Re}\Gamma_{\bot} \left( {k,\tau} \right) - \mathrm{Re}\tilde {\Gamma}_{\bot} \left( {k,l - k,\tau} \right)} \right) - 1\} .
\end{equation}

For the logarithmic damping rate of concurrence we will then write
\begin{equation}
\label{eq165}
d \ln C\left( {l,k,\tau} \right)/d\tau 
= - 3/2d\left( {\mathrm{Re}\Gamma_{\bot} \left( {l,\tau} \right) + \mathrm{Re}\Gamma_{\bot} \left( {k,\tau} \right) + \mathrm{Re}\tilde {\Gamma}_{\bot} \left( {k,l - k,\tau} \right)} \right)d\tau . 
=
\end{equation}
$$ 
= - \frac{{9a^{2}}}{{2\pi} }\left( {R_{\bot} \left( {\Delta b_{l} ,\tau} \right) + R_{\bot} \left( {\Delta b_{k} ,\tau} \right) + R_{\bot} \left( {\Delta b_{k} ,l - k,\tau} \right)} \right). 
$$
For $\Delta b_{k} > 0$ parameter $\Delta b_{l} = \Delta b_{k} - g\left( {l - k} \right)$ changes the sign and parameter $s^{2}$ in denominator may not be omitted. Using next Eq.(\ref{eq133}) we will write
\begin{equation}
\label{eq166}
R_{\bot} \left( {\Delta b_{l} ,\tau} \right) = \int\limits_{0}^{\sqrt {b_{C}^{2} + \pi^{2}/12} - b_{C}}  {\left( {\sqrt {1 + b_{C}^{2}}  + b_{C} + \xi} \right)\,\,Y\left( {\xi + \Delta b_{k} - g\left( {l - k} \right),\tau} \right)\,d\xi}  
\end{equation}

\begin{figure}
   \begin{center}
   \begin{tabular}{c}
   \includegraphics{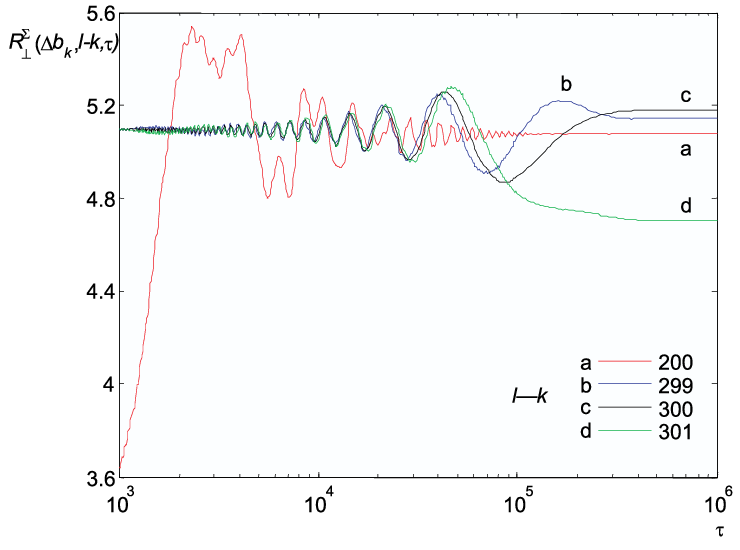}
   \end{tabular}
   \end{center}
   \caption[1]
   {\label{fig:9} 
The $\tau-$dependence of $R_{\bot}^{\Sigma} \left( {\Delta b_{k} ,l - k,\tau} \right)$ for values $\Delta b_{k} = 3.10^{-3}$, $s = 10^{-5}$, $l - k = 200$ (away frow turning point), $l - k = 299, 300, 301$ (in the vicinity of the turning point).
   }
   \end{figure}

Finally, for the value of concurrence-damping rate in the context of second order of permutation theory we will obtain
\begin{equation}
\label{eq167}
dC\left( {l,k,\tau} \right)/d\tau = 
- \frac{{9a^{2}}}{{2\pi} }R_{\bot}^{\Sigma} \left( {\Delta b_{k} ,l - k,\tau} \right),
\end{equation}
\noindent
where (Fig.~\ref{fig:9})
$$ 
 R_{\bot}^{\Sigma} \left( {\Delta b_{k} ,l - k,\tau} \right) 
= R_{\bot} \left( {\Delta b_{k} ,\tau} \right) 
+ R_{\bot} \left( {\Delta b_{l} ,\tau} \right) 
+ R_{\bot} \left( {\Delta b_{k} ,l - k,\tau} \right) 
=
$$
\begin{equation}
\label{eq168}
= \int\limits_{0}^{\sqrt {b_{C}^{2} + \pi^{2}/12} - b_{C}}  {\left( {\sqrt {1 + b_{C}^{2}}  + b_{C} + \xi} \right) \cdot \left[ {Y\left( {\xi + \Delta b_{k} ,\tau} \right) + Y\left( {\xi + \Delta b_{k} - g\left( {l - k} \right),\tau} \right) 
}\right.}
+
\end{equation}
$$ 
+ 
{\left.{
Y\left( {\xi + \Delta b_{k} - g\left( {l - k} \right)/2\,,\,\,\tau} \right)\, J_{0} \left( {\sqrt {12\left[ {\left( {b_{C} + \xi} \right)^{2} - b_{C}^{2} \left( {l - k} \right)} \right]\,\,} }\right) } \right]\,d\xi} .   
$$

The velocity damping of concurrence tends to constant for $\tau \to \infty $:
$$ 
R_{\bot}^{\Sigma} \left( {\Delta b_{k} ,l - k,\infty} \right) 
= 
$$
\begin{equation}
\label{eq169}
= s\int\limits_{0}^{\sqrt {b_{C}^{2} + \pi^{2}/12} - b_{C}}  {\left( {\sqrt {1 + b_{C}^{2}}  + b_{C} + \xi} \right)\left[ {\frac{{1}}{{\left( {\xi + \Delta b_{k}} \right)^{2}}}\, + \frac{{1}}{{\left( {\xi + \Delta b_{k} - g\left( {l - k} \right)} \right)^{2}}}\,\, 
}\right.}
+  
\end{equation}
$$ 
{\left.{
+ \frac{{J_{0} \left( {\sqrt {12\left[ {\left( {b_{C} + \xi} \right)^{2} - b_{C}^{2}} \right]} \left( {l - k} \right)} \right)}}{{\left( {\xi + \Delta b_{k} - g\left( {l - k} \right)/2} \right)^{2}}}\,\,} \right]\,d\xi } > 0. 
$$

\section{Appendixes}

\subsection*{A1. Determination of the coefficient $A$ in Eq.(\ref{eq52})}

Substituting Eqs.(\ref{eq50}), (\ref{eq51}) in condition (\ref{eq32}), we will obtain the equation for determination of the quantity $A$:
$$
\frac{{2EA^{2}}}{{\sqrt {1 + b_{C}^{2}}  + E}} 
\int_{- \infty }^{\infty}  {\exp\left\{ { - \frac{{i}}{{g}}\left( {\left( {E - {E}'} \right)q_{x}} \right)}\right\} \cdot M\left( {{E}',\mathbf{q}_{\bot} } \right) dq_{x}} + O\left( {g} \right) 
= \delta \left( {E - {E}'} \right), 
\eqno (\mathrm{A}1.1)
$$
\noindent
where
$$ 
M\left( {E,\mathbf{q}_{\bot} } \right) 
= \sum\limits_{Q_{x}}  {\exp\left\{ - \frac{{i}}{{g}}\left( {\int\limits_{q_{x}}^{q_{x} + Q_{x}}  {E\left( {\mathbf{q}_{\bot} } \right)\;dq_{x}}  - EQ_{x}} \right)\right\} }
=
$$
$$
= \sum\limits_{m = - \infty}^{\infty}  {\exp\left\{ - \frac{{i}}{{g}}\left( {\int\limits_{q_{x} }^{q_{x} + 2\pi m} {E\left( {\mathbf{q}_{\bot} } \right)\;dq_{x}}  - E2\pi m} \right)\right\}}  .
\eqno (\mathrm{A}1.2)
$$

To achieve the convergence of summing over $m$ in Eq.(A1/2), we will introduce a weak damping:
$$
iEm \to iEm - \nu |m|,\,\,\,\,\nu \to + 0. 
\eqno (\mathrm{A}1.3)
$$

Using Poisson summing formula (Ref.\cite{23}, Eqs.~4.8-9) and performing the rearrangement of this sum to integral over continuous variable $2\pi m \to g\mu$, we obtain the following asymptotical expression
$$ 
 M\left( {E,\mathbf{q}_{\bot} } \right)\, = \frac{{1}}{{2\pi} }\lim\limits_{\nu \to + 0} \,g\sum\limits_{p 
= - \infty}^{\infty}  {\int_{- \infty}^{\infty}  {\exp\{ - i\left( {\int\limits_{q_{x}}^{q_{x} + g\mu}  {E\left( {\mathbf{q}_{\bot} } \right)dq_{x}}  - E\,\mu - p\mu} \right) - \nu |\mu |}}  \} \,d\mu 
\approx 
$$
$$
\approx \frac{{1}}{{2\pi} }\lim\limits_{\nu \to + 0} \,\int_{- \infty}^{\infty}  {\exp\{ - i\left( {E\left( {\mathbf{q}_{\bot} } \right) - E} \right)\,\mu /g - \nu |\mu |/g\} \,d\mu}  + O\left( {g} \right)\} 
= g\left[ {\delta \left( {E - E\left( {\mathbf{q}_{\bot} } \right)} \right) + O\left( {g} \right)} \right]. 
\eqno (\mathrm{A}1.4) 
$$

The last result ignores here the fast oscillating terms with $p \ne 0$.

On the other hand, we may write Eq.(A1.4) in the form
$$ 
M\left( {E,\mathbf{q}_{\bot} } \right) = \lim\limits_{\nu \to + 0} \,\sum\limits_{m = - \infty}^{\infty}  {\exp\{ - \frac{{i}}{{g}}\left( {\int\limits_{q_{x}}^{q_{x} + 2\pi m} {E\left( {\mathbf{q}_{\bot} } \right)\;dq_{x}}  - E2\pi m} \right) - \frac{{1}}{{g}}\nu 2\pi |m|\}}  + O\left( {g} \right) 
=
$$
$$
= 1 + 2\mathrm{Re}\lim\limits_{\nu \to + 0} \,\sum\limits_{m = 1}^{\infty}  {\exp\left\{ { - \frac{{i}}{{g}}\left( {\int\limits_{q_{x}}^{q_{x} + 2\pi m} {E\left( {\mathbf{q}_{\bot} } \right)\;dq_{x}}  - E2\pi m - \nu 2\pi m} \right)} \right\}} + O\left( {g} \right)
\approx 1 + O\left( {g} \right) , 
\eqno (\mathrm{A}1.5) 
$$

\noindent
where the oscillating sum is omitted.

Comparison of Eq.(A1.4) with Eq.(A1.5) allows making the conclusion:
$$
\lim\limits_{
E \to E\left( {\mathbf{q}_{\bot} } \right), g \to 0} g \delta \left( {E - E\left( {\mathbf{q}_{\bot} } \right)} \right) \to 1 
\eqno (\mathrm{A}1.6) 
$$

To achieve the convergence of integrating over $q_{x}$ in Eq.(A1.1) we introduce again a weak damping:
$$
i{E}'q_{x} \to i{E}'q_{x} - s|q_{x} |,\,\,\,\,s \to + 0. 
\eqno (\mathrm{A}1.7) 
$$

Let us next extend the integration limits over ``extended'' variable $\eta = q_{x} /g$ to$ \pm \infty$. Then the expression in the left part of Eq.(A1.1) takes the form
$$ 
\lim\limits_{s \to + 0} \,\int_{- \infty}^{\infty}  {\exp\{ - \frac{{i}}{{g}}\left( {E - {E}'} \right)q_{x} - s|q_{x} |\} \cdot M\left( {{E}',\mathbf{q}_{\bot} } \right)} dq_{x} 
=
$$
$$
= \lim\limits_{s \to + 0} \,g\int_{- \infty}^{\infty}  {\exp\{ i\left( {E - {E}'} \right)\eta - s|\eta |\} \cdot \left( {1 + O\left( {g} \right)} \right)} d\eta 
= 2\pi g\delta \left( {E - {E}'} \right)\left( {1 + O\left( {g} \right)} \right), 
\eqno (\mathrm{A}1.8)
$$

\noindent
where result (A1.5) and the relation
$$
\lim\limits_{s \to + 0} \int_{- \infty}^{\infty}  {\exp\left[ {i\left( {E - {E}'} \right)\eta - s|\eta |} \right]d\eta = 2\pi \delta \left( {E - {E}'} \right)} 
\eqno (\mathrm{A}1.9)
$$

\noindent
was used.

For determination of value of $A^{2}$ we will now obtain from Eqs.(A1.1),(A1.8), the equation:
$$
\frac{{4\pi gEA^{2}}}{{\left( {1 + b_{\mathrm{A}}} \right) + E}} = 1 
\eqno (\mathrm{A}1.10) 
$$

\subsection*{A2. The calculation of indirect nuclear spin interaction}

After commutator transformation in Eq.(\ref{eq65}) we will find
$$ 
h_{II} \left( {k,l} \right) 
\approx 
$$
$$ 
\approx - ia^{2}/4\{ I^{+} \left( {k} \right)I^{-} \left( {l} \right)\int\limits_{- \infty}^{0} {\langle 0|\left[ {S^{-} \left( {\tau ,k} \right),\,\,S^{+} \left( {\tau + {\tau}',l} \right)} \right]|0\rangle \,\,\exp\left( {s{\tau}'} \right)d{\tau}'} 
+ 
$$
$$
+ I^{+} \left( {l} \right)I^{-} \left( {k} \right)\int\limits_{- \infty }^{0} {\langle 0|\left[ {S^{-} \left( {\tau ,l} \right),\,\,S^{+} \left( {\tau + {\tau}',k} \right)} \right]|0\rangle \,\,\exp\left( {s{\tau}'} \right)d{\tau}'} \} \quad + \mathrm{H.c.} 
- 
\eqno (\mathrm{A}2.1) 
$$
$$ 
- ia^{2}/4\sum\limits_{j = k,l} {\int\limits_{- \infty}^{0} {\{ 1/2\,\,\langle 0|\left[ {S^{-} \left( {\tau ,j} \right),\,\,S^{+} \left( {\tau + {\tau}',j} \right)} \right]|0\rangle}  }
+  
$$
$$ 
+ I_{z} \left( {j} \right)\langle 0|S^{-} \left( {\tau ,j} \right)S^{+}\left( {\tau + {\tau}',j} \right) + S^{+} \left( {\tau + {\tau}',j} \right)S^{-} \left( {\tau ,j} \right)|0\rangle \,\exp\left( {s{\tau}'} \right) \} d{\tau}' + \mathrm{H.c.}, 
$$
\noindent
where it was accounted that
$$
I^{\pm} \left( {j} \right)I^{\mp} \left( {j} \right) 
= 1/2 \pm I_{z} \left( {j} \right),\,\,\,\,\,\,\left[ {I^{\pm} \left( {j} \right),I^{\mp }\left( {j} \right)} \right] 
= \pm 2I_{z} \left( {j} \right), 
\eqno (\mathrm{A}2.2) 
$$
$$ 
I_{z} \left( {j} \right)I^{\pm} \left( {j} \right) 
= - I^{\pm} \left( {j} \right)I_{z} \left( {j} \right) 
= \pm 1/2I^{\pm} \left( {j} \right),\,\,\,\,\,\left[ {I_{z} \left( {j} \right),I^{\pm} \left( {j} \right)} \right] 
= \pm I^{\pm} \left( {j} \right).  
$$

The mean value of commutator of electron spin operators takes the form
$$ 
\langle 0|\left[ {S^{-} \left( {\tau ,k} \right),\,\,S^{+} \left( {\tau + {\tau }',l} \right)} \right]|0\rangle 
= 
$$
$$ 
= \frac{{1}}{{\left( {2\pi} \right)^{2}}}\int {\{ u^{*} \left( {\mathbf{q}_{\bot} ,E} \right)\,u\left( {{\mathbf{q}}'_{\bot}  ,{E}'} \right)\,\langle 0|\left[ {\xi^{+} \left( {q_{y} ,E_{-} } \right),\xi \left( {{q}'_{y} ,{E}'_{-} } \right)} \right]|0\rangle \exp\left( { - i{E}'_{-}  {\tau}' + s{\tau}'} \right) }
+ 
$$
$$ 
+ v\left( {\mathbf{q}_{\bot}  ,E} \right)v^{*} \left( {{\mathbf{q}}'_{\bot}  ,{E}'} \right)\langle 0|\left[ {\xi \left( {q_{y} ,E_{+} } \right),\xi^{+}\left( {{q}'_{y} ,{E}'_{+} } \right)} \right]|0\rangle \exp\left( {i{E}'_{+}  {\tau}' + s{\tau}'} \right)\} 
\cdot 
\eqno (\mathrm{A}2.3) 
$$
$$ 
\cdot \exp\left[ {i\left( {q_{x} k - {q}'_{x} l} \right)} \right]\;dEd{E}'d\mathbf{q}_{\bot} d{\mathbf{q}}'_{\bot}  . 
$$

We will use next the mean value on ground state of commutator of magnon operators (\ref{eq21}), written for $\tau = 0$:
$$
\left\langle {0|\left[ {\xi \left( {q_{y} ,E} \right),\,\;\xi^{+} \left( {{q}'_{y} ,{E}'} \right)} \right]|\left. {0} \right\rangle} \right. 
= \left\langle {0|\xi \left( {q_{y} ,E} \right)\xi^{+} \left( {{q}'_{y} ,{E}'} \right)|\left. {0} \right\rangle} \right. 
= \delta \left( {q_{y} - {q}'_{y}} \right)\delta \left( {E - {E}'} \right) 
\eqno (\mathrm{A}2.4) 
$$
\noindent
and thus we will have
$$ 
\langle 0|\left[ {S^{-} \left( {\tau ,k} \right),\,\,S^{+} \left( {\tau + {\tau }',l} \right)} \right]|0\rangle 
= 
$$
$$ 
= \frac{{1}}{{\left( {2\pi} \right)^{2} }} \int {\{ - u^{*} \left( {\mathbf{q}_{\bot}  ,E} \right)\,u\left( {{q}'_{x} ,q_{y} ,E} \right)\,\exp\left( { - iE_{-}  {\tau}' + s{\tau}'} \right) }
+
\eqno (\mathrm{A}2.5) 
$$
$$ 
+ v\left( {\mathbf{q}_{\bot}  ,E} \right)v^{*} \left( {{q}'_{x} ,q_{y} ,E} \right)\exp\left( {iE_{+}  {\tau}' + s{\tau}'} \right)\} \cdot \exp\left[ {i\left( {q_{x} k - {q}'_{x} l} \right)} \right]\;dEd\mathbf{q}_{\bot}  d{q}'_{x} . 
$$

We will find then for mean value of the electron spin operators product the expression:
$$ 
 \langle 0|S^{-} \left( {\tau ,j} \right)S^{+}\left( {\tau + {\tau}',j} \right) + S^{+} \left( {\tau + {\tau}',j} \right)S^{-} \left( {\tau ,j} \right)|0\rangle 
=
$$
$$ 
= \frac{{1}}{{\left( {2\pi} \right)^{2}}}\int {\left\{ { u^{*} \left( {\mathbf{q}_{\bot}  ,E} \right)\,u\left( {{q}'_{x} ,q_{y} ,E} \right)\,\exp\left( { - iE_{-}  {\tau}' + s{\tau}'} \right) } 
\right. }
+
\eqno (\mathrm{A}2.6)
$$
$$ 
{\left.{
+ v\left( {\mathbf{q}_{\bot},E} \right)\,v^{*} \left( {{q}'_{x},q_{y},E} \right) \exp\left( {iE_{+}  {\tau}' + s{\tau}'} \right) }\right\} \exp\left( {i\left( {q_{x} - {q}'_{x}} \right)j} \right)\;dEd\mathbf{q}_{\bot}  d{q}'_{x} 
}
$$
\noindent
and Eq.(A2.1) takes the form:
$$ 
h_{II} \left( {k,l} \right) 
=
$$
$$ 
= - i\frac{{a^{2}}}{{4\left( {2\pi} \right)^{2}}}
\int {\int\limits_{- \infty}^{0} { \{ 
I^{+} \left( {k} \right)I^{-} \left( {l} \right)
\left[ { - u^{*} \left( {\mathbf{q}_{\bot},E} \right)u\left( {{q}'_{x} ,q_{y} ,E} \right)\exp\left( { - iE_{-}  {\tau}' + s{\tau}'} \right) } \right. }}
+ 
$$
$$ 
\left.{
+ v^{*} \left( {\mathbf{q}_{\bot},E} \right)v\left( {{q}'_{x} ,q_{y} ,E} \right)\exp\left( {iE_{+}{\tau}'+s{\tau}'} \right)} \right]\exp\left[ {i\left( {q_{x} k - {q}'_{x} l} \right) } \right]
+
$$
$$
+ I^{+} \left( {l} \right)I^{-} \left( {k} \right)
\left[ { - u^{*} \left( {\mathbf{q}_{\bot},E} \right)u\left( {{q}'_{x} ,q_{y} ,E} \right)\exp\left( { - iE_{-}  {\tau}' + s{\tau}'} \right) } \right. 
+
\eqno (\mathrm{A}2.7)
$$
$$ 
{ \left.{
+ v^{*} \left( {\mathbf{q}_{\bot}  ,E} \right)v\left( {{q}'_{x} ,q_{y} ,E} \right)\exp\left( {iE_{+}  {\tau}' + s{\tau}'} \right)} \right]\exp\left[ {i\left( {q_{x} l - {q}'_{x} k} \right) }\right] \} dEd\mathbf{q}_{\bot}  d{q}'_{x} d{\tau}' 
} 
- 
$$
$$ 
- i\frac{{a^{2}}}{{4\left( {2\pi} \right)^{2}}} 
\int {\int\limits_{- \infty}^{0} { 
\sum\limits_{j = k,l} {
\{ 
1/2 \left[ { - u^{*} \left( {\mathbf{q}_{\bot} ,E} \right)u\left( {{q}'_{x} ,q_{y} ,E} \right)\exp\left( { - iE_{-}  {\tau}' + s{\tau}'} \right) }    
\right. }}} 
+
$$
$$ 
\left. {
+ v\left( {\mathbf{q}_{\bot}  ,E} \right)\,v^{*} \left( {{q}'_{x} ,q_{y} ,E} \right)\,\exp\left( {iE_{+}  {\tau}' + s{\tau}'} \right)} \right] 
+ 
$$
$$ 
+ I_{z} \left[ {u^{*} \left( {\mathbf{q}_{\bot}  ,E} \right)u\left( {{q}'_{x} ,q_{y} ,E} \right)\exp\left( { - iE_{-}  {\tau}' + s{\tau}'} \right) 
} \right.
+
$$
$$ 
{ \left.{ 
+ v\left( {\mathbf{q}_{\bot}  ,E} \right)\,v^{*} \left( {{q}'_{x} ,q_{y} ,E} \right)\,\exp \left( {iE_{+}  {\tau}' + s{\tau}'} \right)} \right] \exp\left[ { i\left( {q_{x} - {q}'_{x}} \right)j} \right]\} dEd\mathbf{q}_{\bot}  d{q}'_{x} d{\tau}' }+ \mathrm{H.c.}
$$

After integrating of Eq.(A2.7) over ${\tau}'$, we will obtain the expression for the effective two nuclear spins Hamiltonian for atoms $k$ and $l$ among the same sublattice \textbf{A} as a correction of the second order perturbation theory to the energy of antiferromagnet ground state:
$$
h_{II} \left( {k,l} \right) 
= 
\eqno (\mathrm{A}2.8) 
$$
$$ 
= - \sum\limits_{j = k,l} {\{ \left[ {\omega_{I} \left( {j} \right) - a/2 - W\left( {j} \right)} \right]I_{z} + \,\,U\left( {j,j} \right)/2\} } 
- U\left( {k,l} \right)\left[ {I^{-} \left( {k} \right)I^{+} \left( {l} \right) + I^{+} \left( {k} \right)I^{-} \left( {l} \right)} \right], 
$$
\noindent
where the expressions for indirect interaction between two separated nuclear spins, belonging to common sublattice and generalizing the known Nakamura's expression for antiferromagnet in homogeneous fields \cite{7} and for correction for nuclear spin resonance frequency have the forms
$$
U\left( {k,l} \right) 
= 
\eqno (\mathrm{A}2.9)
$$
$$ 
= \frac{{a^{2}}}{{2\left( {2\pi} \right)^{2}}}
\mathrm{Re}\int {\left[ {\frac{{u^{*} \left( {q_{x} ,q_{y} ,E} \right)u\left( {{q}'_{x} ,q_{y} ,E} \right)}}{{E_{-}  + is}} 
- \frac{{v\left( {q_{x} ,q_{y} ,E} \right)v^{*} \left( {{q}'_{x} ,q_{y} ,E} \right)}}{{E_{+}  - is}}} \right]\exp\left[ {i\left( {q_{x} k - {q}'_{x} l} \right)} \right]\;dEd\mathbf{q}_{\bot}  d{q}'_{x} },
$$ 
\noindent
and
$$
W\left( {j} \right) 
=
\eqno (\mathrm{A}2.10) 
$$
$$ 
= \frac{{a^{2}}}{{2\left( {2\pi} \right)^{2}}}
\mathrm{Re}\int {\left[ {\frac{{u^{*} \left( {q_{x} ,q_{y} ,E} \right)u\left( {{q}'_{x} ,q_{y} ,E} \right)}}{{E_{-}  + is}} 
+ \frac{{v\left( {q_{x} ,q_{y} ,E} \right)v^{*} \left( {{q}'_{x} ,q_{y} ,E} \right)}}{{E_{+}  - is}}} \right]\exp\left[ {i\left(  {q_{x} - {q}'_{x}} \right) j } \right]\;dEd\mathbf{q}_{\bot}  d{q}'_{x} }.
$$

\section{Conclusion}

In this paper it was considered, as a quantum register, one-dimensional chain of magnetic atoms with nuclear spins, which is placed by regular way in thin plate of easy-axis 3D antiferromagnet.

When the external magnetic field is directed along the easy axis normally to the plane of the plate and has a constant gradient along the nuclear spin chain, antiferromagnet spin Hamiltonian in the case of inhomogeneous external field in spin-wave approximation was obtained. If the field has a weak gradient, asymptotic expression for coefficients of unitary transformations to diagonal form of this Hamiltonian was found.

The expression for indirect inter-spin coupling that is due to hyperfine nuclear-electron coupling in atoms and spin-wave propagation in antiferromagnet was evaluated and it was shown that owing to gradient of external field significantly change. So, in the case that the value of local field in the middle point position for two spins coincides which critical field for homogeneous phase transition, the indirect interaction fast grows and, even so, it takes the oscillating character of distance dependence. We have denoted the corresponding points as ``turning points''. In conditions of homogeneous AFR the indirect interaction has additional turning points, which determined by frequency and microwave power.

The nonadiabatic mechanism of nuclear spin statesdecoherence in quantum register caused by interaction of nuclear spins with magnon excitations was considered and the calculations of one qubit and two qubit relaxation and decoherence rates were made.

It turns out that the character of decoherence processes essentially depends on antiferromagnet anisotropy (parameter $b_{C} $) and on inhomogeneity of external field (parameter $g$). As this takes place, the temperature whereby the thermal magnon excitations are excluded and the two-magnon spin-lattice relaxation is especially suppressed, should be defined by values $T \ll T_{C} \left( {1 - b/b_{C}} \right),\,\,\,T_{C} = \hbar \gamma_{S} B_{E} /k_{\mathrm{B}} B_{C}$. As an example we have also considered decoherenc of pair qubits maximally entanglement state and have calculated the concurrence damping rate.

Due to the availability of inhomogeneous external field, it becomes possible to control not only the individual nuclear spin resonance frequency but also the interaction between spatially separated spins in large-scale quantum register without resort to controlling many gate systems. The switching of interaction between far removed spin-qubit, required for two-qubit operation, in considered register may be performed by the tuning of qubit state to turning point state, where indirect interqubit interaction has a large value. The external magnetic field, its gradient and microwave power may play the role of control parameters.

For the examination of realization problems of considered model for large scaled quantum computer, as in the case of any other model, one should refer to five necessary fundamental requirements given in paper \cite{24}. One of the most difficult problems in the realization of quantum computer is the problem of quantum state initialization. For its solution, it is believed to use the method of dynamical polarization of nuclear spin in antiferromagnet for homogeneous field condition in the manner similar to the method suggested for semiconductors in book \cite{11} and the scheme of optical pumping, as the one discussed in paper \cite{25}. The second difficult problem is the realization of readout processes, for this purpose an ensemble approach scheme like the one considered in book \cite{11} may be suggested.

\section*{Acknowledgments}

We are thankful to K.A.Valiev for attention to this work and for fruitful discussions and V.V.Vyurkov, M.I.Kurkin and E.B.Fel'dman for useful comments. This work was partially supported by Russian Foundation for Basic Research under projects 05-02-17412-a, 06-07-89129-a and 08-07-00481-a.

\end{document}